\documentclass[12pt]{iopart}

\usepackage{iopams}

\usepackage{graphicx}
 \usepackage{latexsym}
    \usepackage{tabularx}
    
    \usepackage{color}
 
\newcommand{\bm}[1]{\boldsymbol{#1}}
\def\l{\left}
\def\r{\right}
\def\DM{\mathrm{d}}
\def\hatxi{\widehat{\xi}}
\def\hatn{\widehat{n}}
\def\eq#1{{Eq.~(\ref{#1})}}

\def\scm{{Schwarzschild\ metric}}
\def\G#1#2#3{{\Gamma^#1_{#2#3}}}
\def\mes#1#2{\int_{#2} d^{#1}x\, \sqrt{-g}\, }
\def\fig#1{{Fig.~\ref{#1}}}

\def\g{\sqrt{-g}\,}
\def\G#1#2#3{{\Gamma^#1_{#2#3}}}

\def\scm{{Schwarzschild\ metric}}

\newcommand{\LL}{Lanczos-Lovelock}
\def\mes#1#2{\int_{#2} d^{#1}x\, \sqrt{-g}\, }

\newcommand{\Cal}[1]{\ensuremath{\mathcal{#1}}}
\newcommand{\ph}[1]{\phantom{#1}}
\newcommand{\D}{\ensuremath{\nabla}}
\newcommand{\AltC}[8]{\ensuremath{\delta^{#1 #2 #3... #4}_{#5 #6 #7
      ... #8}}} 
\newcommand{\Alt}[6]{\ensuremath{\delta^{#1 #2 ... #3}_{#4 #5
      ... #6}}} 
       
\newcommand{\Riem}[4]{\ensuremath{R^{#1 #2}_{#3 #4}}}
\newcommand{\LDm}{\ensuremath{L_{(m)}}}
\newcommand{\sD}[1]{\sum_{m=1}^{K}{#1}}
\newcommand{\dV}{\ensuremath{\partial\Cal{V}}}
\newcommand{\cc}{cosmological constant}

\begin{document}

\review{Thermodynamical Aspects of Gravity:
 New insights}

\author{T.Padmanabhan}

\address{IUCAA, Pune University Campus, Ganeshkhind, Pune 411007, INDIA.}
\ead{paddy@iucaa.ernet.in}

\begin{abstract}
The fact that one can associate thermodynamic properties with horizons brings together
principles of quantum theory, gravitation and thermodynamics and  possibly offers a window to the nature of quantum geometry. This review discusses certain aspects of this topic concentrating on new insights gained from some recent work.
 After a brief introduction of the overall perspective, 
 Sections 2 and 3 provide the pedagogical background on the geometrical features of bifurcation horizons,  path integral derivation of horizon temperature, black hole evaporation, structure of \LL\ models, the concept of Noether charge and its relation to horizon entropy. 
Section \ref{sec:thermosecond} discusses several conceptual issues introduced by the existence of temperature and entropy of the horizons. In Section 5 we take up the connection between horizon thermodynamics and gravitational dynamics and describe several peculiar features
 which have no simple interpretation
in the conventional approach. The next two sections  describe the recent progress achieved in an alternative perspective of gravity. In Section \ref{sec:emergentspacetime} we provide a thermodynamic interpretation of the field equations of gravity in any diffeomorphism invariant theory and  in Section \ref{sec:gravitystory}  we obtain the field equations of gravity from an entropy maximization principle. The last section provides a summary. 
\end{abstract}

\maketitle

\section{Introduction and perspective}\label{sec:intro}

Soon after Einstein developed  the gravitational field equations, Schwarzschild found the simplest
exact solution to these equations, describing a spherically symmetric spacetime. When expressed in a natural coordinate system which makes the symmetries of the solution obvious, it leads to the line interval:
\begin{equation}
ds^2= -f(r)dt^2+\frac{dr^2}{f(r)}+r^2(d\theta^2+\sin^2\theta d\phi^2)
\end{equation} 
where $f(r)\equiv 1-(r_g/r)$ with
$r_g \equiv 2GM/c^2=2M$, in units with $G=c=1$. It was immediately noticed that this metric
exhibits a curious pathology. One of the metric coefficients, $g_{tt}$ 
vanished on a surface $\mathcal{H}$,
of finite area $4\pi r_g^2$, given by  $r=r_g$, while another metric coefficient $g_{rr}$ diverged on the same surface. After some initial confusion, it was realized that
the singular behaviour of the metric is due to bad choice of coordinates and that the spacetime geometry is well behaved at $r=r_g$. However, the surface $\mathcal{H}$ acts as a horizon blocking the propagation of information from the region $r<r_g$ to the region $r>r_g$. This leads to several new features in the theory, many of which, even after decades of investigation, defies a complete understanding. The most important amongst them is the relationship between physics involving horizons and thermodynamics.

This connection, between black hole dynamics involving the horizon and the laws of thermodynamics, became apparent as  a result of the research  in early 1970s.
Hawking proved \cite{areatheorem} that in any classical process involving the black holes, 
the sum of the areas of a black hole horizons cannot decrease which, with hindsight, is reminiscent of the behaviour of entropy in classical thermodynamics. This connection was exploited by Bekenstein in his response to a  conundrum raised by John Wheeler.
Wheeler pointed out  that  an external observer can drop material with non-zero entropy into the inaccessible region beyond the horizon thereby reducing the entropy accessible to outside observers.\footnote{John Wheeler posed this as a question to Bekenstein:  What happens if you mix cold and hot tea and pour it down a horizon, erasing all traces of ``crime" in increasing the entropy of the world? This is based on what Wheeler told me in 1985, from \textit{his} recollection of events; it is also mentioned in his book, see page 221 of Ref. \cite{wheeler}.
 I have heard somewhat different versions from other sources.}
Faced with this difficulty, Bekenstein came up with the idea that the black hole horizon should be attributed an entropy which is proportional to its area \cite{Bekenstein:1972tm,Bekenstein:1973ur,Bekenstein:1974ax}. 
It was also realized around this time that one can formulate four laws of black hole dynamics in a manner analogous to the laws of thermodynamics \cite{Bardeen:1973gs}. In particular, if a physical process (say, dropping of small amount of matter into the Schwarzschild black hole) changes the mass of a black hole by $\delta M$ and the area of the event horizon by
$\delta A$, then it can be proved that
\begin{equation}
 \delta M = \frac{\kappa}{8\pi} \delta A = \frac{\kappa}{2\pi}\delta\left(\frac{A}{4}\right)
 \label{firstlaw}
\end{equation} 
where $\kappa = M/(2M)^2 = 1/4M$ is called the surface gravity of the horizon.
This suggests an analogy with the thermodynamic law $\delta E = T \delta S$ with 
$S\propto A$ and $T\propto \kappa$.  However, classical considerations alone cannot determine the proportionality constants. (We will see later that quatum mechanical considerations suggest $S=(A/4)$ and
$T=\kappa/2\pi$, which is indicated in the second equality in \eq{firstlaw}.) 

In spite of this, 
Bekenstein's idea did \textit{not} find favour with the community  immediately; the fact that laws of black hole dynamics have an uncanny similarity with the laws of 
thermodynamics  
was initially considered to be only a curiosity. 
(For a taste of history, see e.g \cite{kst}.)
The key objection at that time  was the following. If black holes possess entropy as well as energy (which they do), then they must have a non-zero temperature and must radiate --- which seemed to contradict the view that nothing can escape the black hole horizon. Investigations  by Hawking, however, led  to the discovery that a non-zero temperature should be attributed to the 
black hole horizon \cite{Hawking:1975sw}. 
He found that black holes, formed by the collapse of matter, will radiate particles with a thermal spectrum at late times, as detected by a stationary observer at large distances.
This result, obtained from the study of quantum field theory in the black hole spacetime, showed that one can consistently attribute to the black hole horizon an entropy \textit{and} temperature.

 An immediate question that arises is whether this entropy  is the same as the ``usual
   entropy". If so, one should be able to show that, for any processes involving matter and black holes,
   we must have $d(S_{BH}+S_{matter})/dt\geq 0$ which goes under the name generalized 
   second law (GSL).  One simple example in which the area (and thus the entropy) of the black hole 
   \emph{decreases} is in the emission of Hawking radiation itself; but the GSL holds since the thermal radiation produced in the process has entropy.
   It is generally believed that GSL always holds though a completely general proof
   is difficult to obtain. Several thought experiments, when analyzed properly, uphold this law (see, for example, Ref. \cite{Unruh:1982ic})
   and a proof is possible under different sets of assumptions  \cite{tenproofs}.  
   All these suggest that the area of the black hole 
   corresponds to an entropy which is  same  as the ``usual entropy''. 

These ideas  can be extended to black hole solutions in more general  theories   than just Einstein's gravity. 
The temperature $T$  can be determined 
by    techniques like, for example,  analytic continuation to imaginary time (see Sec. \ref{thermalden}) which depends only on the metric and not on the field equations which led to the metric. But the concept of entropy needs to  generalized in these models and will no longer  be one quarter of the area of horizon. 
This could be done by using the  first law of black hole dynamics
itself, say, in the form $TdS = dM$. Since the temperature is known, this equation can be integrated to determine $S$. This was done by Wald and it turns out that the entropy can be related to a conserved charge called Noether charge which arises from the diffeomorphism invariance of the theory \cite{Wald:1993nt}.
Thus,  the notions of entropy and temperature  can be attributed to  black hole solutions in a wide class of theories.

This raises the  question: What are the degrees of freedom responsible for the black hole entropy? 
There have been several attempts in  the literature to answer this question  both with and  without inputs from quantum gravity models.
 A statistical mechanics derivation of entropy was originally attempted in \cite{Gerlach:1976hbh};
 the entropy  has been interpreted as the logarithm of: (a)  the number of 
      ways in which black hole might have been formed \cite{Bekenstein:1973ur,Hawking:1976de};
     (b) the number of internal black hole states consistent with a 
      single black hole  exterior  \cite{Bekenstein:1975tw} and
     (c)  the number of horizon quantum states \cite{wheeler,'tHooft:1990fr,Susskind:1993if}.
      There are also  other ideas which are more formal and geometrical \cite{Jacobson:1994vj,Visser:1993nu,Banados:1994qp,Susskind:1993ws},
       or  based on thermo-field theory \cite{Frolov:1998vs,Frolov:1997up,Frolov:1997aj}
       just to name two possibilities.
      In addition, considerable amount of work has been done
in calculating black hole entropy based on different candidate models for quantum gravity. 
(We will briefly summarize some aspects of these  in Sec. \ref{qgbh};
more extensive discussion as well as references to original literature can be found in the reviews \cite{Rovelli:1998yv,Das:2000su}).

Clearly, there is no agreement in literature as regards the degrees of freedom which contribute
to black hole entropy and the  attempts mentioned above   sample the divergent views.  In fact, once the answer is known, it seems fairly easy to come up with very imaginative derivations
of the result. 
 
 A crucial new dimension is added to this problem when we study horizons  which are \textit{not} associated with black holes. Soon after Hawking's discovery of a temperature associated with the black hole horizon, it was 
realized that this result was not confined to black holes alone. The study of quantum field theory in \textit{any} spacetime with a horizon showed that all horizons possess  temperatures \cite{Fulling:1973md,Davies:1975th,PIdesitter}.
In particular, an observer who is accelerating through the vacuum state in flat spacetime perceives a horizon and will attribute to it \cite{Unruh:1976db} a temperature $T= \kappa / 2\pi$ proportional to her acceleration $\kappa$.
(For a review, see Refs.\cite{Birrel:bkqft,mukhanovwini,Dewitt:1975ys,Takagi:1986kn,Sriramkumar:1999nw,Brout:1995rd,Wald:1999vt}.)
The situation regarding entropy --- especially whether one should attribute entropy to all horizons --- remains  unclear. In fact, many of the attempts    to interpret \textit{black hole} entropy mentioned earlier cannot be generalized to interpret the entropy of other horizons. So a unified understanding of horizon thermodynamics encompassing both entropy  and temperature still remains elusive --- which will be one of the key issues we will  discuss in detail in this review.

This is also closely related to the question of whether gravitational dynamics --- in particular, the field equations --- have any relation to the horizon thermodynamics.
Given a spacetime metric with a horizon (which may or may not be a solution to Einstein's equations)
one can study quantum field theory in that spacetime and discover that the horizon behaves like a black body with a given temperature.  
\textit{At no stage in such an analysis do we need to invoke the gravitational field equations.}
So it is reasonable to doubt whether the dynamics of gravity has anything to do with horizon thermodynamics and one may --- at first --- think that there is no connection between the two.

Several recent investigations have shown, however, that there is indeed a deeper connection between gravitational dynamics and horizon thermodynamics (for a recent review, see Ref.~\cite{tpdialogue}). 
For example, studies have shown that: 
\begin{itemize}
\item 
Gravitational field equations in \textit{a wide variety of theories}, when evaluated on a horizon, reduce to  a thermodynamic identity $TdS=dE+PdV$. This result, first pointed out in Ref.\cite{tdsingr}, has now been demonstrated  in several cases 
like the  stationary
axisymmetric horizons and evolving spherically symmetric horizons
in Einstein gravity, static spherically symmetric
horizons and dynamical apparent horizons in
Lovelock gravity, three dimensional BTZ black hole
horizons, FRW cosmological
models in various gravity
theories and even  in the case Horava-Lifshitz gravity (see Sec.~\ref{sec:connection} for detailed references). 
If horizon thermodynamics has no deep connection with gravitational dynamics, 
it is not possible to understand why the field equations should encode information about horizon thermodynamics.
\item

Gravitational action functionals in a wide class of theories have a a surface term and a bulk term. In the conventional  approach, we \textit{ignore} the surface term completely
(or cancel it with a counter-term) and obtain the field equation
from the bulk term in the action. Therefore, any solution to the field equation obtained 
by this procedure is logically independent of the nature of the surface term.
But  when the \textit{surface term} (which was ignored) is evaluated at the horizon that arises
in any given solution, it  gives the entropy of the horizon! Again, this result extends far beyond Einstein's theory to situations in which the entropy is \textit{not}  proportional to horizon area.
This is possible only because there is a specific holographic relationship \cite{TPhol002,TPgravquantum,ayan}
between the 
surface term and the bulk term which, however, is  an unexplained feature in the conventional
approach to gravitational dynamics.
Since the surface term has the thermodynamic interpretation as the entropy of horizons,
and is related holographically to the bulk term, we are again led to suspect 
an indirect connection between spacetime dynamics and horizon thermodynamics.
\end{itemize}

Based on these features --- \textit{which have no explanation in the conventional approach} --- one can argue  that there  is a  conceptual reason  to
 revise  our perspective towards spacetime
(Sec.~\ref{sec:emergentspacetime} and \ref{sec:gravitystory}) and
 relate horizon thermodynamics with gravitational dynamics. This approach
 should work for 
  a wide class of theories far more general than just Einstein gravity. This will be the new insight which we will focus on in this review.

To set the stage for this future discussion,  let us briefly describe
 this approach and summarize the conclusions. We begin by examining
 more closely the implications of the existence of 
temperature for horizons.

In the study of normal macroscopic systems --- like, for example, a solid or a gas --- one can \textit{deduce} the existence of microstructure just from the fact that the \textit{object can be heated}. The supply of energy in the form of heat needs to be stored in some form in the material which is not possible unless the material has microscopic degrees of freedom. This was  the insight of Boltzmann which led him to suggest that heat is essentially a form of motion of the microscopic constituents of matter. That is, the existence of temperature  is sufficient for us to infer the existence of microstructure without any direct experimental evidence. 

The thermodynamics of the horizon shows that we can actually heat up a spacetime, just as one can heat up a solid or a gas. An unorthodox way of doing this would be to take some amount of matter and arrange it to collapse and form a black hole. The Hawking radiation emitted by the black hole can be used to heat up, say, a pan of water just as though the pan was kept inside a microwave oven.
In fact the same result can be achieved by just accelerating through the inertial vacuum carrying the pan of water which will eventually be heated to a temperature proportional to the acceleration. These processes show that the temperatures of the horizons are as ``real'' as any other temperature. Since they arise in a class of hot  spacetimes, 
it follows \textit{\`{a} la} Boltzmann that 
the spacetimes should possess microstructure.

In the case of a solid or gas, we know the nature of this microstructure from atomic and molecular physics. Hence, in principle, we can work out the thermodynamics of these systems from the underlying statistical mechanics. This is not possible in the case of spacetime because we have no clue about its microstructure. However, one of the remarkable features of thermodynamics
--- in contrast to statistical mechanics ---
is that the thermodynamic description is fairly insensitive to the details of the microstructure and can be developed as a fairly broad frame work. For example, a thermodynamic identity like $TdS = dE+PdV$ has a universal validity and the information about a \textit{given} system is only encoded in the form of the entropy functional $S(E,V)$. In the case of normal materials, this
entropy arises because of our coarse graining over microscopic degrees of freedom which are not tracked in the dynamical evolution. In the case of spacetime, the existence of horizons for a particular class of observers makes it \textit{mandatory} that these observers  integrate out degrees of freedom hidden by the horizon. 

To make this notion clearer, let us start from the principle of equivalence which allows us to construct local inertial frames (LIF), around any event in an arbitrary curved spacetime.
  Given the LIF, we can next construct a local Rindler frame (LRF)
  by  boosting along one of the directions  with an acceleration $\kappa$.
The observers at rest  in the LRF will perceive  a patch of null surface in LIF as  a horizon $\mathcal{H}$ with temperature $\kappa/2\pi$.
 These local Rindler observers and the freely falling inertial observers will attribute
   different thermodynamical properties to matter in the spacetime. For example, they will attribute different temperatures and entropies to the vacuum state
   as well as excited states of matter fields.
  When some matter with energy $\delta E$ moves close to the horizon --- say, within a few Planck lengths because,  formally, it takes infinite Rindler time for matter to actually cross $\mathcal{H}$ --- the local Rindler observer will consider it to have  transfered an entropy  $\delta S = (2\pi/\kappa) \delta E$ 
   to the horizon degrees of freedom. We will  show (in Sec.~\ref{sec:emergentspacetime}) that, when the metric satisfies the field equations of any diffeomorphism invariant theory, this transfer of entropy can be given  
   \cite{tp09papers}
   a  geometrical interpretation as the change in the  entropy of the horizon.

This result allows us to associate an entropy functional with the null surfaces which the local Rindler observers perceive as  horizons. 
We can now demand that the sum of the horizon entropy and the entropy of matter that flows across the horizons (both as perceived by the local Rindler observers), should be an extremum for all observers in the spacetime.
This leads \cite{TPgravitystory} to a constraint on the geometry of spacetime which can be stated, in $D=4$, as 
\begin{equation}
 (G_{ab} - 8\pi T_{ab}) n^a n^b =0
 \label{Eenn}
\end{equation} 
for all null vectors $n^a$ in the spacetime.
The general solution to this equation is given by $G_{ab} = 8\pi T_{ab} + \rho_0 g_{ab}$
where $\rho_0$ has to be a constant because of the conditions $\nabla_a G^{ab} =0 = \nabla_aT^{ab}$.
Hence the thermodynamic principle leads uniquely to Einstein's equation with a cosmological constant in 4-dimensions. Notice, however, that \eq{Eenn} has a new symmetry and is invariant \cite{TPgravimmune,TPgravijtmp} under the transformation $T_{ab} \to T_{ab} + \lambda g_{ab}$ which 
the standard Einstein's theory does not posses. (This has important implications for the cosmological constant problem  \cite{TPadvscilett} which we will discuss in Sec. \ref{sec:cc}.)
In $D>4$, the same entropy maximization 
 leads to a more general class of theories called \LL\ models (see Sec. \ref{sec:llgravity}).

We can now remedy another conceptual shortcoming of the conventional approach.
An unsatisfactory feature of all theories of gravity is that the field
equations do not have any direct physical interpretation. 
The lack of an elegant principle which can lead to the dynamics of gravity
(``how matter tells spacetime to curve'')
is quite striking when we compare this situation with the kinematics of gravity (``how spacetime makes the matter move''). The latter can be determined through the principle of equivalence by demanding that all freely falling observers, at all events in spacetime, must find that the equations of motion for matter  reduce to their special relativistic form.
 
In the alternative perspective, \eq{Eenn} arises from our demand that the thermodynamic extremum principle should hold for \textit{all} local Rindler observers.   
This is identical  to the manner in which  freely falling observers are used to determine how gravitational field influences matter. Demanding the validity of special relativistic laws for the 
 matter variables, as determined by all the freely falling observers, allows us to 
determine the influence of gravity on matter. Similarly, demanding the maximization of entropy of horizons (plus matter), as measured by all local Rindler observers, leads to the dynamical equations of gravity.

In this review, we will examine several aspects of these features and will try to provide, in the latter part of the review, a synthesis of ideas which offers an interesting new perspective on the nature of gravity that makes the connection between horizon thermodynamics and gravitational dynamics obvious and natural. In fact, we will show that the thermodynamic underpinning goes far beyond  Einstein's theory and encompasses a wide class of gravitational theories. 

The review is organized as follows. In Section \ref{sec:gravhorizon}
we will review several features of horizons which arise in different contexts in gravitational theories.  In particular, we will describe some generic features of the spacetimes with horizons which will be important in the later discussions. 
In Section \ref{sec:thermofirst} we shall provide a simple derivation
of the temperature of a (generic) horizon using path integral methods.
Section \ref{sec:pireview} introduces the basic concepts related to path integrals and
Section \ref{thermalden} applies them to a spacetime with horizon to obtain the temperature.
Some alternate ways of obtaining the temperature of horizons are discussed in 
Section \ref{sec:complext}. The origin of Hawking radiation from  matter that collapses to form a black hole is discussed in Section \ref{sec:hawrad}. 
The next two subsections describe the generalization of these ideas to theories
other than Einstein's general relativity. The relationship between horizon entropy and
the Noether charge is introduced in Section \ref{sec:noetherent} and
the structure of \LL\ models is summarized in Section \ref{sec:llgravity}.
Section \ref{sec:thermosecond} discusses several conceptual issues raised by the existence of temperature and entropy of the horizons, concentrating on the nature of degrees of freedom which contribute to the entropy and the observer dependence of the concept of entropy. This discussion is continued in the next section where we take up the connection between horizon thermodynamics and gravitational dynamics and describe several peculiar features
 which have no simple interpretation
in the conventional approach. In Section \ref{sec:emergentspacetime} we provide a thermodynamic interpretation of the field equations of gravity in any diffeomorphism invariant theory. This forms the basis for the alternative perspective of gravity described in Section \ref{sec:gravitystory} in which we obtain the field equations of gravity from an entropy maximization principle. The last section provides a summary. 

We will use the signature $(- +++)$ and units with $G=\hbar=c=1$. The Greek superscripts and subscripts will run over the spatial coordinates while the Latin letters will cover time coordinate as well as spatial coordinates.

\section{Gravity and its horizons}\label{sec:gravhorizon}

\subsection{The Rindler horizon in flat spacetime}

\begin{figure}[htbp] 
\begin{center}
\includegraphics[scale=0.5]{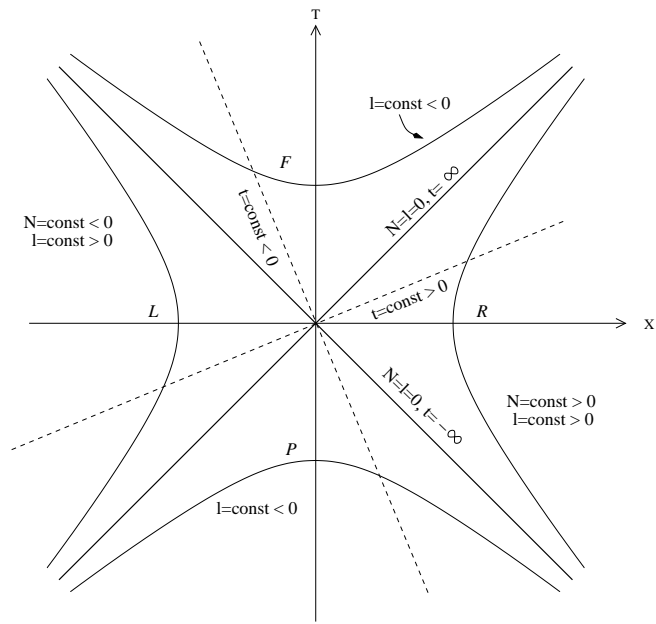}
  \caption{The global manifold with different coordinate systems in the four quadrants. See 
  text for discussion.}
\label{hyperfig}
\end{center}
\end{figure}

The simplest context in which a horizon arises for a class of observers occurs in the flat spacetime itself. Consider the standard flat spacetime metric with Cartesian coordinates in the $X-T$ plane given by 
\begin{equation}
 ds^2 = -dT^2 + dX^2 + dL_\perp^2
\end{equation} 
where $dL_\perp^2$ is the line element in the transverse space.
The lines $X=\pm T$ divide the $X-T$ plane into four quadrants (see \fig{hyperfig})
marked the right ($\mathcal{R}$) and left ($\mathcal{L}$) wedges  as well as
the past ($\mathcal{P}$) and future ($\mathcal{F}$) of the origin. We now introduce two new coordinates ($t,l$) in place of $(T,X)$ in all the four quadrants through the transformations:
   \begin{equation}
   \kappa T=\sqrt{2\kappa l} \sinh (\kappa t); \quad \kappa X=\pm \sqrt{2\kappa l} \cosh (\kappa t)
   \label{expone}
   \end{equation}
   for $|X|>|T|$ with the positive sign in $\mathcal{R}$ and negative sign in 
   $\mathcal{L}$ and 
 \begin{equation}
   \kappa T=\pm \sqrt{-2\kappa l} \cosh (\kappa t); \quad \kappa X= \sqrt{-2\kappa l} \sinh (\kappa t)
   \label{exptwo}
   \end{equation}
   for $|X|<|T|$ with the positive sign in $\mathcal{F}$ and negative sign in 
   $\mathcal{P}$.
   Clearly, $l<0$ is used in $\mathcal{F}$ and  $\mathcal{P}$.
   With these transformations, the metric in all the four quadrants can be expressed in the form
   \begin{equation}
   ds^2 = - 2 \kappa l \ dt^2 + \frac{dl^2}{2\kappa l}  + dL_\perp^2
   \label{standardhorizon}
   \end{equation}
   Figure \ref{hyperfig} shows the geometrical features of the coordinate systems from which we see
   that: (a) The coordinate $t$ is timelike and $l $ is spacelike in Eq.~(\ref{standardhorizon}) only in $\mathcal{R}$  and $\mathcal{L}$ where $l>0$
  with their roles reversed in $\mathcal{F}$ and $\mathcal{P}$ with  $l<0$. (b) A given value of $(t,l)$ corresponds to a \textit{pair}
  of points in $\mathcal{R}$ and $\mathcal{L}$ for $l>0$ and to a \textit{pair} of points in
  $\mathcal{F}$ and $\mathcal{P}$ for $l<0$. 
    (c) The surface $l=0$ acts as a horizon for observers in $\mathcal{R}$. In particular, observers who are stationary in the new coordinates with $l=$ constant, $\mathbf{x}_\perp =$ constant will follow a trajectory $X^2 - T^2 = 2l/\kappa$ in the $X-T$ plane. 
    These are trajectories of observers moving with constant proper acceleration in the inertial frame who perceive a horizon at $l=0$. Such observers are usually called Rindler observers and the metric in \eq{standardhorizon}
    is called the Rindler metric. 
    (The label $N$ in \fig{hyperfig} corresponds to $N=\sqrt{2\kappa |l|}$.)

\subsection{The Rindler frame as the near-horizon limit}

The Rindler frame will play a crucial role in our future discussions for two reasons.
First, one can introduce Rindler observers even in curved spacetime in any local region.
To do this, we 
 first transform to the locally inertial frame (with coordinates $T,X$) around that event and then introduce
the local Rindler frame with coordinates $(t,l)$ by the  transformations in \eq{expone} and \eq{exptwo}. Such a local notion is approximate but will prove to be valuable in our future discussions because it can be introduced around any event in any curved spacetime.
Second, the Rindler (like) transformations work for a wide variety of spherically symmetric solutions to gravitational field  equations.
This general class of spacetimes can be expressed by a metric of the form 
\begin{equation}
 ds^2 =- f(r)  dt^2 + \frac{dr^2}{f(r)}  + dL_\perp^2
 \label{fbyf}
\end{equation} 
where the function $f(r)$ has a simple zero at some point $r=a$ with a non-zero first derivative $f'(a)\equiv 2\kappa$. A Taylor series expansion of $f$ near $r=a$ 
gives $f \approx 2\kappa l$ with $l=r-a$. It is, therefore, obvious that near the horizon, located at $r=a$, all these metrics can be approximated by the Rindler metric in 
\eq{standardhorizon}. Hence Rindler metric is useful in the study of spacetime near the horizon in several exact solutions. (We are assuming that $f'(a)\neq 0$; there are certain solutions --- called extremal horizons --- in which this condition is violated and $f'(a)=0$; we will not discuss them in this review.)

In the above analysis we started from a flat spacetime expressed in standard inertial coordinates and then introduced the transformation to Rindler coordinates. This transformation, in turn, brought in a pathological behaviour for the metric at $l=0$.
Alternatively, if we were given the metric in Rindler coordinates in the form of 
\eq{standardhorizon} we could have used the transformation in \eq{expone} and
\eq{exptwo} to remove the pathological behaviour of the metric. In such a process
we would have also discovered that a given value of ($t,l$) actually corresponds to a \textit{pair} of events in the full spacetime thereby `doubling up' the manifold. This process is called analytic extension. 

In the case of  metrics given by \eq{fbyf} the pathology at $f =0$ is similar to the pathology
of the Rindler metric at $l=0$. Just as one can eliminate the latter by analytic extension,
one can also eliminate the singularity at $r=a$ in the metric in \eq{fbyf} by suitable coordinate transformation. Consider for example, the transformations from $(t,r)$ to
$(T,X)$ by the equations
\begin{equation}
  \kappa X=e^{ \kappa\xi} \cosh  \kappa t; \  \kappa T=e^{ \kappa\xi} \sinh  \kappa t;
  \qquad \xi \equiv \int \frac{dr}{f(r)}
  \label{keytransform}
  \end{equation}
  This leads to a metric of the form 
  \begin{equation}
  ds^2 =\frac{f}{ \kappa^2(X^2 - T^2)} (-dT^2 +dX^2) + dL_\perp^2
  \end{equation}
  where $f$ needs to be expressed in terms of $(T,X)$ using the coordinate transformations.
  The horizon $r=a$ now gets mapped to $X^2 = T^2$; but it can be shown that the factor
  $f/(X^2 - T^2)$ remains finite at the horizon.

  The similarity between the coordinate transformations in \eq{keytransform}
  and \eq{expone}
is obvious. (As in the case of \eq{expone}, one can introduce another set of transformations to cover the remaining half of the manifold by interchanging $\sinh$ and $\cosh$ factors.)
The curves of constant $r$ in the original spherically symmetric metric in \eq{fbyf}
become hyperbolas in the $T-X$ plane, just as in the case of transformation from Rindler to inertial coordinates.

\subsection{Horizons in static spacetimes}

In the Rindler frame (as well as  near the horizon in a curved spacetime), one can introduce another coordinate system which often turns out to be useful.
This is done by transforming from ($t,l$) to $(t, x)$ where 
$ l = (1/2) \kappa x^2$. Then the Rindler metric in \eq{standardhorizon} reduces to the form
\begin{equation}
   ds^2=-\kappa^2 x^2 dt^2 + dx^2 + dL_\perp^2
   \label{dsfirst}
   \end{equation}
and the coordinate transformation transformation corresponding to \eq{expone}
becomes 
  \begin{equation}
  T=x \sinh (\kappa t); \quad X=\pm x \cosh (\kappa t)
   \label{expthree}
   \end{equation}

The form of the metric in \eq{dsfirst} also arises in a wide class of static (and stationary, though we will not discuss this case) spacetimes with the following properties:
(i) The metric is static  in the given coordinate system, $g_{0\alpha} =0, g_{ab} (t,{\bf x}) = g_{ab}({\bf x})$;
   (ii) $g_{00}({\bf x}) \equiv -N^2({\bf x})$ vanishes on some 2-surface $\mathcal{H}$ defined
   by the equation $N^2 =0$, (iii) $\partial_\alpha N$ is finite and non zero on $\mathcal{H}$
   and (iv) all other metric components and curvature
   remain finite and regular on $\mathcal{H}$.  
 The line element will now be:
 \begin{equation}
   ds^2=-N^2 (x^\alpha)  dt^2 +  \gamma_{\alpha\beta} (x^\alpha) dx^\alpha dx^\beta
   \label{startmetric}
   \end{equation}
The comoving observers in this frame have trajectories ${\bf x}=$
    constant, four-velocity $u_a=-N\delta^0_a$ and four acceleration $a^i=
    u^j\nabla_ju^i=(0,{\bf a})$ which has the purely
    spatial components $a_\alpha=(\partial_\alpha N)/N$.
 The unit normal $n_\alpha$ to the $N=$ constant surface is given by
    $n_\alpha =\partial_\alpha N (g^{\mu\nu}\partial_\mu N  \partial_\nu N)^{-1/2} =
    a_\alpha (a_\beta a^\beta)^{-1/2}$. A simple computation now shows that 
    the normal component of the acceleration $a^i n_i = a^\alpha n_\alpha$, `redshifted'
    by a factor $N$, has the value
\begin{equation}
    N(n_\alpha a^\alpha) = ( g^{\alpha\beta} \partial_\alpha N \partial_\beta N)^{1/2}\equiv Na({\bf x})
\label{defkappa}
\end{equation}
where the last equation defines the function $a$. From our assumptions, it follows that on the horizon $N=0$,
this quantity has a finite limit $Na\to \kappa$; the $\kappa$ is called the surface gravity of the horizon.

These static spacetimes, however,  have a more natural coordinate system defined in terms of the level surfaces of $N$. That is, we transform from the original space coordinates $x^\mu$ in Eq.(\ref{startmetric}) to the set $(N,y^A), A=2,3$ by treating $N$ as one of the  spatial coordinates  ---  which is always possible locally. The $y^A$ denotes the two
transverse coordinates on the $N=$ constant surface. The line element in the new coordinates will be:
\begin{equation}
ds^2=-N^2dt^2+ \frac{dN^2}{(Na)^{2}}+
\sigma_{AB}(dy^A-\frac{a^A dN}{Na^2})(dy^B-\frac{a^BdN}{Na^2})
\label{iso}
\end{equation}
where $a^A$ etc. are the components of the acceleration in the new coordinates.
The original 7 degrees of freedom in $(N,\gamma_{\mu\nu})$ are now reduced to 6 degrees of freedom in $(a,a^A,
\sigma_{AB})$, because of our choice for $g_{00}$.  In  Eq.(\ref{iso})  the spacetime
is described
in terms of
the magnitude  of acceleration $a$, the transverse components $a^A$ and the  metric $\sigma_{AB}$ on 
the two surface and maintains the $t-$independence. 
The $N$ is now merely a coordinate and the spacetime geometry is  described in terms of 
$(a,a^A,\sigma_{AB})$
all of which are, in general, functions of $(N,y^A)$. In  spherically symmetric spacetimes with horizon, for example, we will have $a=a(N),a^A=0$ if we choose $y^A=(\theta,\phi)$. Important features of dynamics are usually encoded in 
the function $a(N,y^A)$.
Near the $N\to 0$ surface, $Na\to \kappa$, the surface gravity, and the metric reduces to 
   \begin{equation}
   ds^2=-N^2dt^2+ \frac{dN^2}{(Na)^{2}}+dL_\perp^2
   \simeq-N^2 dt^2 + \frac{dN^2}{\kappa^2} +dL_\perp^2
   \label{dssecond}
   \end{equation}
    where   the second equality is applicable 
   close to  $\mathcal{H}$.
This  is the same metric as in  
 Eq.(\ref{dsfirst}) if we set $N=\kappa x$.
 Therefore a wide class of metrics with horizon can be mapped to the Rindler form near the horizon.
 
 The form of the metric in \eq{dsfirst} is particularly useful to study the analytic continuation to imaginary values of time coordinate. If we denote $T_E = i T, t_E = i t$, then, in the right wedge $\mathcal{R}$, 
the transformations become
\begin{equation}
  T_E=x \sin (\kappa t_E); \quad X= x \cos (\kappa t_E)
   \label{expeuclid}
   \end{equation}
which are just the coordinate transformation from the Cartesian coordinates $(T_E, X)$
to the polar coordinates $(\theta=\kappa t_E, x)$ in a two dimensional plane. To avoid a conical singularity at the origin, it is necessary that $\theta$ is periodic with period $2\pi$, which --- in turn --- requires $t_E$ to be periodic with period $2\pi/\kappa$. We will see later that such a periodicity in the imaginary time signals the existence of non-zero temperature.

\subsection{Exponential redshift and thermal power spectrum}
 
 Another generic feature of the horizons we have defined  is that they act as surfaces of infinite redshift.
 To see this, consider the redshift of  a photon 
 emitted at $(t_e, N_e, y^A)$, 
   where $N_e$ is close to the horizon surface $\mathcal{H}$,  and is observed
   at $(t, N, y^A)$.  The frequencies at emission $\omega(t_e)$ and detection $\omega(t)$  are related by
   $[\omega(t)/\omega(t_e)]=[N_e/N] $. 
   The radial trajectory of the out-going photon is given by $ds^2=0$ which integrates to 
   \begin{equation}
   t-t_e =  \int_{N_e}^N \frac{dN}{N^2a} \simeq -\frac{1}{\kappa} \ln N_e + \textrm{constant}
   \label{photonpath}
   \end{equation}
  where we have approximated the integral by the dominant 
   contribution near $N_e =0$. This gives $N_e  \propto \exp(- \kappa t)$,
   leading to the exponentially redshifted frequency 
  \begin{equation}
 \omega(t)\propto N_e\propto \exp(-\kappa t)                                              \end{equation} 
 as detected by an observer at a fixed $N$ as a function of $t$.
 
Such an exponential redshift is also closely associated with the emergence of a temperature in the presence of a horizon. To see this, let us consider how an observer in Rindler frame (or, more generally, in the spherically symmetric frame with the metric given by \eq{fbyf})
will view a monochromatic plane wave moving along the $X-$axis in the inertial frame (or, more generally, in the analytically extended coordinates). Such a scalar wave can be represented by 
$\phi(T,X)=\exp[-i\Omega(T- X)]$ with $\Omega>0$. 
  Any other observer who is inertial with respect to the $X=$ constant observer will see this as a 
  {\it monochromatic} wave, though  with a different (Doppler-shifted) frequency. But an accelerated observer,
  at $N=N_0 =$ constant  
   using her proper
  time  $\tau \equiv N_0 t$ will see the same mode as varying  as 
  \begin{equation}
  \phi = \phi(T(t),X(t))=\exp[i\Omega q e^{- \kappa t}]
  = \exp[i\Omega q \exp - (\kappa/N_0)\tau]
\label{expodamp}
  \end{equation}
  where we have used Eq.~(\ref{keytransform}) and defined $q\equiv \kappa^{-1} \exp( \kappa\xi) $.
  This is clearly not monochromatic and has a frequency which is being exponentially redshifted in time. The power spectrum   of this wave 
  is given by $P(\nu) = |f(\nu)|^2$ where $f(\nu)$ is the Fourier transform of $\phi(\tau)$ with
  respect to $\tau$:
  \begin{equation}
  \phi(\tau) = \int_{-\infty}^\infty \frac{d\nu}{2\pi} f(\nu) e^{-i\nu \tau}
  \equiv \int_0^\infty \frac{d\nu}{2\pi} \left[ A(\nu) e^{-i\nu \tau} + B(\nu)e^{i\nu \tau}\right]
  \label{ftf}
  \end{equation}
  with $A(\nu)= f(\nu)$ and $B(\nu)=f(-\nu)$.
   Because of the exponential redshift, this power spectrum will \emph{not} vanish
  for $\nu<0$ leading to $B \neq 0$. Evaluating this Fourier transform (by changing to the variable 
  $\Omega q \exp[-( \kappa/N_0) \tau ] =z$
  and analytically continuing to Im $z$) one gets:
  \begin{equation}
 f(\nu) =(N_0/ \kappa) (\Omega q)^{i\nu N_0/\kappa } \Gamma(-i\nu N_0/ \kappa)
  e^{\pi \nu N_0/2 \kappa} 
 \label{powernu}
 \end{equation}
  This leads to the 
   the remarkable result
  that the power, per logarithmic band in frequency, at negative frequencies  
  is a Planckian at temperature $T= (\kappa/2\pi N_0)$: 
  \begin{equation}
\nu|B(\nu)|^2 = \nu\vert f (-\nu)\vert^2 =  \frac{\beta}{e^{\beta \nu} - 1 } ; \quad \beta = \frac{2\pi N_0}{\kappa }
  \label{planck}
  \end{equation}
  Though $f(\nu)$ in Eq.~(\ref{powernu})
   depends on $\Omega$, the power spectrum $|f(\nu)|^2$ is independent of 
  $\Omega$; monochromatic plane waves of any frequency (as measured by the freely falling observers with $X=$ constant) will appear to have Planckian
  power spectrum in terms of the (negative) frequency $\nu$, defined with respect to the proper time of 
   the accelerated observer located at $N=N_0=$ constant.  The scaling of the temperature $\beta^{-1}\propto N_0^{-1}\propto |g_{00}|^{-1/2}$ is precisely what is expected in general relativity for temperature.
   (Similar results also arise in the case of \textit{real} wave with $\phi
   \propto \cos \Omega(T-X)$; see ref. \cite{tplsksrinia,tplsksrinib}.)  
   
   We saw earlier (see Eq.~(\ref{photonpath})) that  waves propagating 
   from a region near the horizon will undergo
   exponential redshift.
  An observer detecting this  exponentially redshifted radiation 
   at late times $(t\to \infty)$, originating from  a region close 
   to $\mathcal{H}$ will attribute to this radiation a Planckian power spectrum given by  Eq.~(\ref{planck}). 
   This 
  result lies at the foundation of associating a temperature with a horizon.

   The Planck spectrum in  Eq.~(\ref{planck})
  is in terms of the  frequency and $\beta\propto c/\kappa$ has the  (correct) dimension of  time;   no $\hbar$
appears in the result. If we now switch  the variable to energy
   and write $\beta \nu = (\beta/\hbar) (\hbar \nu)
  = (\beta/\hbar) E$, then one can identify a temperature $k_BT =( \kappa\hbar /2\pi c)$ which
  scales with $\hbar$. This ``quantum mechanical'' origin of temperature 
  is superficial because it arises merely because of a  change of  units from $\nu $ to $E$.
  An astronomer measuring frequency rather than photon energy  will see the spectrum in 
  Eq.~(\ref{planck}) as Planckian without any quantum mechanical input. 
The real role of quantum theory is not in  the conversion of frequency to energy but in providing the complex wave in the inertial frame. It represents  the vacuum fluctuations of the quantum field.

  \subsection{Field theory near the horizon: Dimensional reduction}\label{sec:ftdim}
  
  The fact that $-g_{00} =N^2 \to 0$ on the horizon leads to several interesting conclusions regarding
  the behaviour of  any classical (or quantum)
  field  near the horizon. Consider, for example, an interacting scalar field in a background spacetime described by the metric in Eq.(\ref{iso}),
  with the action: 
  \begin{eqnarray}
  A &=& -\int d^4x \sqrt{-g} \left( \frac{1}{2} \partial_a \phi \partial^a \phi +V \right)\\
 &=& \int dt dN d^2y\, \left(\frac{\sqrt{\sigma}}{N^2a}\right)
 \  \left[ \frac {\dot \phi^2}{2} -N^4a^2
 \left(  \frac{\partial \phi}{\partial N}\right)^2- N^2\left[\frac
  {(\partial_\perp \phi)^2}{2} + V\right] \right]\nonumber
  \end{eqnarray}
where $(\partial_\perp \phi)^2$ denotes the contribution from the derivatives in the transverse directions
including cross terms of the type $(\partial_N \phi \partial_\perp  \phi)$.
  Near $N=0$, with $Na\to \kappa$, the action reduces to the form 
  \begin{equation}
  A\approx   \int \sqrt{\sigma} d^2x_\perp \int dt \int d\xi \,  \left\{ \frac{1}{2} \left[ \dot \phi^2 - 
  \left(  \frac{\partial \phi}{\partial \xi}\right)^2 \right]  \right\}
  \end{equation}
  where we have changed variable to $\xi$ defined in Eq.~(\ref{keytransform}) (which behaves as
$\xi\approx (1/\kappa)\ln N$ near $N\simeq 0$) and ignored terms that vanish as 
  $N\to 0$. Remarkably enough this action represents a two dimensional free field theory
  in the $(t,\xi)$ coordinates  which has the enhanced symmetry of invariance under
  the conformal transformations $g_{ab} \to f^2(t,\xi) g_{ab}$ (see e.g., Section 3 of \cite{Padmanabhan:2002ha},\cite{solo}). 
The solutions to the
  field equations near $\mathcal{H}$  are plane waves in the $(t,\xi)$ coordinates:
   \begin{equation}
   \phi_\pm = \exp[-i\omega (t \pm  \xi)] 
   = N^{\pm i\omega/\kappa}e^{-i\omega t}
   \label{twotwo}
   \end{equation}
   These modes are the same as $\phi=\exp iA$ where $A$ is the solution 
    to the classical  Hamilton-Jacobi equation; this equality arises because the divergence of $(1/N)$ factor near the horizon 
   makes the WKB approximation almost exact near the horizon. The mathematics involved in this phenomenon
   is fundamentally the same as the one which leads to the ``no-hair-theorems" 
  (see, eg.,  \cite{Bekenstein:1998aw}) for the black hole.
These solutions possess several other symmetry properties  which are worth mentioning:

    To begin with, the  metric and the solution near $\mathcal{H}$ are  invariant under the rescaling
   $N\to \lambda N$,  in the sense that this transformation merely adds a phase  to 
   $\phi$. This scale invariance can also be demonstrated by studying the spatial part of the 
   wave equation \cite{Srinivasan:1998ty} near $\mathcal{H}$,  where the equation reduces to a
    Schrodinger equation
  for the zero energy eigenstate in
  the potential $V(N) = - \omega^2/N^2$ .
   This Schrodinger equation has the natural scale invariance with respect to $N\to \lambda N$
   which is reflected in our problem. 

Second, the relevant metric $ds^2=-N^2 dt^2+(dN/\kappa)^2$ in the $t-N$
plane is also invariant, up to a conformal factor, to the metric obtained by $N\to \rho=1/N$:
\begin{equation}
ds^2=-N^2 dt^2+\frac{dN^2}{\kappa^2}=\frac{1}{\rho^4}(-\rho^2 dt^2+\frac{d\rho^2}{\kappa^2})
\end{equation}
Since the two dimensional field theory is conformally invariant, if $\phi(t,N)$ is a solution, then
$\phi(t,1/N)$ is also a solution. This is clearly true for the solution in Eq.~(\ref{twotwo}). Since $N$ is a coordinate in our description, this connects up the infrared behaviour of the field theory with the ultraviolet behaviour.   

  More directly, we note that the symmetries of the theory enhance significantly near the
$N=0$ hypersurface.
Conformal invariance,  similar to the one found above, occurs in the  gravitational sector as well.
Defining  $q=-\xi$ by $dq=-dN/N(Na)$, we see that $N\approx\exp(-\kappa q)$ near the horizon, where $Na\approx \kappa.$ The space part of the metric in Eq.(\ref{iso}) becomes, near the horizon
$dl^2=N^2(dq^2+e^{2\kappa q}dL_\perp^2)$
which is conformal to the metric of the anti-De Sitter (AdS) space. The horizon
becomes the $q\to\infty$ surface of the AdS space. These results hold in any dimension.
There is a strong indication that most of the results related to horizons
 will arise from the enhanced symmetry of the theory near
the $N=0$ surface (see e.g. \cite{Carlip:1999cy,Park:1999tj,Park:2001zn,grumiller07} and references cited therein).
One can construct the metric in the bulk by a Taylor
series expansion, from the form of the metric near the horizon,
along the lines of Exercise 1 (page 290) of \cite{lltwo} to demonstrate the enhanced symmetry.
These results arise  because, algebraically, $N\to 0$ makes certain terms
in the diffeomorphisms vanish and increases the symmetry. This fact will prove to be useful in Sec. \ref{sec:gravthermoiden}.

\subsection{Three specific examples of horizons}

For the sake of reference, we briefly describe three specific solutions to Einstein's equations with horizons having a metric of the form in \eq{fbyf},  viz. the Rindler, Schwarzschild and de Sitter spacetimes. In each of these cases, the metric can be expressed in the form of
Eq.~(\ref{fbyf}) with different forms of $f(l)$ given in the Table \ref{table:metricprop}. 
All these cases  have only one horizon at
some surface $l=l_H$ and the surface gravity $\kappa$ is well defined. 
(We have relaxed the condition that the horizon occurs at $l=0$; hence $\kappa$ is 
defined as $(1/2) |f'|$ evaluated at the location of the horizon, $l=l_H$.) The  coordinate transformations relevant for analytic extension of  these three spacetimes are also given in Table \ref{table:metricprop}. 
The coordinates $(T,X)$ are well behaved near the 
horizon while the original coordinate system $(t,l)$ is singular at the horizon.
  Figure \ref{hyperfig} describes all the three cases of horizons 
  which we are interested in, with suitable definition for the coordinates.

\begin{table*}[b]

\begin{tabular*}{\linewidth}{>{$}l<{$}>{$}c<{$}>{$}c<{$}>{$}c<{$}}
   \hline
  \textrm{Metric} & \textrm{Rindler} & \textrm{Schwarzschild} & \textrm{De Sitter}\\
    \hline\hline
    \noalign{\medskip} 
  f(l) & 2\kappa l & \left[1 - \frac{2M}{l}\right] & ( 1 - H^2 l^2)\\
   \noalign{\medskip}
  \kappa=\frac{1}{2} |f'(l_H)| &  \kappa & \displaystyle{ \frac{1}{4 M }} &   H \\
     \noalign{\medskip} 
     \xi & \frac{1}{2\kappa} \ln\kappa l &  l+ 2M \ln \left[ \frac{l}{2M} - 1\right] &  \frac{1}{2H}
     \ln \left( \frac{1-Hl}{1+Hl} \right) \\
      \noalign{\medskip}
     \kappa X& \sqrt{2\kappa l} \cosh\kappa t & \ \  e^{\frac{l}{4M}} \left[ \frac{l}{2M} - 1\right]^{1/2} 
               \cosh \left[\frac{t}{4M}\right] & \ \left( \frac{1-Hl}{1+Hl} \right)^{1/2} \cosh Ht \\	       
     \noalign{\medskip}         
   \kappa T & \sqrt{2\kappa l} \sinh\kappa t & \ \   e^{\frac{l}{4M}} \left[ \frac{l}{2M} - 1\right]^{1/2} 
              \sinh\left[\frac{t}{4M}\right] &\  \left( \frac{1-Hl}{1+Hl} \right)^{1/2} \sinh Ht\\	      
  \noalign{\bigskip}
   \hline
   \noalign{\bigskip}
   \end{tabular*}
   \caption{Properties of Rindler, Schwarzschild and De Sitter metrics}
    \label{table:metricprop}
   \end{table*}

 The horizons with the above features arise  in Einstein's theory as well as in more general theories of gravity. While the detailed properties of the spacetimes in which these horizons occur are widely different, there are some key features  shared by all the horizons we are interested in, which is worth summarizing:
  
  In all these cases, there exists
 a Killing vector field $\xi^a$ which is timelike  in part of the manifold with the components
  $\xi^a =(1,0,0,0)$ in the Schwarzschild-type static coordinates.
   The norm of this field $\xi^a\xi_a$ vanishes on the horizon that acts as a bifurcation surface $\mathcal{H}$.
  Hence, the points of $\mathcal{H}$ are fixed points of the Killing field. 
  Further the surface gravity $\kappa$ of the horizon can be defined using the `acceleration' of the Killing vector by:
  \begin{equation}
   \xi^b\nabla_b\xi^a=\kappa\xi^a 
   \label{defkappa1}
  \end{equation} 
  When defined in this manner, the value of $\kappa$ depends on the normalization chosen for $\xi^a$. (If we rescale $\xi^a\to \mu\xi^a$, the surface gravity also scales as $\kappa\to\mu\kappa$). Very often, however, we will be interested in the combination $\xi^a/\kappa$, which is invariant under this scaling.
  
  In these spacetimes, there exists a spacelike hypersurface $\Sigma$
  which includes $\mathcal{H}$ and is divided by $\mathcal{H}$ into two pieces $\Sigma_R$ and 
 $ \Sigma_L$, the intersection of which is in fact $\mathcal{H}$. 
  In the case of black hole manifold, for example,  $\Sigma$ is the $T=0$ surface, $\Sigma_R$ and $\Sigma_L$
  are parts of it in the right and left wedges and $\mathcal{H}$ corresponds to the $l=2M$ surface.
  The topology of $\Sigma_R$ and $\mathcal{H}$ depends on the details of the spacetime
  but $\mathcal{H}$ is assumed to have a non-zero surface gravity. Given this structure
  it is possible to generalize most of the results we will be discussing in the later sections.
  
  Finally, to conclude this section, we shall summarize a series of geometrical facts related to the Rindler frame and the Rindler horizons which will turn out to be
  useful in our future discussions. Though we will present the results in the context of a 
  2-dimensional Rindler spacetime, most of the ideas have a very natural generalization to other bifurcation horizons. 
We begin with the metric for the Rindler spacetime expressed in different sets of coordinates:
\begin{equation}
ds^2 =-dT^2 + dX^2 =-dU dV = - e^{\kappa(v-u)}du dv =-2\kappa l \, dt^2 + \frac{dl^2}{2\kappa l} 
\end{equation} 
The coordinate transformations relating these have been discussed earlier.
 In particular, note that 
the null coordinates in the two frames are related by
$U = T-X =-\kappa^{-1} e^{-\kappa u} , \ V =  T + X = \kappa^{-1} e^{\kappa v}$.
We will now introduce several closely related vectors and  their properties.

(i) Let $k^a$ be a future directed null vector with components proportional to $(1,1)$
in the inertial frame.  The corresponding  affinely parameterized null curve can be taken
to be $x^a = \kappa X(1,1)_I$ with $X$ being the affine parameter and subscript $I$ (or $R$)
indicates the components in the inertial (or Rindler) frame.

(ii) We also have the natural   Killing vector $\xi^a$ corresponding to translations in the 
Rindler time coordinate. This vector has the components, 
$ \xi^a  = (1,0)_R = \kappa (X,T)_I $ and $ \xi^a \xi_a= -2\kappa l =-N^2$.
This shows that the bifurcation horizon $\mathcal{H}$ is at the location where $\xi^a\xi_a=0$.
The ``acceleration'' of this Killing vector is given by 
$a^i = \xi^b \nabla_b \xi^i = \kappa^2 (T,X)_I $ 
and hence, on the horizon, $a^i = \kappa \xi^i $ consistent with \eq{defkappa1}. 
It is also easy to see that, on the horizon, $\xi^a \to \kappa X k^a$ with $k^a = (1,1)_I$.

(iii) Another natural vector which arises in the Rindler frame is the  four-velocity $u^a$ of observers, moving along the orbits of the Killing vector 
$\xi^a $.  On the horizon $\mathcal{H}$,
this four-velocity has the limiting behaviour
$N u^a \to \kappa Xk^a$.

(iv) Lastly, we introduce  the unit normal $r_a$ to $l=$ constant surface, 
which also has the limiting behaviour
$Nr_a\to \kappa X k^a$
when we approach the horizon. 
It therefore follows that 
  $Nu^i, Nr^i, a^i$ and $\xi^i$
all tend to vectors proportional to $k^i$ on the horizon.
These  facts will prove to be useful   in our later discussions.

  \section{Thermodynamics of horizon: A first look}\label{sec:thermofirst}
  
  We shall now provide a general argument which associates a non-zero temperature with a bifurcation horizon. This argument, originally due to T.D. Lee, (\cite{lee}; also see ref. \cite{Unruh:1983ac}) is quite powerful and elegant and applies to all the horizons which we will be interested in. It uses techniques from path integral approach to quantum field theory, which we shall first review briefly.
  
\subsection{Review of Path Integral approach}\label{sec:pireview}  

  It is known in quantum mechanics that the 
net probability amplitude $K(2;1)$ for the particle to 
go from the event ${\cal P}_1$ to the event ${\cal P}_2$ is obtained by adding
up the amplitudes for all the paths connecting the events:
\begin{equation} 
K(\mathcal{P}_2;\mathcal{P}_1)\equiv K(t_2,q_2;t_1,q_1)=\sum_{{\rm paths}} \exp[i {\cal A}({\rm path})] 
\label{kerdef} 
\end{equation}
where $\mathcal{A}$(path) is the action evaluated for a given path connecting the end points
$\mathcal{P}_1$ and $\mathcal{P}_2$.
The addition of the {\it amplitudes} allows for the quantum mechanical interference
between the paths.  
The quantity $K(t_2,q_2;t_1,q_1)$ contains the full dynamical information about the quantum mechanical system.
Given $K(t_2,q_2;t_1,q_1)$ and the initial amplitude $\psi(t_1,q_1)$ for the particle to be 
found at $q_1$, we can compute the wave function $\psi(t,q)$ at any later time
by the usual rules for combining the amplitudes:
\begin{equation} \psi(t,q)=\int dq_1 K(t,q;t_1,q_1) \psi(t_1,q_1) 
\label{qwave}
\end{equation}
The above expressions continue to hold even when we deal with several degrees of freedom
$q_1, q_2, ...$ which may still be collectively denoted as $q=\{q_i\}$; it is understood that  the integral  in \eq{qwave} has to be performed over all the degrees of freedom.

To obtain the corresponding results in field theory, one needs to go from a discrete set of degrees of freedom (labeled by $i = 1,2,....$) to a continuum of variables denoting the coordinates $\mathbf{x}$ in a spacelike hypersurface. In this case the dynamical variable at time $t=t_1$ is the field configuration $q(\mathbf{x})$. (For every value of $\mathbf{x}$ we have one degree of freedom.) The integral in \eq{qwave} now becomes a functional integral over the initial field configuration and \eq{qwave} becomes 
\begin{equation} \psi(t,q(\mathbf{x}))=\int \mathcal{D}q_1 K(t,q(\mathbf{x});t_1,q_1(\mathbf{x})) \psi(t_1,q_1(\mathbf{x})) 
\label{qwave1}
\end{equation}
We shall, however, not bother to indicate this difference between field theory and point quantum mechanics and will continue to work with latter since the generalizations will be quite obvious by context.

We will next obtain a relation between the ground state wave function of the system and the path integral
kernel which will prove to be useful. 
In the conventional approach to quantum mechanics, using the Heisenberg
picture, we will   describe the system in terms of the position and momentum operators $\hat q$
and $\hat p$. Let $|q,t\rangle  $ be the eigenstate of the operator $\hat q (t)$ with eigenvalue
$q$.
The kernel --- which represents the probability amplitude for a particle
 to propagate
from $(t_1,q_1)$ to $(t_2,q_2)$ --- can be expressed, in a more conventional
notation, as the matrix element: 
\begin{equation}K(t_2,q_2;t_1,q_1)=\langle q_2,t_2|t_1,q_1\rangle  
=\langle q_2,0|\exp [-{i}\hat H(t_2-t_1)]|0,q_1\rangle  .
\end{equation}
where $\hat H$ is the time-independent Hamiltonian describing the system.
This relation allows one to represent the kernel in terms of the energy eigenstates
of the system. We have
\begin{eqnarray} 
K(T,q_2;0,q_1)&=& \langle q_2,0|\exp -{i}HT|0,q_1\rangle  \nonumber\\
&=&\sum_{n,m}
\langle q_2|E_n\rangle  \langle E_n|\exp -{i}HT|E_m\rangle  \langle E_m|q_1\rangle  \nonumber\\
          &=&\sum_n\psi_n(q_2)\psi^*_n(q_1)\exp (-{i}E_nT)
	  \label{ASninety}
	  \end{eqnarray}
where $\psi_n(q)=\langle q|E_n\rangle  $ is the n-th energy eigenfunction of the system under consideration.
Equation~(\ref{ASninety}) allows one to express the kernel in terms of the eigenfunctions of the Hamiltonian.
For any Hamiltonian which is bounded from below it is convenient to add  a constant to the 
Hamiltonian so that the ground state --- corresponding to the $n=0$ term in the 
above expression ---  has zero energy. We shall assume that this is done.
Next, we will analytically continue the expression
in \eq{ASninety} to imaginary values of $T$ by writing $iT = T_E$.
The Euclidean kernel obtained from \eq{ASninety}  has the form 
\begin{equation}
K_E(T_E, q_2; 0, q_1) = \sum_n \psi_n (q_2)\psi_n^* (q_1) \exp(-E_n T_E)
\end{equation} 
Suppose we now set $q_1 =q$, $q_2 =0$ in the above expression and take the limit
$T_E\to \infty$. In the large time limit, the exponential will suppress all the terms
in the sum except the one with $E_n=0$ which is the ground state for which the wave function is real.
We, therefore, obtain the result
\begin{equation}
\lim_{T\to\infty} K(T,0;  0,q)\approx
\psi_0(0)\psi_0(q_1)\propto \psi_0(q)
\label{euclzero}
\end{equation}
Hence the ground state wave function can be obtained
by analytically continuing the kernel into imaginary time and
taking a suitable limit. The proportionality constant in \eq{euclzero} is irrelevant since it can
always be obtained by normalizing the wave function $\psi_0(q)$. Hence we have
\begin{equation}
  \psi(q) \propto K(\infty,0; 0,q) = K(0,q;-\infty,0)
  \label{twoways}
 \end{equation} 
 where, in the arguments of $K$, the first one refers to Euclidean time and the second one refers to the dynamical variable. The last equality is obtained by noting that in
 \eq{euclzero} we can take the limit $T\to \infty$ either by $(t_2\to \infty, t_1=0)$
 or by $(t_2 = 0, t_1 \to -\infty)$. 
This result holds for any closed system with bounded Hamiltonian. 
Expressing the kernel as a path integral we can write this result in the form
\begin{equation}
 \psi_0(q) = \int_{T_E=0,q}^{T_E=T,0} \mathcal{D}q\, e^{-A} 
 \label{gsfrpi}
\end{equation} 
This formula is also valid in field theory if $q$ is replaced by the field configuration $q(\mathbf{x})$.

The analytic continuation to imaginary values of time also has close mathematical
connections with the description of systems in thermal bath.
To see this, consider the mean value of some observable $\mathcal{O}(q)$ 
of a quantum mechanical system. If the system is in an energy eigenstate described
by the wave function $\psi_n(q)$, then the expectation value of $\mathcal{O}(q)$   can 
be obtained by integrating $\mathcal{O}(q)|\psi_n(q)|^2$ 
over $q$. If the system is in a thermal bath at temperature
$\beta^{-1}$, described by a canonical ensemble, 
then the mean value has to be computed by averaging over all the energy eigenstates \textit{as well}
with a weightage $\exp(-\beta E_n)$. In this case, the mean value
can be expressed as 
\begin{equation}
\langle \mathcal{O} \rangle = \frac{1}{Z} \sum_n \int dq\, \psi_n(q) \mathcal{O}(q) \psi^*_n(q)\,  e^{-\beta E_n} \equiv \frac{1}{Z}\int dq\, \rho (q,q) \mathcal{O}(q)
\label{expO}
\end{equation} 
where $Z$ is the partition function and we have defined a \textit{density matrix} $\rho(q,q') $ by
\begin{equation}
\rho(q,q') \equiv \sum_n \psi_n(q)\psi^*_n(q')\, e^{-\beta E_n}
\label{denmatdef}
\end{equation} 
in terms of which  we can rewrite \eq{expO} as
\begin{equation}
\langle \mathcal{O} \rangle = \frac{\mathrm{Tr}\, (\rho \mathcal{O})}{\mathrm{Tr}\, (\rho)}
\end{equation} 
where the trace operation involves setting $q=q'$ and integrating over $q$.
This standard result shows how $\rho(q,q')$ contains information about 
both thermal and quantum mechanical averaging.
Comparing \eq{denmatdef} with \eq{ASninety} 
we find that the density matrix can be immediately obtained from the Euclidean
kernel by:
\begin{equation}
\rho(q,q')  = K_E(\beta,q; 0,q')
\end{equation} 
with the Euclidean time acting as inverse temperature.

  \subsection{Horizon temperature from a path integral}\label{thermalden}
  
  We shall now consider the quantum field theory in a spacetime with a horizon which  can be described in two different coordinate systems. The first one ($T,X,\mathbf{x}_\perp$) is a global coordinate system which covers the entire spacetime manifold (which could be the inertial Cartesian coordinate system 
  in flat spacetime or the Kruskal-like coordinate system in the case of spherically symmetric metrics with  horizon).  The second one ($t,x,\mathbf{x}_\perp$)
  covers the four different quadrants of the spacetime and is related to the first set  
 by a set of transformations similar to \eq{keytransform}. 
 We shall now show  that the global vacuum state defined on the surface $T=0$ appears as a thermal state to observers confined to the right wedge $\mathcal{R}$ with a temperature $\beta^{-1} = (2\pi/\kappa)$.

\begin{figure}[htbp]
\begin{center}
\includegraphics[scale=0.5]{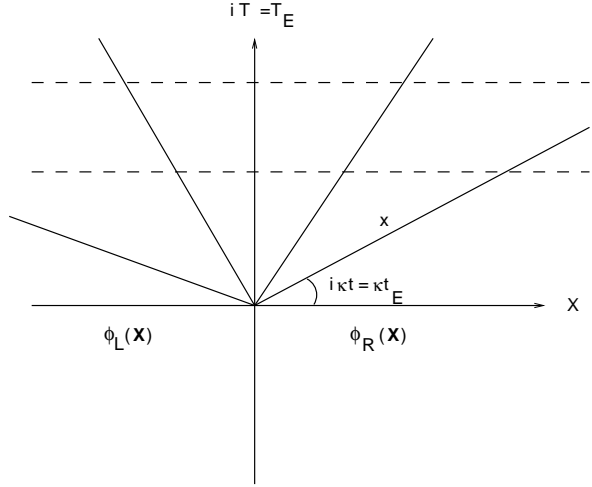}
\caption{Thermal effects due to a horizon; see text for a discussion.}
\label{fig:horizon}
\end{center}
\end{figure}

 On analytic continuation to imaginary time, the two sets of coordinates behave as shown in \fig{fig:horizon}. The key new feature is that $t_E$ becomes an angular coordinate having a periodicity $(2\pi/\kappa)$.
 While the evolution in $T_E$
  (effected by the inertial Hamiltonian $H_I$)
  will take the field configuration from 
 $T_E =0$ to $T_E\to \infty$, the same time evolution gets mapped in terms of 
 $t_E$ into evolving the ``angular'' coordinate $t_E$ from $0$ to $2\pi/\kappa$ 
  and is effected by the Rindler Hamiltonian $H_R$. (This 
  should be  clear from
 Fig.~\ref{fig:horizon}.) It is obvious that the entire upper half-plane
 $T>0$ is covered in two completely different ways in terms of the evolution
 in $T_E$ compared to evolution in $t_E$. 
 In $(T_E,X)$ coordinates, we vary $X$ in the range $(-\infty,\infty)$ for each $T_E$ 
 and vary $T_E$ in the range $(0,\infty)$. In $(t_E,x)$ coordinates, we vary $x$ in the range $(0,\infty)$ for each $t_E$ 
 and vary $t_E$ in the range $(0,\pi/\kappa)$.
 This fact allows us to prove that 
 \begin{equation}
   \langle {\rm vac} |\phi_L , \phi_R \rangle \propto \langle \phi_L \vert e^{-\pi H_R/\kappa } \vert \phi_R \rangle
   \label{qkeyvac}
   \end{equation} 
   as we shall see below.
    
To provide a simple  proof of \eq{qkeyvac}, let us consider
the ground state wave functional $\langle {\rm vac} |\phi_L , \phi_R\rangle $ in the extended spacetime
expressed as a path integral. From \eq{gsfrpi} we know that the ground state wave functional can be represented
as a Euclidean path integral of the form
\begin{equation}
\langle {\rm vac} |q\rangle \propto \int_{T_E=0;\phi=q}^{T_E=\infty;\phi=0}
\mathcal{D}\phi e^{-A}\label{euclpath1}
\end{equation}
where $T_E=iT $ is the Euclidean time coordinate and
 we have denoted the field configuration on the $T=0$ hypersurface by
$q(\mathbf{x})$. But we know that this field configuration can also be specified uniquely by specifying $\phi_L(\mathbf{x})\equiv q(\mathbf{x})$ with $X<0$ and $\phi_R(\mathbf{x})\equiv q(\mathbf{x})$ with $X>0$. Hence we can write the above result in terms of $\phi_L$ and $\phi_R$ as 
\begin{equation}
\langle {\rm vac} |\phi_L , \phi_R \rangle \propto \int_{T_E=0;\phi=(\phi_L,\phi_R)}^{T_E=\infty;\phi=(0,0)}
\mathcal{D}\phi e^{-A}\label{euclpath}
\end{equation}
   From Fig.~\ref{fig:horizon}  it is obvious that
  this path integral could also be evaluated in the polar coordinates by varying the angle
  $\theta =\kappa  t_E$ from 0 to $\pi$. When $\theta =0$ the field configuration corresponds to 
  $\phi = \phi_R$ and when $\theta = \pi$ the field configuration corresponds to $\phi = \phi_L$.
  Therefore \eq{euclpath} can also  be expressed as
  \begin{equation}
\langle {\rm vac} |\phi_L , \phi_R \rangle \propto \int_{\kappa t_E=0;\phi=\phi_R}^{\kappa t_E=\pi;\phi=\phi_L}
\mathcal{D}\phi e^{-A}
\label{rindact}
\end{equation}
   But in the
    Heisenberg picture, `rotating' from $\kappa t_E =0$ to 
    $\kappa t_E =\pi$ is a time evolution governed by the Rindler Hamiltonian $H_R$. So the path integral \eq{rindact} can be represented  as a matrix element of the Rindler Hamiltonian $H_R$
    giving us the result:   
    \begin{equation}
\langle {\rm vac} |\phi_L , \phi_R \rangle \propto \int_{\kappa t_E=0;\phi=\phi_R}^{\kappa t_E=\pi;\phi=\phi_L}
\mathcal{D}\phi e^{-A} 
= \langle \phi_L|e^{-(\pi/\kappa )H_R}|\phi_R\rangle
\end{equation}
proving \eq{qkeyvac}.

If we denote the proportionality constant in \eq{qkeyvac} by $C$, then the normalization condition
   \begin{eqnarray}
 1&=& \int \mathcal{D} \phi_L\, \mathcal{D} \phi_R\, \big|\langle {\rm vac} |\phi_L , \phi_R \rangle \big|^2 
= \int \mathcal{D} \phi_L\, \mathcal{D} \phi_R \, \langle {\rm vac} |\phi_L , \phi_R \rangle  
\langle \phi_{L} , \phi_{R} |{\rm vac}\rangle\nonumber\\
&=& C^2 \int \mathcal{D} \phi_L\, \mathcal{D} \phi_R\, \langle\phi_L|e^{-\pi H_R/\kappa} | \phi_R \rangle \, \langle \phi_R|e^{-\pi H_R/\kappa} | \phi_L\rangle
= C^2 \, {\rm Tr}\, \left(e^{-2\pi H_R/\kappa}\right)\nonumber\\
\end{eqnarray}  
     fixes the proportionality constant $C$, allowing us to write \eq{qkeyvac} in the  form: 
\begin{equation}
\langle {\rm vac} |\phi_L , \phi_R \rangle = \frac{\langle \phi_L \vert e^{-\pi H/\kappa } \vert \phi_R \rangle}{ \left[ {\rm Tr}(e^{-2 \pi H/\kappa })\right]^{1/2} } \label{eqn:fifteen}
\end{equation}
From this result, we can compute the density matrix for observations confined to the Rindler wedge $\mathcal{R}$ by tracing out the field configuration $\phi_L$ on the left wedge.
We get:
\begin{eqnarray}
\rho (\phi_R,\phi'_R)
 &=& \int \mathcal{D}\phi_L \langle {\rm vac} \vert \phi_L, \phi_R \rangle  \langle   \phi_L,\phi'_R \vert {\rm vac} \rangle 
 \nonumber \\
&&\hskip -5em = \int \mathcal{D}\phi_L  \frac{  \langle \phi_R \vert e^{-(\pi/\kappa) H_R } \vert \phi_L \rangle \langle \phi_L \vert e^{-(\pi/\kappa) H_R}\vert \phi'_R \rangle}{Tr(e^{-2\pi H_R/\kappa })}\nonumber\\
&&\hskip -5em= \frac{\langle \phi_R \vert e^{-(2\pi/\kappa) H_R } \vert \phi'_R \rangle  }{Tr(e^{-2\pi H_R/\kappa })} \label{eqn:twentytwo}
\end{eqnarray}
Thus, tracing over the field configuration $\phi_L$ in the region behind the horizon leads to a thermal density
matrix $\rho \propto \exp[-(2\pi/\kappa )H_R]$ for the observables in $\mathcal{R}$.

The main ingredients which have gone into this 
   result are the following. (i) The singular behaviour of the $(t,x)$ coordinate system
   near $x=0$ divides the $T=0$ hypersurface into two separate regions.
   (ii) In terms of \emph{real} $(t,x)$ coordinates, it is not possible to distinguish between the points
   $(T,X)$ and $(-T,-X)$ but the \emph{complex} transformation $t\to t\pm i\pi$ maps the point
   $(T,X)$ to the point $(-T,-X)$. That is, a rotation in the complex  plane (Re $t$, Im $t$)
   encodes the information contained in the full $T=0$ plane. 
 
 In fact, one can  obtain the expression for the density matrix directly from path integrals along the following lines. We begin with the standard relation in \eq{twoways} which gives
 \begin{equation}
  \langle \mathrm{vac}\vert q \rangle  = K(\infty,0; 0,q) = K(0,q;-\infty,0)= 
  \langle \mathrm{vac}\vert \phi_L,\phi_R\rangle
 \end{equation} 
 where, in the arguments of $K$, the first one refers to Euclidean time and the second one refers to the dynamical variable.  The density matrix used by the observer in the right wedge can be expressed as the integral 
 \begin{eqnarray}
 \label{keyintegral}
  \rho (\phi_R, \phi_R') &=& \int \mathcal{D}\phi_L 
  \langle \mathrm{vac}\vert \phi_L,\phi_R\rangle \langle \mathrm{vac}\vert \phi_L,\phi_R'\rangle\\
  &=& \int \mathcal{D}\phi_L 
  K(\infty,(0,0); 0,(\phi_L,\phi_R))K(0,(\phi_L,\phi_R'); -\infty,(0,0))\nonumber
 \end{eqnarray} 
where we explicitly decomposed $q(\mathbf{x})$ into the set $\phi_L(\mathbf{x}), \phi_R(\mathbf{x})$ everywhere.
The expression in the right hand side evolves the system from $T_E = -\infty$ to
$T_E=+\infty$ with some specific restrictions on the field configuration on the $T_E=0$ hypersurface. Since we are integrating out all the field configurations $\phi_L$, it follows that there is no restriction on the field along $X<0$. To handle the field configurations on $X>0$, we can proceed as follows. We first note that $T_E=0$ is the same as $t_E=0$
when $X>0$. Instead of considering $T_E=t_E=0$, let us consider an infinitesimally displaced hypersurface $t_E=\epsilon$ in one of the kernels and $t_E=(2\pi/\kappa) -\epsilon$
in the second kernel. That is, instead of specifying $\phi_R(\mathbf{x})$
and $\phi_R'(\mathbf{x})$ at $t_E=0$ we will specify $\phi_R(\mathbf{x})$
at $t_E=\epsilon$ and $\phi_R'(\mathbf{x})$ at $t_E=(2\pi/\kappa) -\epsilon$. It is then obvious that the result of the integral in \eq{keyintegral} is the propagation kernel
that propagates $\phi_R(\mathbf{x})$ at $t_E=\epsilon$ to  $\phi_R'(\mathbf{x})$ at $t_E=(2\pi/\kappa) -\epsilon$  in the Rindler time. Taking the $\epsilon \to 0$ limit is now trivial and we get 
\begin{eqnarray}
\label{denmat}
  \rho (\phi_R, \phi_R') &=& \int \mathcal{D}\phi_L 
  \langle \mathrm{vac}\vert \phi_L,\phi_R\rangle \langle \mathrm{vac}\vert \phi_L,\phi_R'\rangle\\ 
 & =& K((2\pi/\kappa), \phi_R'; 0, \phi_R)
  = \langle \phi_R'|\exp[-(2\pi/\kappa)H_R] |\phi_R\rangle\nonumber
  \end{eqnarray}
  where $H_R$ is the Rindler  Hamiltonian.
  Comparing the first and last expressions we find that the operator corresponding to the density matrix is just $\rho = \exp[-(2\pi/\kappa)H_R]$. (This is an unnormalized
  density matrix since we have not bothered to normalize the wavefunctions.)

\subsection{Complex time and the region beyond the horizon}\label{sec:complext}

There are two points that need to be stressed regarding the above derivation. First,
the light cones described by $X^2 - T^2=0$ get mapped  to the origin of the Euclidean sector through the equation $X^2 + T_E^2 =0$. Consequently the quadrants $\mathcal{F}$ and $\mathcal{P}$ \textit{disappear} from the Euclidean sector --- or rather, they collapse into the 
origin. This implies that the region beyond the horizon is not covered by the Euclidean coordinates. The second point is that even though  the horizon 
collapses to a single point in the Euclidean sector, the \textit{Euclidean Rindler time} $t_E$
contains information about the left quadrant $\mathcal{L}$. To see this, we only have to compare \eq{expthree} taken with the positive sign  and \eq{expeuclid}.
When  $t$ ranges from $-\infty$ to $+\infty$, the coordinate $X=x\cosh \kappa t$ remains positive. On the other hand, when the corresponding Euclidean time $t_E$ varies from $0$ to 
$2\pi/\kappa$, the coordinate  $X=x\cos\kappa t_E$ covers both the right wedge $\mathcal{R}$
and the left wedge $\mathcal{L}$. Therefore the range of Euclidean Rindler time from  $t_E=\pi/2\kappa$ to $t_E=-\pi/2\kappa$
covers the region beyond the horizon \cite{loctemp}. It is because of this peculiar feature (which, of course, is closely related to periodicity in the imaginary time) that we can obtain
the thermal effect due to horizon from the Euclidean approach. The conclusions are strengthened by  a few other considerations which are worth mentioning briefly.

 To begin with, note that similar results arise in a more general context for \emph{any} system described by 
     a wave function $\Psi(t,l; E) = \exp[iA(t,l; E)]$ in the WKB approximation \cite{Padmanabhan:2004tz}.
      The dependence of the quantum mechanical
     probability $P(E) =|\Psi|^2$ on the energy $E$ can be quantified in terms of the derivative
     \begin{equation}
     \frac{\partial \ln P}{\partial E} \approx -\frac{\partial}{\partial E}2(\textrm{Im} A) = 
     -2 \textrm{Im} \left(\frac{\partial A}{\partial E}\right)
     \label{pofe}
     \end{equation}
     in which the dependence on $(t,l)$ is suppressed. 
     Under normal circumstances, action will be real in the leading order approximation
     and the imaginary part will vanish. (One well known counter-example is in the case of tunneling
     in which the action acquires an imaginary part; Eq.~(\ref{pofe}) then correctly describes
     the dependence of tunneling probability on the energy.)
     For any Hamiltonian system, the quantity $(\partial A/\partial E)$ can be set to a constant
     $t_0$ to determine the trajectory of the system: $(\partial A/\partial E)=-t_0$.
     Once the trajectory is known, this equation determines $t_0$ as a function of 
     $E$ [as well as $(t,l)$]. Hence we can write 
     \begin{equation}
     \frac{\partial \ln P}{\partial E} \approx 2\textrm{Im} \left[ t_0 (E)           \right]
     \label{imto}
     \end{equation}
     From the trajectory in  Eq.~(\ref{photonpath}) which is valid near the horizon,  we note that $t_0(E)$
     can pick up an imaginary part if the trajectory of the system crosses the 
     horizon. In fact, since $\kappa t \to\kappa  t-i\pi$ changes $X$ to $-X$ 
     [see Eqs.~(\ref{expone},\ref{exptwo},\ref{keytransform})], the imaginary
     part is given by $(-\pi/\kappa  )$ leading to  $(\partial \ln P/\partial E) = -2\pi/\kappa $.
     Integrating, we find that  the probability for the trajectory of any system to cross the horizon,
     with the energy $E$, will be given by the Boltzmann
     factor 
     \begin{equation}
     P(E) \propto \exp \left[- \frac{2\pi}{\kappa } E\right] = P_0\exp \left[- \beta  E\right] 
     \end{equation}
      with temperature $T=\kappa /2\pi$. (For special cases of this general result see \cite{Keski-Vakkuri:1997xp} and references
      cited therein.)

 It is also interesting to examine how these results relate to the more formal approach to quantum field theory.
       The relation between quantum field theories in 
   two sets of coordinates $(t,{\bf x})$ and $(T,{\bf X})$,
   related by  
        Eq.~(\ref{keytransform}),  with the metric being 
     static in the $(t,{\bf x})$ coordinates can be 
     described as follows:  
  Static nature suggests a natural decomposition of wave modes
     as 
     \begin{equation}
     \phi(t,{\bf x}) = \int d\omega  [a_\omega  f_\omega  ({\bf x}) e^{-i\omega t} 
     + a_\omega ^\dagger f_\omega ^* ({\bf x}) e^{i\omega t}]
     \end{equation}
     in $(t,{\bf x})$ coordinates.
     These modes, however,  behave badly (as $x^{\pm i\omega/\kappa }$; see \eq{twotwo})
     near the horizon since the metric is singular near the horizon in these coordinates. 
     We could, however, expand $\phi(t,{\bf x})$ in terms of some other set of 
     modes $F_\nu (t,{\bf x})$ which are well behaved at the horizon. This could, for example, 
     be done by solving the wave equation in $(T,{\bf X})$ coordinates and rewriting the solution
     in terms of $(t,{\bf x})$. This gives an alternative expansion for the field:
     \begin{equation}
     \phi(t,{\bf x}) = \int d\nu [A_\nu F_\nu (t,{\bf x})  + A_\nu^\dagger F_\nu^* (t,{\bf x}) ]
     \end{equation}
     Both these sets of creation and annihilation operators define two 
     different vacuum states $a_\omega|0\rangle _a =0, A_\nu |0\rangle _A=0$.
     The modes $F_\nu(t,{\bf x})$ will contain both positive and negative frequency components
     with respect to $t$
     while the modes $f_\omega ({\bf x}) e^{-i\omega t}$ are pure positive frequency components.
     The positive and negative frequency components of $F_\nu(t,{\bf x})$ 
      can be extracted through the Fourier transforms:
        \begin{equation}
     \alpha_{\omega \nu} = \int_{-\infty}^\infty dt \ e^{i\omega t} F_\nu (t, {\bf x}_f); \quad
      \beta _{\omega \nu} = \int_{-\infty}^\infty dt \ e^{-i\omega t} F_\nu (t, {\bf x}_f)
     \label{bogo}      
     \end{equation}
     where ${\bf x}_f$ is some convenient fiducial location far away from the horizon.
     One can think of $|\alpha_{\omega \nu}|^2$ and $|\beta_{\omega \nu}|^2$ as
     similar to  unnormalized transmission and 
     reflection coefficients.  
    (They are very closely related to the Bogoliubov coefficients
     usually used to relate two sets of creation and annihilation operators.)
     The $a-$particles in the $|0\rangle _A$ state is determined by the quantity
     $|\beta_{\omega \nu}/\alpha_{\omega \nu}|^2$. If the particles are uncorrelated,
      then the standard relation $N|\alpha|^2 = (N+1)|\beta|^2$ between absorption and emission leads to the flux of 
      out-going particles:
     \begin{equation} 
     {N} = \frac{|\beta_{\omega \nu}/\alpha_{\omega \nu}|^2}{1-|\beta_{\omega \nu}/\alpha_{\omega \nu}|^2}
     \label{bogon}
     \end{equation}
     If the $F$ modes are chosen to be regular near the horizon, varying as
     $\exp(-i \Omega  U)$ etc., then Eq.~(\ref{keytransform}) shows that
     $F_\nu(t,{\bf x}_f) \propto \exp(-i\Omega q e^{-\kappa t})$ etc.  The integrals
     in Eq.~(\ref{bogo}) again reduces to the Fourier transform of an exponentially
     redshifted wave and we get $|\beta_{\omega \nu}/\alpha_{\omega \nu}|^2 = e^{-\beta \omega}$
     and Eq.~(\ref{bogon}) leads to the Planck spectrum. 
     This is the quantum mechanical version of Eq.~(\ref{expodamp}) and Eq.~(\ref{planck}).

Finally, one can relate the above result to the  analyticity properties of wave modes of a scalar field in the $U =(T-X)$ coordinates
and $u=(t-x)$ coordinates.  Since the positive frequency mode
solution to the wave equation in the $(T,X)$ coordinates has the form 
$\exp(-i\Omega U)$ (with $ \Omega>0$)  and is analytic in the lower half of complex $U$ plane,
     any arbitrary superposition of  such modes with different (positive) values of $\Omega$ 
     will also be analytic
     in the  lower half of complex $U$ plane. Conversely, if we construct a mode which is 
     analytic in the lower half of complex $U$ plane, it can be expressed as a superposition of purely positive 
frequency modes \cite{Unruh:1976db}.
   From the transformations in Eq.~(\ref{keytransform}), we find that the 
    positive frequency  wave mode near the horizon, $\phi=\exp(-i\omega u)$ can be expressed as
    $\phi\propto U^{i\omega/\kappa }$ for $U<0$. If we interpret this mode as $\phi\propto (U-i\epsilon)^{i\omega/\kappa }$
    then, this mode is analytic throughout the lower half of complex $U$ plane.  We can then interpret the mode as
    \begin{equation}
     (U-i\epsilon)^{i\omega/\kappa } =
     \cases{
    e^{[i(\omega/\kappa)\ln U]}&  (for $U>0$)\\
   e^{\pi\omega/\kappa}e^{[(i\omega/\kappa)\ln |U|) }&   (for
      $U<0$)\\
    }
    \label{analyticity}
    \end{equation}  
    This interpretation
    of $\ln (-U)$ as $\ln|U| -i\pi=\kappa u-i\pi=\kappa t-\xi-i\pi$ is consistent with 
    the procedure of replacing  $\kappa t\to \kappa t-i\pi$ to go from $X>0$
    to $X<0$. This is precisely what happens in the Euclidean continuation. The factor
    $e^{\pi\omega/\kappa}$ in the second line of the \eq{analyticity} leads to the thermal effects in the conventional picture.

\subsection{Hawking radiation from black holes}\label{sec:hawrad}

The description in the previous sections shows that
the vacuum state defined in a coordinate system ---  which covers the full manifold --- appears
as a thermal state to an observer who is confined to part of the manifold
partitioned by a horizon.\index{black hole!Hawking radiation}\index{Hawking radiation!black hole} 
This result (and the analysis)  will hold for
any static spacetime 
with a bifurcation horizon, like the Schwarzschild spacetime, de Sitter spacetime etc.
All these cases describe a situation in    thermal \textit{equilibrium} at a temperature 
$T= \kappa/2\pi$ (where $\kappa$ is the surface gravity of the horizon) as far as an 
observer confined to the region $R$ is concerned.

A completely different phenomenon arises in the case of a dynamical situation like, for example, the collapse of  a spherically symmetric massive body to form a black hole.
In this case, time reversal invariance is explicitly broken.
The study of a quantum field theory in such a context shows that, at late times,
there will be a flux of radiation flowing towards the future null infinity
with a Planckian spectrum corresponding to a temperature $\kappa/2\pi$.
 This process is called \textit{black hole evaporation}.

 This result is conceptually different from associating a temperature with the horizon.
In the case of a Rindler spacetime, for example, there is no steady flux of radiation
propagating towards future null infinity even though an observer confined to
the region $\mathcal{R}$ will interpret the expectation values of operators as thermal averages
corresponding to a temperature $\kappa/2\pi$. This corresponds to a situation
which \textit{is} time reversal invariant characterized by thermal equilibrium.
The black hole evaporation, in contrast, is an irreversible process.

We shall now work out the corresponding result for  a scalar field in the
time dependent  metric generated by
 collapsing matter. The scalar field can be decomposed into positive and negative frequency
modes in the usual manner. 
We choose these modes in such a way that at early times they  correspond to a vacuum state.
 In the presence of collapsing matter, these modes evolve 
at late times to those with exponential redshift, thereby
 leading 
   to  thermal behaviour. 

To do this, we need an explicit model
for the collapsing matter. 
  Since only the exponential redshift of the modes at late times is relevant as far as the thermal spectrum
  is concerned, the result should be independent of the detailed nature of the collapsing matter.
  So  
we shall choose a simple model for the formation of the black hole, 
 based on a  spherical shell  of mass
  $M$ that collapses under its own weight. 
  The metric inside the shell will be flat while the one outside will be Schwarzschild.
  Further, the angular coordinates do not play a significant role
  in this analysis, allowing us to work in the 2-dimensional $(t,r)$ subspace. 
  
  The line element  outside and inside the collapsing,
   spherically symmetric, collapsing shell
  is taken to be 
  \begin{equation}
  ds^2 = 
  \cases{
  -C(r)\, du dv = - \left( 1- \frac{2M}{r}\right) du dv& ({exterior})\\
  - dU dV & ({interior})\\
        }
  \end{equation}
  where
  \begin{equation}
   u=t-\xi +R^*_0;\quad v=t+\xi -R^*_0; \quad \xi  = \int dr\, C^{-1}; 
   \label{defuv}
  \end{equation} 
  \begin{equation}
U=\tau- r+R_0,\qquad V=\tau+ r-R_0
\label{defUV}
\end{equation}  
   The $R_0$ and $R_0^*$ are constants related in the same manner as
   $r$ and $\xi $.

     Let us assume that, for $\tau\le 0$, matter was at rest with its surface at 
   $r=R_0$ and for $\tau>0$, it collapses inward along the trajectory $r=R(\tau)$. The coordinates
   have been chosen so that at the onset of collapse ($\tau=t=0$) we have $u=U=v=V=0$
   at the surface. Let the  coordinate transformations between the interior and exterior be given by
   the functional forms $U=f(u)$ and $v=h(V)$. Matching the geometry along the trajectory
   $r=R(\tau)$, 
   requires the condition 
   \begin{equation}
    C\frac{du}{dU} =  \frac{dV}{dv}; \qquad (\mathrm{on}\  r=R(\tau)) 
   \end{equation} 
   Using \eq{defuv} and \eq{defUV} along the trajectory, this equation can be 
   simplified to give 
   \begin{equation}
    \left( \frac{dt}{d\tau}\right)^2 = \frac{\dot R^2}{C^2}+ \frac{1}{C} \left( 1 - \dot R^2\right)
   \end{equation} 
   where $\dot R $ denotes $dR/d\tau$ and $U,V$ and $C$ are evaluated along $r=R(\tau)$.
  Using this again in the definition of $u,v$ etc., it is easy to show that
   \begin{equation}
   \frac{dU}{du} = \frac{df}{du} = (1-\dot R)C \left( \left[ C ( 1-\dot R^2) + \dot R^2\right]^{1/2} - \dot R\right)^{-1}
   \label{fofu}
   \end{equation}
   \begin{equation}
  \frac{dv}{dV} = \frac{dh}{dV} = \frac{1}{C(1+\dot R)}\left( \left[ C ( 1-\dot R^2) 
  + \dot R^2\right]^{1/2} + \dot R\right)
  \label{hofv}
   \end{equation}
   Since $\dot R < 0$ for the collapsing shell, we should take
   $(\dot R^2)^{1/2} = - \dot R$.

  We  now introduce a massless scalar field in this spacetime  which satisfies the equation $\square \phi =0$.
  As the modes of the scalar field propagate inwards they will reach $r=0$ and re-emerge as out-going modes.
  In the $(t,r)$ plane, this requires reflection of the modes on the $r=0$ line, which 
  corresponds to $V=U-2R_0$. 
  Since the modes vanish for $r<0$, continuity requires $\phi=0$ at $r=0$.
  The solutions to the 2-dimensional wave equations $\Box \phi =0$
  which (i) vanish on the line $V=U-2R_0$ and (ii) reduce to standard exponential form in the 
  remote past, can be determined by noting that, along $r=0$ we have
  \begin{equation}
  v=h(V)=h[U-2R_0] = h[f(u)-2R_0]
  \end{equation}
  where the square bracket denotes functional dependence of $h$ on its argument.
  Hence the solution is 
  \begin{equation}
  \Phi = \frac{i}{\sqrt{4\pi \omega}} \left( e^{-i\omega v} - e^{-i\omega h[f(u)-2R_0]}\right)
  \label{modes}
  \end{equation}
  Given the trajectory $R(\tau)$, one can integrate  Eq.~(\ref{fofu}) to obtain $f(u)$ and use
  Eq.~(\ref{modes}) to completely solve the problem. This will describe time-dependent particle production
  from some collapsing matter distribution and --- in general --- the results will depend on the details of the collapse \cite{Brout:1995rd,aseemtpbh09}.
  
  The analysis, however, simplifies considerably and a universal
  character emerges if the collapse proceeds to form a horizon on which $C\to 0$. Near $C=0$, equations
  (\ref{fofu}) and (\ref{hofv}) simplify to 
  \begin{equation}
  \frac{dU}{du} \approx\frac{\dot R -1}{2\dot R} C(R); \quad \frac{dv}{dV} \approx\frac{(1-\dot R )}{2\dot R} 
  \label{nearhori}
  \end{equation}
  where we have used the fact that $(\dot R^2)^{1/2}=-\dot R$ for the collapsing
  solution. Further, near $C=0$, we can expand $R(\tau)$ as 
  $R(\tau) = R_h +\nu(\tau_h - \tau) + \mathcal{O} [(\tau_h - \tau)^2]$
  where $R=R_h$ at the horizon and $\nu=-\dot R(\tau_h)$. 
  We have denoted by $\tau_h$ the time at which the shell crosses the horizon.
  Integrating Eq.~(\ref{nearhori})  we get
  \begin{equation}
   \kappa  u \approx - \ln |U+R_h - R_0 -\tau_h| + \textrm{const}
  \label{uofU}
  \end{equation}
  where $\kappa =(1/2)(\partial C/\partial r)_{R_h}$ is the surface gravity
  and
  \begin{equation}
  v \approx \textrm{constant} -  V(1+ \nu) / 2\nu 
  \label{vofV}
  \end{equation}
  It is now clear that: (i) The relation between $v$ and $V$ is linear and hence holds no surprises. (ii) The relation between $U$ and $u$, which can be written as $U\propto \exp(- \kappa   u)$   signifies the exponential  redshift we have alluded to several times. The late time
  behaviour of out-going modes can now be determined using Eq.~(\ref{uofU}) and Eq.~(\ref{vofV}) in Eq.~(\ref{modes}).
  We get:
  \begin{equation}
  \Phi \cong  \frac{i}{\sqrt{4\pi \omega}} \left( e^{-i\omega v} 
  - \exp\left(i\omega\left[ ce^{- \kappa  u} + d\right]\right)\right)
  \label{latephi}
  \end{equation}
  where $c,d$ are constants. This mode with exponential redshift, can be expressed in terms of the modes $\exp(\mp \nu u)$ as
  \begin{equation}
\Phi_\omega(u) = \int_0^\infty \frac{d\nu}{2\pi} \left[ \alpha_{\omega\nu} \, e^{-i\nu u}+ \beta_{\omega\nu} e^{i\nu u}\right]
\end{equation}
Determining $\alpha_{\omega\nu},\ \beta _{\omega\nu}$ by Fourier transforming this relation, we get
\begin{equation}
  \alpha_{\omega\nu} =- \frac{i \nu e^{-i\omega d } }{4\pi\kappa \sqrt{\nu \omega}} 
  \left( \frac{-e^{-\kappa d}}{\omega c} \right)^{- i \nu/\kappa} 
   \  e^{\pi \nu/2\kappa} \Gamma (- i \nu/\kappa); \quad 
   \beta _{\omega\nu} = e^{- \pi \nu/\kappa } \alpha^*_{\omega \nu}
  \end{equation}  
  Note that these quantities {\it do}  depend on $c,d$ etc; but the modulus 
  \begin{equation}
  |\beta_{\omega\nu}|^2=\frac{1}{2\kappa} \frac{1}{\left[\exp(2\pi \nu/\kappa) -1)\right]}
  \end{equation} 
  is independent of these factors. (The mathematics is essentially the same as in Eqs.(\ref{powernu}), (\ref{planck})).
 This shows that the vacuum state at early times will be interpreted as containing 
  a thermal spectrum of particles at late times with temperature $T= \kappa/2\pi$.
  In the case of a black hole, $\kappa = (1/4M)$ and the temperature turns out to be
  $T= (1/8\pi M)$.

Our result implies that when spherically symmetric configuration of matter collapses to form a black hole, observers at large distances will receive a thermal radiation of particles from the black hole at late times. (It is possible to prove this more formally by considering the expectation values of the energy-momentum tensor of the scalar field; this will demonstrate the flux of energy to large distances). 
It seems natural to \textit{assume} that the source of this energy radiated to infinity is the mass of the collapsing structure. Strictly speaking, this is an extrapolation from our result because it involves changes in the background metric --- which was parametrized by $M$ --  due to the effect of the radiation   while our original result was based on a \textit{test} scalar field in a \textit{fixed} background metric. We shall, nevertheless, make this assumption and explore its consequences. Given the temperature of the black hole $T(E)=1/8\pi M$ as a function of the energy $E=M$, we can integrate the expression $dS=dE/T(E)$ to define an `entropy' $S(E)$ for  the black hole:\index{black hole!entropy}
\begin{equation}
S=\int_0^M(8\pi E)dE=4\pi M^2=\frac{1}{4}\left(\frac{A_{hor}}{L_P^2}\right)
\label{entbyint}
 \end{equation} 
where $A_{hor}=4\pi(2M)^2$ is the area of the $r=2M,t=$ constant surface and $L_P^2=G\hbar/c^3$ is the square of the Planck length. This result shows that the entropy obtained by integrating the $T(E)$ is proportional to the area of the horizon. 

This result connects up with several classical features of black holes. We know that the area of the horizon does not decrease during classical processes involving the black holes which suggests an analogy between horizon area and  entropy.  The \eq{firstlaw} for classical processes involving black holes  now acquires a direct thermodynamic interpretation. The factor $(\kappa/2\pi)$ and $(A/4)$ can indeed be identified with physical temperature and entropy. Note that classical analysis  can only identify these quantities up to a multiplicative factor.  On the other hand, the analysis of quantum fields in the \scm\ allows us to determine the temperature and entropy uniquely and the entropy turns out to be one-quarter of the area of the horizon expressed in Planck units. \footnote{There is  an ambiguity in the overall additive constant to entropy which is settled in  \eq{entbyint} by assuming that $S=0$ for $M=0$. This might appear reasonable but recall that $T\to\infty$ when $M\to0$; 
flat spacetime, treated as the $M\to 0$ limit of \scm, has infinite temperature rather than zero temperature. 
Hence, it is worth emphasizing that choosing $S=0$ for $M=0$ is a specific assumption.}

\subsection{Horizon entropy in generalized theories of gravity}\label{sec:noetherent}

 There has been considerable amount of work in analyzing the nature of horizon thermodynamics in theories different from general relativity. In a wide class of such theories, one does get solutions with horizons and one can associate a temperature and entropy with them. While the temperature can be identified from the periodicity of Euclidean time, determining the correct form of entropy is more non-trivial. We shall now briefly describe how these results arise in a class of theories which are natural generalizations of Einstein gravity. 

Consider a theory for gravity described by the metric $g_{ab}$ coupled to matter. We will take the action describing such a theory in $D-$dimensions to be
\begin{equation}
A=\int d^Dx \sqrt{-g}\left[L(R^{ab}_{cd}, g^{ab})+L_{matt}(g^{ab},q_A)\right]
\label{genAct}
\end{equation}  
where $ L$ is any scalar built from metric and curvature  and $L_{matt}$ is the matter Lagrangian depending on the metric and some matter variables $q_A$. (We have assumed that   $L$ does not involve derivatives of curvature tensor, to simplify the discussion; a more general structure is explored e.g., in ref.\cite{derofR}) Varying $g^{ab}$ in \eq{genAct} we get $\delta (L_{\rm matt} \sqrt{-g}) = - (1/2) \sqrt{-g} T_{ab} \delta g^{ab}$ and
 \begin{equation}
\delta(L\sqrt{-g }) =\sqrt{-g }\left( \mathcal{G}_{ab} \delta g^{ab} + \nabla_{a}\delta v^a\right). \label{variationL} 
\end{equation}
The variation of the gravitational Lagrangian
density generically leads to a surface term  which is   expressed 
by the $\nabla_a (\delta v^a) $ term. Ignoring this term for the moment (we will comment on this later) we get equations of motion (see e.g. Refs. \cite{ayan,mohut}) to be  $2\mathcal{G}_{ab}=T_{ab}$ where the explicit form of  $\mathcal{G}_{ab}$ is 
\begin{equation}
\mathcal{G}_{ab}=P_a^{\phantom{a} cde} R_{bcde}- \frac{1}{2} L g_{ab} - 2 \nabla^c \nabla^d P_{acdb} 
\equiv \mathcal{R}_{ab}-2 \nabla^c \nabla^d P_{acdb}
\label{genEab}
\end{equation}
where
\begin{equation}
P^{abcd} \equiv \frac{\partial L}{\partial R_{abcd}}
\end{equation}
(Our notation is based on the fact that $\mathcal{G}_{ab}=G_{ab},\mathcal{R}_{ab}=R_{ab}$ in Einstein's gravity.)
For any Lagrangian $L$, the functional
derivative $\mathcal{G}_{ab}$ satisfies the generalized off-shell Bianchi identity:
$
\nabla_a \mathcal{G}^{ab} = 0. 
$

Many such models have been investigated in the literature and  most of these models  have  black hole solutions. Whenever the black hole metric can be approximated by a Rindler metric near the horizon, it is possible to associate a temperature with the horizon, using the procedures described earlier, e.g. in Sec.(\ref{thermalden}). Associating the entropy is more nontrivial and we shall now indicate how this is usually done.
(For a rigorous proof, which we shall \textit{not} provide, see e.g., ref \cite{Wald:1993nt}. We will, however, provide an alternative route to this result in Section~\ref{sec:gravactfree}.)

In any generally covariant theory, the infinitesimal coordinate transformations $x^a \to x^a + \xi^a$ lead
to conservation of a Noether current which depends on $\xi^a$. 
To derive the expression for the Noether current, let us consider 
 the variations in $\delta g_{ab}$ which arise through the diffeomorphism $x^a \rightarrow x^a + \xi^a$. In this case, $\delta (L\sqrt{-g} ) = -\sqrt{-g} \nabla_a (L \xi^a)$, with $\delta g^{ab} = (\nabla^a \xi^b + \nabla^b \xi^a)$. Substituting these in \eq{variationL} and using $
\nabla_a \mathcal{G}^{ab} = 0 
$, we obtain the
conservation law $\nabla_a J^a = 0$, for the current,
\begin{equation}
J^a \equiv \left(  2\mathcal{G}^{ab} \xi_b + L\xi^a + \delta_{\xi}v^a \right)
=2\mathcal{R}^{ab} \xi_b+\delta_{\xi}v^a
\label{current}
\end{equation}
where $\delta_{\xi}v^a$ represents the boundary term which arises for the specific variation
of the metric in the form $ \delta g^{ab} = ( \nabla^a \xi^b + \nabla^b \xi^a$). 
Quite generally, the boundary term can be expressed as \cite{mohut},
\begin{equation}
\delta v^a = \frac{1}{2} \alpha ^{a (b c)} \delta g_{ bc} + \frac{1}{2} \beta ^{a (b c)}_{d} \delta \Gamma^{d}_{bc}
\label{genv}
\end{equation}
where, we have used the notation $Q^{i j} = Q^{i j} + Q^{j i}$. The coefficient $\beta ^{a b c d}$ arises from the derivative of $L_{grav}$ with respect to $R^{ a b c d}$ and hence possess all the algebraic symmetries of curvature tensor. 
In the special case of diffeomorphisms, $x^a \rightarrow x^a + \xi^a$, the variation
$\delta_\xi v^a$ is given by \eq{genv} with:
\begin{equation}
\delta g_{a b} = -\nabla_{( a} \xi_{b)}; \quad\delta \Gamma^{d}_{bc} = 
 -\frac{1}{2} \nabla_{(b} \nabla_{c)} \xi^d 
 + \frac{1}{2} R^{d}_{(b c) m} \xi^m
 \label{variations}
\end{equation}
Using the above expressions in \eq{current}, it is possible to write an explicit expression for the current $J^a$ for any diffeomorphism invariant theory.
 It is also convenient to introduce an anti-symmetric tensor $J^{ab}$ by $J^a = \nabla_b J^{ab}$. 
 For the general
class of theories we are considering, the $J^{ab}$ and $J^a$ can be expressed in the form
\begin{equation}
J^{ab} = 2 P^{abcd} \nabla_c \xi_d - 4 \xi_d \left(\nabla_c P^{abcd}\right)
\label{noedef}
\end{equation} 
\begin{equation}
J^a = - 2 \nabla_b \left (P^{adbc} + P^{acbd} \right ) \nabla_c \xi_d + 2 P^{abcd} \nabla_b \nabla_c \xi_d - 4 \xi_d \nabla_b \nabla_c P^{abcd} 
\end{equation} 
where $P_{abcd}\equiv (\partial L/\partial R^{abcd})$.
(The expression for $J_a, J_{ab}$ are  not unique. This ambiguity has been extensively discussed in the literature but for our purpose we will use the $J^a$ defined as above.)

 We shall see that, for most of our discussion,  we will not require the explicit form of $\delta_{\xi}v^a$ except for one easily proved  result:
$
\delta_{\xi}v^a=0
$
when $\xi^a$ is a Killing vector and satisfies the conditions
\begin{equation}
 \nabla_{( a} \xi_{b)} = 0;\quad  \nabla_a \nabla_b \xi_c = R_{c b a d} \xi^d
 \label{cond1}
\end{equation}
The expression for Noether current simplifies considerably when $\xi^a$ 
satisfies \eq{cond1}  and is given by
\begin{equation}
 J^a = \left( 2\mathcal{G}^{ab} \xi_b + L\xi^a  \right)
 =2\mathcal{R}^{ab} \xi_b
 \label{current1}
\end{equation}
The integral of $J^a$ over a spacelike surface defines the conserved Noether charge, $\mathcal{N}$. 

To obtain a relation between the horizon entropy and Noether charge, we first note that
 on-shell i.e., when field equations hold ($2\mathcal{G}_{ab} = T_{ab}$), we can write:
 \begin{equation}
J^a = \left( T^{aj} + g^{aj} L \right) \xi_j
\label{nontriviala}
\end{equation}
Therefore,  for any vector $k_a$ which satisfies $k_a \xi^a =0$, we get the result:
\begin{equation}
(k_a J^a) =T^{aj}k_a\xi_j.
\label{nontrivial}
\end{equation}  
The change in this quantity, when
$T^{aj}$ changes by a small amount $\delta T^{aj}$, will be $\delta (k_aJ^a) = k_a\xi_j\delta T^{aj}$. It is this relation which  can be used to obtain an expression
for horizon entropy in terms of the Noether charge.  When some amount of matter with energy-momentum tensor $\delta T^{aj}$ crosses the horizon, the corresponding energy flux can be thought of as
given by integral of  $k_a\xi_j\delta T^{aj}$ over the horizon
where $\xi^a$ is the Killing vector field corresponding to the bifurcation horizon
and $k_a$ is a vector orthogonal to $\xi^a$ which can be taken as the normal
to a timelike surface, infinitesimally away from the horizon. (Such a surface is sometimes called a `stretched horizon' and is defined by the condition $N=\epsilon$ where $N$ is the lapse function with $N=0$ representing the horizon.) In the $(D-1)$  dimensional integral  over this surface, one coordinate is just time; since we are dealing with an approximately stationary situation, the time integral reduces to multiplication by the range of integration. Based on our discussion earlier
we will assume that the time integration can be restricted to the range  $(0,\beta)$   where $\beta=2\pi/\kappa$ and $\kappa$ is the surface gravity of the horizon. 
(The justification for this requires a much more detailed mathematical analysis which we shall not get into.)
Thus, on integrating
$\delta (k_aJ^a)$ over the horizon we get 
\begin{eqnarray}
\delta \int_\mathcal{H} d^{D-1} x \sqrt{h}  (k_aJ^a) &=&  \int_\mathcal{H} d^{D-1} x \sqrt{h}  k_a\xi_j\delta T^{aj}\nonumber\\
&=& \beta  \int_\mathcal{H} d^{D-2} x \sqrt{h}  k_a\xi_j\delta T^{aj}
\end{eqnarray} 
where the integration over time has been  replaced  by a multiplication
by $\beta = (2\pi/\kappa)$ assuming approximate stationarity of the expression. The integral over $\delta T^{aj}$  is  
the flux of energy $\delta E$ through the horizon so that $\beta\delta E$
can be interpreted as the rate of change of the entropy associated with this energy flux.
 One can obtain, using these facts, an
 expression for entropy,  given by
\begin{equation}
S_{\rm Noether} \equiv \beta \mathcal{N} = \beta\int d^{D-1}\Sigma_{a} J^{a}= \frac{\beta}{2}  \int d^{D-2}\Sigma_{ab} J^{ab}
\label{noetherint}
\end{equation} 
where $ d^{D-1}\Sigma_{a} = d^{D-1} x \sqrt{h}  k_a$, the Noether charge is $\mathcal{N}$
 and we have  introduced the antisymmetric tensor $J^{ab}$ by $J^a = \nabla_b J^{ab}$.
 In the final expression the integral is over any surface with $(D-2)$ dimension which is a spacelike cross-section of the Killing horizon on which the norm of $\xi^a$ vanishes.

As an example, consider  the special case of  Einstein gravity in which \eq{noedef}  reduces to
\begin{equation}
J^{ab} = \frac{1}{16\pi} \left( \nabla^a \xi^b - \nabla^b \xi^a\right)
\end{equation}
If $\xi^a$ be  the timelike Killing vector in the  spacetime describing  a Schwarzschild black hole., we can compute the Noether charge $\mathcal{N}$  as an integral of $J^{ab}$  over any
two surface  which is a spacelike cross-section of the Killing horizon on which 
the norm of $\xi^a$ vanishes. The area element on the horizon can be taken to be  $d\Sigma_{ab}=  (l_a\xi_b-l_b\xi_a)\sqrt{\sigma}d^{D-2}x$ in \eq{noetherint} with $l_a$ being an auxiliary vector field satisfying the condition $l_a\xi^a=-1$. Then the integral in \eq{noetherint} reduces to
\begin{eqnarray}
 S_{Noether}&=& -\frac{\beta}{8\pi}\int\sqrt{\sigma}d^{D-2}x
 (l_a\xi_b)\nabla^b \xi^a\nonumber\\
 &=&\frac{\beta\kappa}{8\pi}\int\sqrt{\sigma}d^{D-2}x
 =\frac{1}{4}A_H
\end{eqnarray} 
where we have used \eq{defkappa1}, the relation $l_a\xi^a=-1$, and the fact that $\xi^a$ is a Killing vector. The result, of course, agrees with the standard one.\footnote{There is, however, a subtlety that needs to be stressed regarding this derivation. In Einstein's theory, $J^a = 2 R^a_b \xi^b$. Hence, for  any static vacuum solution to Einstein's theory with $\xi^a$ being the Killing vector, the Noether current $J^a$ vanishes identically! A direct integration should therefore give zero entropy. This difficulty is circumvented by first obtaining $J^{ab}$ and performing the integral over it on a \textit{single} 2-surface rather than integrating $J^a $ over a compact region in spacetime. The same situation arises in the calculation of Komar mass integrals for vacuum spacetimes.}

 It is also possible to show, using the expression for $J^{ab}$ that the entropy in \eq{noetherint} is also equal to 
\begin{equation}
S_{Noether} = \frac{2\pi}{\kappa} \oint_\Sigma \left( \frac{\delta L}{\delta R_{abcd}}\right)\epsilon_{ab}\epsilon_{cd} d\Sigma
\label{SNoether}
\end{equation} 
where $\kappa $ is the surface gravity of the horizon and the $(D-2)$-dimensional
integral is on a spacelike bifurcation surface with $\epsilon_{ab}$ denoting
the  bivector normal to the bifurcation surface. The variation is performed
as if $R_{abcd}$ and the metric are independent and the whole expression is
evaluated on a solution to the equation of motion.
A wide class of theories have been investigated using such a generalization in order to identify the thermodynamic variables relevant to the horizon.

\subsection{The \LL\ models of gravity}\label{sec:llgravity}

Among the class of theories described by the field equations $2\mathcal{G}_{ab} = T_{ab}$ with $\mathcal{G}_{ab}$ given by \eq{genEab},
one subset deserves special mention. These are the theories for which the Lagrangian 
satisfies the condition $\nabla_a P^{abcd} =0$.
(Since $P^{abcd}$ has the symmetries of the curvature tensor, it follows that it will be divergence-free in all the indices.) In this case, \eq{genEab} simplifies considerably and we get 
\begin{equation}
\mathcal{G}_{ab}=P_a^{\phantom{a} cde} R_{bcde}  - \frac{1}{2} L g_{ab}
=\mathcal{R}_{ab}-\frac{1}{2} L g_{ab};\quad
P^{abcd} \equiv \frac{\partial L}{\partial R_{abcd}}
\label{genEab1}
\end{equation}
The crucial difference between \eq{genEab} and  \eq{genEab1} 
is the following. Since $L$ does not depend on the derivatives of the curvature tensor,
it contains at most second derivatives of the metric;  therefore, $P^{abcd}$ also contains only up to the second derivatives of the metric. It follows that
\eq{genEab1} will not lead to derivatives of the metric higher than second order  in the field equations. In contrast, \eq{genEab} can contain up to fourth order in the derivatives of the metric. 
 Though sometimes explored in the literature, field equations with derivatives higher than second order create several difficulties. (For example, when the equations are second order, the boundary condition in a variational principle is more easily defined than when higher order terms occur in the equations of motion;  in such a case, the 
 variation of the action functional requires very special procedures.)  In view of this, there is a strong theoretical motivation to consider theories in which the $\mathcal{G}_{ab}$ is of the form in \eq{genEab1}.

 The Lagrangians which lead to the expression in \eq{genEab1}
are, of course, quite special and are known as \LL\ Lagrangians. They can be expressed as a sum of terms, each involving products of  curvature tensors with the $m-$th term being a product of $m$ curvature tensors. The  
general \LL\ Lagrangian has the form \cite{lovelock},
\begin{equation}
{L} = \sD{c_m\LDm}\,~;~{L}_{(m)} = \frac{1}{16\pi}
2^{-m} \Alt{a_1}{a_2}{a_{2m}}{b_1}{b_2}{b_{2m}}
\Riem{b_1}{b_2}{a_1}{a_2} \cdots \Riem{b_{2m-1}}{b_{2m}}{a_{2m-1}}{a_{2m}}
\,,  
\label{twotw}
\end{equation}
where the $c_m$ are arbitrary constants and \LDm\ is the $m$-th
order \LL\ Lagrangian.
The $m=1$ term is proportional to $\delta^{ab}_{cd}R^{cd}_{ab} \propto R$ and leads
to Einstein's theory. The $m=2$ term gives rise to what is known as Gauss-Bonnet theory.
Because of the determinant tensor, it is obvious that in any given dimension $D$ we can only have $K$ terms where 
$D\geq 2K$.
It follows that, if $D=4$, then only the $K=1, 2$ are non-zero.
Of these, the Gauss-Bonnet term corresponding to $K=2$ gives, on variation of the
action, a vanishing bulk contribution.
In dimensions $D=5$ to $8$, one can have both the Einstein-Hilbert term
and the Gauss-Bonnet term etc. and so on for higher dimensions.
 It is conventional to take $c_1 =1$ so that
the ${\ensuremath{{L}_{(1)}}}$, which gives Einstein gravity, reduces to $(R/16\pi)$.
The normalizations $m>1$ are somewhat ad-hoc for individual \LDm\ since the $c_m$s 
are unspecified at this stage.

The \LL\ models possess black hole solutions and their thermodynamic properties have been investigated quite extensively. It can be shown, for example, that the entropy of a black hole horizon $\mathcal{H}$ in \LL\ models (determined using \eq{SNoether}) is given by
\begin{equation}
S|_{\Cal{H}} = \sD{4\pi m c_m \int_{\Cal{H}}{d^{D-2}x_{\perp} 
  \sqrt{\sigma}\,L_{(m-1)}}} 
  \,,      
\label{ent-limit-2}
\end{equation} 
where $x_{\perp}$ denotes the transverse coordinates on \Cal{H},
$\sigma$ is the determinant of the intrinsic metric on \Cal{H}.  It is interesting to observe that in these models, the entropy for the $m$th order theory is given by a surface integral involving the Lagrangian in the $(m-1)$th order theory. We will now indicate how this result arises.

To do this we need  to evaluate the Noether charge $\mathcal{N}$ corresponding to the current $J^a$, for a static metric with a bifurcation horizon and a Killing vector field $\xi^a = (1, \textbf{0}); \xi_a = g_{a0}$. The location of the horizon is given by the vanishing of the norm $\xi^a\xi_a = g_{00},$ of this  Killing vector.
The  Noether charge is given by
\begin{eqnarray}
\mathcal{N} &=& \mes{D-1}{t} J^0 =\int_{t} d^{D-1}x\, \partial_b(\sqrt{-g} J^{0b})\nonumber\\
&=&\mes{D-2}{t,r_H} J^{0r}
\end{eqnarray} 
in which we have ignored the contributions arising from $b$ when it ranges over the transverse directions. This is
justifiable when transverse directions are compact or in the case of Rindler approximation
when nothing changes along the transverse direction. In the radial direction, the integral picks out the contribution at $r=r_H$ which is taken to be the location of the horizon.
Using  
$J^{ab} =2 P^{abcd}\nabla_c \xi_d $ (see \eq{noedef}) and $\xi_d = g_{da} \xi^a = g_{d0}$ we get:
\begin{equation}
J^{0r}= 2P^{0rcd} \nabla_c \xi_d = 2 P^{0rcd} \partial_c\xi_d = 2 P^{0rcd} \partial_c g_{d0} = 2 P^{cdr0} \partial_d g_{c0}
\end{equation} 
where we have used the symmetries of 
 $P^{abcd} $   which are the same as those of the curvature tensor.
 So
\begin{eqnarray}
\mathcal{N} &=& 2 \mes{D-2}{t,r_H} P^{cdr0}\partial_d g_{c0}\nonumber\\
&=&2m \mes{D-2}{t,r_H} Q^{cdr0}\partial_d g_{c0}
\label{n}
\end{eqnarray} 
where $Q^{abcd}\equiv(1/m)P^{abcd}$. Therefore the entropy is  given by
\begin{equation}
S_{Noether}=\beta\mathcal{N} = 2\beta m \mes{D-2}{t,r_H} Q^{cdr0}\partial_d g_{c0}
\label{s}
\end{equation}  
When the near-horizon geometry has a Rindler limit, the $r$ coordinate becomes the $x$ coordinate and only $g_{00}=-\kappa^2x^2$ contributes. Then this expression reduces to
\begin{equation}
S_{Noether} = 8\pi m \int_\mathcal{H} d^{D-2}x_\perp\, \sqrt{\sigma}\, \left( Q_{0x}^{0x}\right)
\label{res1}
\end{equation} 
where $\sigma$ is the determinant of the metric in the transverse space.

 Let us consider this 
  quantity $Q_{x 0}^{x 0}$ for the $m$-th
order \LL\ action,  given by :
\begin{equation}
Q_{x 0}^{x 0} = \frac{1}{16\pi}\frac{1}{2^m}
\AltC{x }{0}{a_3}{a_{2m}}{x }{0}{b_3}{b_{2m}}
\left(\Riem{b_3}{b_4}{a_3}{a_4}
... \Riem{b_{2m-1}}{b_{2m}}{a_{2m-1}}{a_{2m}}\right)\Bigg|_{x =\,\epsilon}
\,. 
\label{app-ent-4}
\end{equation}
where we have added a normalization which gives Einstein's action for $m=1$ and will
 \textit{define} $Q_{x 0}^{x 0} = 1/16\pi $ for the $m=0$ case.
We have also indicated that we are evaluating this expression in Rindler limit 
of the horizon, as in \eq{res1}.
The presence of both $0$ and $x $ in each row of the alternating tensor
forces all other indices to take the values $2,3,...,D-1$. In fact, we
have $\AltC{x }{0}{a_3}{a_{2m}}{x }{0}{b_3}{b_{2m}} =
\Alt{A_3}{A_4}{A_{2m}}{B_3}{B_4}{B_{2m}}$ with $A_i, B_i =
2,3,...,D-1$ (the remaining combinations of Kronecker deltas on
expanding out the alternating tensor are all zero since
$\delta^0_A=0=\delta^x _A$ and so on). Hence $Q_{x 0}^{x 0}$
reduces to\index{Lanczos-Lovelock Lagrangian!horizon entropy} 
\begin{equation}
Q_{x 0}^{x 0} =
\frac{1}{2}\left(\frac{1}{16\pi}\frac{1}{2^{m-1}}\right) 
\Alt{A_3}{A_4}{A_{2m}}{B_3}{B_4}{B_{2m}}
\left(\Riem{B_3}{B_4}{A_3}{A_4}
... \Riem{B_{2m-1}}{B_{2m}}{A_{2m-1}}{A_{2m}}\right)\Bigg|_{x =\,\epsilon}
\,. 
\label{app-ent-5}
\end{equation}
Therefore, in the $\epsilon\to 0$ limit, recalling that
$\Riem{A}{B}{C}{D}\mid_\Cal{H}
=\,^{(D-2)}\Riem{A}{B}{C}{D}\mid_\Cal{H}$, we find that $Q_{x 0}^{x 0}$ is essentially the \LL\ Lagrangian of order $(m-1)$:
\begin{equation}
Q_{x 0}^{x 0}=
\frac{1}{2}{L}_{(m-1)}\,,
\label{app-ent-6}
\end{equation} 
where we have restored the subscript  giving the order of the Lagrangian.
The entropy  becomes
\begin{equation}
S^{(m)}_{\rm Noether} =  4\pi m
\int_{\Cal{H}}{d^{D-2}x_{\perp}\sqrt{\sigma}{L}_{(m-1)}}
\,,   
\label{app-ent-8}
\end{equation}
 This  entropy in the $m-$th order \LL\ theory is an integral over the \textit{Lagrangian} of $(m-1)$th order. For $m=1$ (Einstein gravity),
the $L_{(0)}$  is a constant giving an entropy proportional to transverse area; for $m=2 $ (Gauss-Bonnet gravity), the entropy is proportional to  integral of $R$ over transverse direction. 

While these results are  satisfactory at a formal level, one must stress that the explicit value of the entropy  depends on the nature of the theory decided by the parameters $c_i$ and may not have simple interpretation for certain range of parameters.
For example,  it is known that the on-shell entropy of the \LL\ models will not be positive definite for all range of parameters
\cite{entneg}.
More seriously, the study of these models  also raises several new conceptual conundrums for which it is difficult to find satisfactory answer. We shall now describe several of these issues.

\section{Thermodynamics of horizons: A deeper look}\label{sec:thermosecond}

\subsection{The degrees of freedom associated with black hole entropy}

We have seen that there is a natural way of associating a temperature with \textit{any}
horizon including, for example, the Rindler horizon in flat spacetime. But the arguments given in the last section leading to the association of \textit{entropy} --- in contrast to the association of a temperature --- cannot be easily generalized from black hole horizon to other horizons. While there is general agreement that all horizons have a temperature,
very few people \cite{Carlip:1999cy,tpted1,TPgravcqg04,tworeviews} have taken a firm stand as regards the question of associating an entropy with a horizon. To certain extent this ambivalence among researchers has led to most of the work being concentrated on analyzing \textit{black hole} entropy (rather than 
\textit{horizon} entropy) and we shall start our discussion with issues connected with black hole 
entropy.

   In the case of normal matter, entropy can be provided
   a statistical interpretation  as the logarithm of the number of 
   available microstates that are consistent with the macroscopic parameters which are
   held fixed. That is,  $S(E)$ is related to the 
degrees of freedom (or phase volume) $g(E)$ by $S(E)=\ln g(E)$. Maximization of the phase 
volume for systems which can exchange energy
will then lead to equality of the quantity $T(E)\equiv (\partial S/\partial E)^{-1}$ for the systems. 
It is usual to identify this variable as the thermodynamic temperature. 
(This definition works even for self-gravitating systems in microcanonical ensemble; 
see eg., \cite{Padmanabhan:1990gm}.)

Assuming that the entropy of the black hole should have a similar interpretation, one is 
led to the conclusion that the density of states for a black hole of energy $E = M$ should vary as
   \begin{equation}
   g(E) \propto \exp \left(\frac{1}{4}\frac{\mathcal{A}_H}{L_P^2}\right)
   \label{gbhofe}
   \end{equation}
  Such a growth implies \cite{TPgravcqg04}, among other things, that the Laplace transform of $g(E)$ 
  does not exist so that  canonical partition function cannot be defined
  (without some regularization). 
  That brings us to the crucial question: What are the microscopic states  which
 one should count to obtain the result in Eq.~(\ref{gbhofe}) ? That is, what are the degrees of freedom
   which lead to this entropy ?

 To begin with, the thermal radiation surrounding the black hole has an entropy 
   which one can  compute. It is fairly easy to see that this entropy will 
   proportional to the horizon area but will
   diverge quadratically.
   Near the horizon  the field becomes free 
and solutions are simple plane waves (see Section \ref{sec:ftdim}). It is the  existence of such a continuum
of wave modes which leads to infinite phase volume for the system.
More formally, the number of modes  $n(E)$ for a scalar field $\phi$ with vanishing
boundary conditions at two radii $r=R$ and $r=L$ is given by
\begin{equation}  
n(E) 
 \simeq \frac{2}{ 3\pi} \int^L_R \frac{r^2 dr}{ \left( 1 - 2M/r\right)^2}\left[ E^2 - 
 \left( 1 -\frac{2M}{ r}\right)  m^2\right]^{3/2} \label{eqn:qnofe}
\end{equation}
in the WKB limit (see
\cite{'tHooft:1985re,Padmanabhan:1986rs}). This expression diverges as $R\to 2M$ showing that
a scalar field propagating  in a black hole spacetime has infinite phase volume. The corresponding entropy
computed using the standard relations:
\begin{equation} 
S = \beta \left[ \frac{\partial}{ \partial \beta} - 1 \right] F; \qquad
F = - \int_0^\infty dE \frac{n(E)}{e^{\beta E} - 1 },
\end{equation}
is quadratically divergent:  $S = (\mathcal{A}_H/l^2)$ with $l \to 0$. 
The divergences described above occur around any infinite
redshift surface and is a geometric (covariant) phenomenon. 
  
   The same result can also be obtained from what is known as
   ``entanglement entropy'' arising from the quantum correlations 
    which exist across the horizon. (For a review, see \cite{shanki}). 
    We saw in Section \ref{thermalden} that if
     the field configuration inside the horizon
    is  traced over in the vacuum functional of the theory, then one obtains a density matrix $\rho$
    for the field configuration outside [and vice versa]. The entropy $ S=-Tr(\rho \ln \rho)$
    is usually called the entanglement entropy
    \cite{Dowker:1994fi,Israel:1976ur,Callan:1994py}. This is essentially the same as the previous 
    calculation and, of course, $S$ diverges quadratically on the horizon
    \cite{Frolov:1993ym,Zurek:1985gd}. In fact, much of this can be done 
    without actually bringing in gravity into the picture; all that is required is a spherical
    region inside which the field configurations are traced out \cite{Bombelli:1986rw,Srednicki:1993im}.  
Physically, however,  it does not seem reasonable to integrate over all modes without any cut off in these calculations.
By cutting off the modes at 
   $l \approx L_P$ one can obtain the ``correct'' result but in the absence of a 
   more fundamental argument for regularizing the modes, this result is not of 
   much significance.  The cut off can be introduced in a more sophisticated manner by changing the dispersion relation near Planck energy scales but again there are different prescriptions that are available 
\cite{Padmanabhan:1998jp,Padmanabhan:1998vr,Unruh:1994zw,Corley:1996ar} and none of them are really convincing.

\subsection{Black hole entropy in quantum gravity models}\label{qgbh}

There have also been attempts to compute black hole entropy in different models of quantum gravity \cite{Peet:2000hn}.
In standard string theory this is done as follows: There are certain special states in string theory, called BPS states,
 that contain electric and magnetic charges which are equal to their mass. Classical supergravity has these states as classical solutions, among which are the extremal black holes with electric charge equal to the mass (in geometric units). These solutions can be expressed as a Reissner-Nordstrom metric with both the roots of $g_{00}=0$ coinciding: obviously, the surface gravity at the horizon, proportional to $g_{00}'(r_H)$ vanishes though the horizon has finite area. Therefore, these black holes
 have zero temperature but finite entropy. For certain compactification schemes in string theory (with
$d=3,4,5$ flat directions), in the limit of
$G\to 0$, there exist BPS states which have the same mass, charge and angular momentum of an extremal black hole in $d$
dimensions. One can explicitly count the number of such states in the appropriate limit and one finds that the result gives the density of states in \eq{gbhofe}  with correct numerical factors \cite{Strominger:1996sh,Das:2000su,Breckenridge:1996sn}.
 This is done in the weak coupling limit
and a duality between strong coupling and weak coupling limits 
is used to argue that the same result will arise in the 
strong gravity regime. Further, if one perturbs the state slightly away from the BPS limit, to get a {\it near} extremal black hole
and construct a thermal ensemble, one obtains the standard Hawking radiation from the corresponding near extremal black hole
 \cite{Breckenridge:1996sn}.

While these results are encouraging, there are several issues which are intriguing: First, the extremality (or near extremality) was used crucially in obtaining these results. We do not know how to address the entropy of a normal Schwarzschild black hole which is far away from the extremality condition. Second, in spite of significant effort, we do not still have a clear idea of
 how to handle the classical singularity or issues related to it. This is disappointing since one might have hoped that these problems are closely related. Finally, the result is very specific to black holes. One does not get any insight into the structure of other horizons, especially De Sitter horizon, which does not fit the string theory structure in a natural manner.

The second approach  in which some success related to black hole entropy is claimed, is in the  loop
quantum gravity (LQG). While string theory tries to incorporate all interactions in a unified manner, loop quantum gravity 
\cite{Rovelli:1998gg,Thiemann:2001yy}
 has the limited goal of providing a canonically quantized version of Einstein gravity. One key result which emerges from
this program is a quantization law for the areas. The variables used in this approach are like
a gauge field $A^i_a$ and the Wilson lines associated with them. The open Wilson lines carry a quantum number $J_i$ with them
and the area quantization law can be expressed in the form: $\mathcal{A}_H=8\pi G\gamma\sum \sqrt{J_i(J_i+1)}$ where $J_i$ are spins defined on the links $i$ of a spin network and $\gamma$ is free parameter called Barbero-Immirizi parameter.
The $J_i$ take half-integral values if the gauge group used in the theory is SU(2) and take integral values if the gauge group is SO(3).
These quantum numbers, $J_i$, which live on the links that intersect a given area, become undetermined if the area refers to a horizon. Using this, one can count the number of microscopic configurations contributing to a given horizon area and estimate the entropy.
One gets the correct numerical factor (only) if $\gamma=\ln m/2\pi \sqrt{2}$ where $m=2$ or $m=3$ depending on whether the gauge
group SU(2) or SO(3) is  used in the theory \cite{Krasnov:1998wc,Rovelli:1996dv,Ashtekar:2000eq,Ashtekar:1998yu}.

Again there are several unresolved issues. To begin with, it is not clear how exactly the black hole solution arises in this approach since it has been never easy to arrive at the low energy limit of gravity in LQG. Second, the answer depends on the Immirizi parameter $\gamma$ which needs to be adjusted to get the correct answer, after we know the correct answer from elsewhere.
Even then, there is an ambiguity as to whether one should have SU(2) with  $\gamma=\ln 2/2\pi \sqrt{2}$ or SO(3) with
 $\gamma=\ln 3/2\pi \sqrt{2}$. The SU(2) was the preferred choice for a long time, based on its close association with fermions
which one would like to incorporate in the theory. However,  there has also been some occasional rethinking on this issue due to the following consideration: For a classical black hole, one can define a class of solutions to wave equations called quasi normal modes [see e.g.,\cite{Corichi:2002ty,Kokkotas:1999bd,Berti:2003jh,Cardoso:2003vt}]. 
These modes have discrete frequencies 
\cite{Padmanabhan:2003fx,Choudhury:2003wd,Motl:2003cd,Motl:2002hd} which are complex, given by
\begin{equation}
\omega_n=i\frac{n+(1/2)}{4M}+\frac{\ln(3)}{8\pi M}+\mathcal{O}(n^{-1/2})
\end{equation}
The $\ln(3)$  in the above equation is not negotiable. If one chooses SO(3) as the gauge group, then one can connect up the frequency 
of quanta emitted by a black hole when the area changes by one quantum in LQG with the quasi normal mode frequency
\cite{Dreyer:2002vy,Hod:1998vk}. It is not clear whether this is a coincidence or a result of some significance. 
If one assumes that this result is of some fundamental significance, the SO(3) gains preference.

The short description given above shows that candidate models for quantum gravity are not yet developed well enough to provide a clear physical picture of horizon thermodynamics. (This is true even as regards numerous other approaches like e.g., those based on noncommutative geometry \cite{nicolini}). Given such a  situation even in the well studied  case 
of black hole horizon it is no suprise that we have virtually no quantum gravitational insight of other horizons. This fact gives additional impetus for studying the thermodynamic approach which we described in Section~\ref{sec:intro}.
There is, however, one central issue brought to the forefront by the quantum gravitational models of black hole entropy which we shall now discuss further.

\subsection{Black hole entropy: Bulk versus surface degrees of freedom}

Two obvious choices for the degrees of freedom contributing to the black hole entropy are
 those  associated with the
 bulk volume inside the black hole (including those related to 
 matter which collapses to form the black hole) or
 degrees of freedom associated with the horizon. 
One would have normally thought that the bulk degrees of freedom hidden by the horizon
should scale as the volume $V\propto M^3$ of the black hole. 
In that case, we would expect to
 get an entropy proportional to the volume rather than area. It is clear that, near a horizon, only a region of length $L_P$ across the horizon contributes to the microstates so that in 
the expression $(V/L_P^3)$, the relevant $V$ is $M^2 L_P$ rather than $M^3$. It is possible to interpret this as due to the entanglements of modes across the horizon over a length scale of $L_P$, which --- in turn --- induces a nonlocal coupling between the modes on the surface of the horizon. Such a field will have one-particle excitations, which have the same density of states as black hole \cite{Padmanabhan:1998jp,Padmanabhan:1998vr}. While this is suggestive of why we get the area scaling rather than volume scaling, a complete understanding is lacking.

In fact, it is fairly easy to obtain an area scaling law for entropy if we assume that the degrees of freedom are on the horizon.  Suppose we have {\it any} formalism of quantum gravity in
which there is a minimum quantum for length or area, of the order of $L_P^2$. 
We have, for example,  considerable evidence of very different nature to suggest Planck length acts as lower bound to the length scales that can be operationally defined and that no measurements can be ultra sharp at Planck scales \cite{tplimitations}.
Then, the horizon area $\mathcal{A}_H$ 
can be divided into $n=(\mathcal{A}_H/c_1L_P^2)$ patches where $c_1$ is a numerical factor. If each patch has $k$ degrees of freedom
(due to the existence of a surface field, say), then the total number of microscopic states are $k^n$ and the resulting entropy is
$S=n\ln k=(4\ln k/c_1)(\mathcal{A}_H/4L_P^2)$ which will give the standard result if we choose $(4\ln k/c_1)=1$. The essential ingredients
are only discreteness of the area and existence of certain degrees of freedom in each one of the patches.   

On the other hand, one cannot completely dismiss the degrees of freedom in the bulk as playing no role. Recall that the thermal density matrix for an eternal black hole can be  obtained by explicitly integrating out the degrees of freedom on the left Rindler wedge (see Sec. \ref{thermalden}). If integrating out certain degrees of freedom leads to a thermal density matrix, it makes sense to identify the same degrees of freedom as also contributing to the entropy. In which case, the bulk degrees of freedom has to play a role and still lead to an entropy that is proportional to the area. 
 One possible way this can arise is through
 some kind of holographic relationship built in the theory of gravity so that the entropy of the bulk region $\mathcal{V}$ can be computed in terms of variables on its boundary $\partial\mathcal{V}$. We shall see that the thermodynamic perspective of  gravity,
described later in this review, does lead to this possibility.

\subsection{Observer dependence of horizons and entropy}\label{sec:obsdependence}

We shall next address some  conceptual issues brought about by the existence of horizons and entropy. We recall that the mathematical formulation leading to the association of temperature with any horizon is fairly universal and it does not distinguish between different horizons, like, for example, Rindler horizon in flat space or a Schwarzschild black hole event horizon or a de Sitter horizon. Assuming that temperature and entropy arise for fundamentally the same reason, it would be extremely unnatural \textit{not} to associate entropy with \textit{all} horizons. 

In this context, a distinction is sometimes made by arguing  that, while the event horizon of a black hole can be given a purely geometrical definition,  the Rindler horizon is observer dependent.
This  is irrelevant for the purpose of associating an entropy with the horizon.
An observer plunging into a black hole will have access to different amount of information (and  will attribute different thermodynamic properties to the black hole) compared to an observer who is remaining stationary outside the horizon. This is similar to what happens in the case of Rindler frame as well; an observer who stops accelerating or an inertial observer, will certainly have access to different regions of spacetime compared to the Rindler observer. In both the cases the physical effect of horizon in blocking information depends on the class of world lines one is considering. In that sense, all horizons are observer dependent. (There exists another example of a horizon --- viz., the de Sitter horizon --- the location of which depends on the observer but is considered ``more real'' than the Rindler horizon; those who object to assigning entropy to Rindler horizon are usually quite ambivalent about de Sitter
or Schwarzschild-de Sitter  \cite{tirthtpgrg} horizons!)
It seems necessary to assign an (observer dependent) entropy to \textit{all} horizons.

This feature, however, brings in a totally new layer of observer dependent  thermodynamics into the theory which
--- though it  need not come as a surprise ---
has to be tackled head-on. 
We know that an inertial observer will attribute
  zero temperature and zero entropy to the inertial vacuum. But a Rindler observer will attribute a finite temperature and non-zero (formally divergent) entropy to the same vacuum state. So entropy is indeed an observer dependent concept \cite{marolf}. When one does  quantum field theory in curved spacetime, it is 
  not only that particles become an observer dependent notion but also the  temperature and entropy. 
  
 The observer in the Rindler wedge $\mathcal{R}$ will also perceive  that the observables exhibit standard thermodynamic properties like entropy maximization, equipartition, thermal fluctuations etc., because the physics is governed by a thermal density matrix. But all these thermodynamical features arise because the  Rindler observer attributes a density matrix to a \textit{pure} quantum state after integrating out the unobservable modes. From this point of view, all these thermal effects are intrinsically quantum mechanical --- which is somewhat different from the `normal' thermal behaviour.  Our results suggest that this distinction between quantum fluctuations and thermal fluctuations is artificial (like e.g., the distinction between energy and momentum of a particle in nonrelativistic mechanics) and should fade away in the correct description of spacetime, when one properly takes into account the fresh observer dependence induced by the existence of  horizons.   

To see what all these imply in a concrete fashion,  consider an excited state of a quantum field with energy  $\delta E$ above the ground state in an inertial spacetime. When we integrate out  the unobservable modes for the Rindler observer, we will get a density matrix $\rho_1$ for this state
  and the corresponding entropy will be $S_1 = - {\rm Tr}\ (\rho_1 \ln \rho_1)$.
  The inertial vacuum state itself has the density matrix $\rho_0$ and the entropy $S_0 = - {\rm Tr}\ (\rho_0 \ln \rho_0)$ in the Rindler frame. The difference $\delta S = S_1 - S_0$ is finite and 
  represents the entropy attributed to this state by the Rindler observer. (This is
  finite though $S_1$ and $S_0$ can be divergent.) In the limit of $\kappa \to \infty$,
  which would correspond to  a Rindler observer who is very close to the horizon,
   we can actually compute it and show that 
  \begin{equation}
\delta S = \beta \delta E = \frac{2\pi}{\kappa} \delta E
\label{delS}
\end{equation}
To prove this, note that if we write $\rho_1=\rho_0 + \delta \rho$, then in the limit of
$\kappa \to \infty$ we can concentrate on states for which $\delta \rho/\rho_0\ll 1$.
Then we have
\begin{eqnarray}
-\delta S &=& {\rm Tr}\ (\rho_1 \ln \rho_1) - {\rm Tr}\ (\rho_0 \ln \rho_0) 
\simeq {\rm Tr}\ (\delta \rho \ln \rho_0)\nonumber\\
&=& {\rm Tr}\ (\delta \rho (-\beta H_R)) = - \beta {\rm Tr}\ \left((\rho_1 -\rho_0)H_R\right) \equiv -\beta \delta E
\end{eqnarray} 
where we have used the facts Tr $\delta \rho \approx 0$ and
$\rho_0 =Z^{-1}\exp(-\beta H_R)$ where $H_R$ is the Hamiltonian for the system in the 
Rindler frame. The last line defines the $\delta E$ in terms of the difference in 
the expectation values of the Hamiltonian in the two states. 
This is the amount of entropy a Rindler observer would claim to be lost when the matter
disappears into the horizon.
(This result can be explicitly proved for, say, one particle excited states of the field
\cite{kkp}.) 

The above result is true in spite of the fact that, formally, matter takes  an infinite amount of
coordinate time to cross the horizon as far as the outside observer is concerned.
This is essentially because
we know that quantum gravitational effects will smear the location of the horizon by $\mathcal{O} (L_P)$ effects \cite{tplimitations}.
 So one cannot really talk about the location of the event horizon ignoring fluctuations of this order. From the operational point of view, we only need to consider matter reaching within few Planck lengths of the horizon to talk about entropy loss.
 In fact, physical processes very close to the horizon must play an important role
in order to provide a complete picture of the issues we are discussing. There is 
already some evidence \cite{Padmanabhan:1998jp,Padmanabhan:1998vr} that the infinite redshift induced by the horizon plays a crucial
role in this though a mathematically rigorous model is lacking.

One might  have naively thought that the expression for entropy of matter
crossing the horizon 
should consist of its energy $\delta E$ and \textit{its own} temperature $T_{\rm matter}$
rather than the horizon temperature. 
But the correct expression is $\delta S = \delta E/T_{\rm horizon}$;
  the  horizon acts as a system with some internal degrees of freedom and temperature $T_{\rm horizon}$ \textit{as far as Rindler observer is concerned} so that when one adds an energy $\delta E$ to it, the entropy change is $\delta S = (\delta E/ T_{\rm horizon})$.

All these are not \textit{new} features but only the consequence of 
result that 
a Rindler observer attributes a non-zero temperature to inertial vacuum. This temperature influences every other thermodynamic variable. 
Obviously, a  Rindler
observer (or an observer outside a black hole horizon) will attribute all kinds of  entropy changes to the horizons she perceives while an inertial observer
(or an observer falling through the Schwarzschild horizon) will see none of these phenomena.
This requires us  to accept the fact that  many of the thermodynamic phenomena needs to be now thought of as specifically observer dependent.

\section{Action functionals for gravity and horizon thermodynamics}

While deriving the expression for the temperature associated with a horizon we stressed the fact that it was completely independent of the dynamical equations satisfied by the metric. For example, we never needed to use Einstein's equations in the case of Schwarzschild black hole, say,  to obtain the expression for temperature. The situation as regards Hawking evaporation is also similar; given a specific form of the  metric, one could obtain Hawking evaporation without demanding that the metric should be a solution to any specific field equation. 

It therefore comes as a surprise that there is a deep and curious connection 
between the field equations of gravity and the horizon thermodynamics. We shall first provide a simple illustration of this result and will then consider a more general approach.

\subsection{An unexplained connection between horizon thermodynamics and gravitational dynamics}\label{sec:connection}

 To illustrate this result in the simplest context \cite{tdsingr}, let us consider 
 a static, spherically symmetric horizon, in a spacetime described by a metric:
\begin{equation}
ds^2 = -f(r) c^2 dt^2 + f^{-1}(r) dr^2 + r^2 d\Omega^2. \label{spmetric}
\end{equation}
(These results can be easily generalized to the case with $g_{00} \ne - g^{rr}$ 
by changing $r^2 d\Omega^2 $ by $R^2(r) d \Omega^2$ where $R(r)$ is an arbitrary function. We will not bother to do this.)
Let the location of  the horizon be given by the simple zero of the function $f(r)$, say at $r=a$. The Taylor series expansion of $f(r)$ near the horizon $f(r)\approx f'(a)(r-a)$ shows that the metric reduces to the Rindler metric near the horizon in the $r-t$ plane  with the surface gravity $\kappa = (c^2/2) f'(a)$. Then, an analytic continuation to imaginary time allows us to identify the temperature associated with the horizon to be 
\begin{equation}
k_BT=\frac{\hbar c f'(a)}{4\pi}
\label{hortemp1}
\end{equation} 
where we have introduced the normal units. 
The association of temperature in \eq{hortemp1} with the metric in \eq{spmetric} only requires the conditions $f(a)=0$ and $f'(a)\ne 0$.
The discussion so far did not assume anything about the dynamics of gravity or Einstein's field equations.  

We shall now take the next step and write down the
Einstein equation for the metric in \eq{spmetric}, which is given by 
$(1-f)-rf'(r)=-(8\pi G/c^4) Pr^2$ where $P = T^{r}_{r}$ is the radial pressure. When evaluated on the horizon $r=a$ we get the result:
\begin{equation}
\frac{c^4}{G}\left[\frac{1}{ 2} f'(a)a - \frac{1}{2}\right] = 4\pi P a^2
\label{reqa}
\end{equation}
If we now consider two solutions to the Einstein's equations differing infinitesimally in the parameters such that horizons occur at two different radii $a$ and $a+da$,
then multiplying the \eq{reqa} by $da$, we get: 
\begin{equation}
\frac{c^4}{2G}f'(a) a da - \frac{c^4}{2G}da = P(4\pi a^2 da)
\label{reqa1}
\end{equation}
The right hand side is just $PdV$ where $V=(4\pi/3)a^3$ is what is called the areal volume which is the relevant quantity when we consider the action of pressure on a surface area. In the first term, we note that $f'(a)$ is proportional to horizon temperature in \eq{hortemp1}. Rearranging this term slightly
and introducing a $\hbar$ factor \textit{by hand} into an otherwise classical equation
to bring in the horizon temperature, we can rewrite \eq{reqa1} as
\begin{equation}
   \underbrace{\frac{{{\hbar}} cf'(a)}{4\pi}}_{\displaystyle{k_BT}}
    \ \underbrace{\frac{c^3}{G{{\hbar}}}d\left( \frac{1}{ 4} 4\pi a^2 \right)}_{
    \displaystyle{dS}}
  \ \underbrace{-\ \frac{1}{2}\frac{c^4 da}{G}}_{
    \displaystyle{-dE}}
 = \underbrace{P d \left( \frac{4\pi}{ 3}  a^3 \right)  }_{
    \displaystyle{P\, dV}}
\label{EHthermo}
\end{equation}
The labels below the equation indicate a natural --- and unique --- interpretation for each of the terms and the whole equation now becomes $TdS=dE+PdV$ allowing us to read off the expressions for entropy and energy:
\begin{equation}
 S=\frac{1}{ 4L_P^2} (4\pi a^2) = \frac{1}{ 4} \frac{A_H}{ L_P^2}; \quad E=\frac{c^4}{ 2G} a
    =\frac{c^4}{G}\left( \frac{A_H}{ 16 \pi}\right)^{1/2}
\end{equation}
where $A_H$ is the horizon area and $L_P^2=G\hbar/c^3$. The result shows that Einstein's equations can be re-interpreted as a thermodynamic
identity for a virtual displacement of the horizon by an amount $da$.

The uniqueness of the factor $P(4\pi a^2) da$, where $4\pi a^2$ is the 
proper area of a surface of radius $a$ in spherically symmetric spacetimes, 
implies that we cannot carry out the same exercise by multiplying \eq{reqa} by   some other arbitrary factor 
$F(a) da$ instead of just $da$ in a natural fashion. This, in turn, uniquely fixes both $dE$ and the 
combination $TdS$. The product $TdS$ is  classical and is independent of $\hbar$ and hence we can  determine $T$ and $S$ only within
a multiplicative factor. The only place we introduced $\hbar$ by hand is in
 using the Euclidean extension of the metric to fix the form of $T$ and thus $S$.  
  The fact that $T\propto \hbar$ and $S\propto 1/\hbar$ is  analogous to the situation in classical thermodynamics  in contrast with statistical mechanics. The $TdS$ in thermodynamics is independent of Boltzmann's constant while statistical mechanics will lead to  $S\propto k_B$ and $T\propto 1/k_B$.

It must be stressed that this result is quite different from the conventional first law of black hole dynamics, a simple version of which was mentioned in \eq{firstlaw}. 
The difference  is easily seen, for example,
in the case of Reissner-Nordstrom black hole for which $T^r_r = P$ is non-zero due to the presence of nonzero electromagnetic energy-momentum tensor in the right hand 
side of Einstein's equations. If a
\textit{chargeless} particle of mass $dM$ is dropped into a
Reissner-Nordstrom black hole, then 
the standard first law of black hole thermodynamics 
 will give $TdS=dM$. But in \eq{EHthermo}, the energy term,
defined as  $E\equiv a/2$, changes by $dE= (da/2)
=(1/2)[a/(a-M)]dM\neq dM$. 
 It is easy to see, however, that for the Reissner-Nordstrom black hole, the combination 
 $dE+PdV$  is precisely
equal to $dM$ making sure $TdS=dM$. So we need the $PdV$ term to get
$TdS=dM$ from \eq{EHthermo}  when a \textit{chargeless} particle is dropped into a
Reissner-Nordstrom black hole. More generally, if $da$ arises due to
changes $dM$ and $dQ$, it is easy to show  that \eq{EHthermo} gives
$TdS=dM -(Q/a)dQ$ where the second term arises from the electrostatic
contribution. 
This ensures that \eq{EHthermo} is perfectly consistent
with the standard first law of black hole dynamics in those contexts
in which both are applicable but  $dE \ne dM$ in general.
It may also be noted that   the way 
\eq{EHthermo}
was derived is completely local and quite different from the way one obtains
first law of black hole thermodynamics.

 It is quite surprising that the Einstein's field equations evaluated on the horizon reduces to a thermodynamic identity. More sharply stated, we have no explanation as to why an equation like \eq{EHthermo} should hold in classical gravity, if we take the conventional route. This strongly suggests that the association of entropy and temperature with a horizon
is quite fundamental and is actually connected with the dynamics (encoded in Einstein's equations)
of the gravitational field. The fact that quantum field theory in a spacetime
with horizon exhibits thermal behaviour should then be thought of as a \textit{consequence} 
of a more fundamental principle. 

\subsection{Gravitational field equations as a thermodynamic identity on the horizon}\label{sec:gravthermoiden}

If this conjecture is correct, the equality --- between field equations on the horizon and the thermodynamic identity --- should have a more general validity. This has now been demonstrated for an impressively wide class of models 
like the cases of stationary
axisymmetric horizons and evolving spherically symmetric horizons
in Einstein gravity~\cite{KSP}, static spherically symmetric
horizons~\cite{KP} and dynamical apparent horizons~\cite{CCHK} in
\LL\ gravity, 
generic, static horizon in \LL\ gravity \cite{tpdawoodgentds},
 three dimensional BTZ black hole
horizons~\cite{Akbar1}, FRW cosmological
models in various gravity
theories~\cite{CK,AC07,CC2,AC1,AC,CC1,GW,Others1,CCH} and even in  \cite{hwg}  the case Horava-Lifshitz gravity.

We shall describe briefly how this is achieved in the case of an arbitrary, static, horizon in Einstein's theory and in \LL\ models (for more details, see ref. \cite{tpdawoodgentds}). Consider a static spacetime with the metric
\begin{eqnarray}
{\DM}s^2 = - N^2 {\DM}t^2 + {\DM}n^2 + \sigma_{A B} {\DM}y^A {\DM}y^B
\label{static-metric}
\end{eqnarray}
where $\sigma_{A B}(n, y^A)$ is the transverse metric, and the Killing horizon, generated by the timelike Killing vector field $\bm{\xi} = \bm{\partial}_t$, is approached as $N^2 \rightarrow 0$. Near the horizon, $N \simeq \kappa n + \mathcal{O}(n^3)$ where $\kappa$ is the surface gravity \cite{dirtyBH}. The $t=$ constant part of the metric is written by employing Gaussian normal coordinates for the spatial part of the  metric spanned by $\left( {n, y^A} \right)$
with  $n$ being the normal distance to the horizon. By manipulating the Einstein's equations evaluated on the horizon, one can prove \cite{tpdawoodgentds} the following relation:
\begin{eqnarray}
\frac{\kappa}{2 \pi} \frac{\partial}{\partial \lambda} \left( \frac{1}{4} \sqrt{\sigma} \right) {\delta \lambda} - \left\{ \frac{1}{8 \pi} R_{\parallel} \sqrt{\sigma} \right\} \; \frac{\delta \lambda}{2} &=& \frac{1}{8 \pi} G^{\hatxi}_{\hatxi} \sqrt{\sigma} \; \delta \lambda 
= \frac{1}{8 \pi} G^{\hatn}_{\hatn} \sqrt{\sigma} \; \delta \lambda
\nonumber\\
 &=& T^{\hatn}_{\hatn} \sqrt{\sigma} \; \delta \lambda 
 \label{relation1}
\end{eqnarray}
where $\lambda$ is the affine parameter along the outgoing null geodesics and  the $R_{\parallel}$ is the Ricci scalar of the on-horizon transverse metric, $\left[ \sigma_H \right]_{A B}$. The Einstein tensor components are evaluated in an orthonormal tetrad appropriate for a timelike observer moving along the orbit of the Killing vector field generating the Killing horizon. This is denoted by a hat on the indices; for example, $\bm{\hatxi} = \left( - g_{t t} \right)^{-1/2}~\bm{\partial}_t$ etc., and $-G^{\hatxi}_{\hatxi} = G_{{\hatxi} {\hatxi}} = G( \bm{\hatxi}, \bm{\hatxi} )$.  We have used $G^{\hatxi}_{\hatxi} \;|_H = G^{\hatn}_{\hatn} \;|_H$ in the second equality
 in \eq{relation1}
 and Einstein's equation in the third one. (The fact that $G^{\hatxi}_{\hatxi}  = G^{\hatn}_{\hatn}$ on the horizon is crucial;  more details regarding this symmetry can be found
 in Ref.~\cite{dirtyBH}.)

Multiplying \eq{relation1} by ${\DM}^2 y$, and integrating over the horizon 2-surface, we  obtain
\begin{eqnarray}
T \frac{\partial}{\partial \lambda} \l[ \int \frac{1}{4} \sqrt{\sigma} \; {\DM}^2 y \r]_\mathcal{H} {\delta \lambda} &-& \l[ \int_\mathcal{H} \frac{1}{8 \pi} R_{\parallel} \sqrt{\sigma} \; {\DM}^2 y \r]~ \frac{\delta \lambda}{2}\nonumber \\
&=& \int_\mathcal{H} P_{\perp} \sqrt{\sigma} \; {\DM}^2 y \; \delta \lambda
\label{eq:connect-sph-symm}
\end{eqnarray}
where we have identified $ T = \kappa / 2 \pi $ as the horizon temperature, and used the interpretation of $T^{\hatn}_{\hatn}$ as the normal pressure, $P_{\perp}$, \textit{on the horizon}. We can therefore interpret 
\begin{equation}
\overline{F} = \int_\mathcal{H} P_{\perp} \sqrt{\sigma} \; {\DM}^2 y 
\end{equation} 
as the the average normal force over the horizon ``surface"  and $\overline{F} ~ {\delta} \lambda$ as the (virtual) work done in displacing the horizon by an affine distance ${\delta} \lambda$. Equation~(\ref{eq:connect-sph-symm}) can now be written as 
\begin{eqnarray}
T {\delta}_{\lambda} S - {\delta}_{\lambda} E = \overline{F} ~ {\delta} \lambda \label{first-law-2}
\end{eqnarray}
where 
\begin{eqnarray}
S = \frac{1}{4} \int \sqrt{\sigma} \; {\DM}^2 y
\end{eqnarray}
is (a priori) just a function of $\lambda$; in particular, the derivative of $S$ with respect to $\lambda$ is well-defined and finite on the horizon. We only need the expression for $S$ very close to the horizon. The value of $S$ at $\lambda=\lambda_H$,
\begin{eqnarray}
S\l( \lambda=\lambda_H \r) = \frac{1}{4} \int_\mathcal{H} \sqrt{\sigma} \; {\DM}^2 y
\end{eqnarray}
is equal to the Bekenstein-Hawking entropy of the horizon.  We  also identify the energy $E$ associated with the horizon as
\begin{eqnarray}
E = \l( \frac{\chi}{2} \r) \frac{\lambda_H}{2} \label{energy}
\end{eqnarray}
where $\chi$ is the Euler characteristic of a 2-dimensional compact  manifold $\mathcal{M}_2$ which in this case would be the horizon 2-surface, given by
\begin{eqnarray}
\chi \left( \mathcal{M}_2 \right) = \frac{1}{4 \pi} \int_{\mathcal{M}_2} ~ R ~ \mathrm{d[vol]}
\end{eqnarray}
(If the manifold has a boundary, then the expression for Euler characteristic will have additional boundary terms.)
Thus Einstein's equations  evaluated on the horizon  can be expressed as a thermodynamic identity and --- as a bonus --- we get a geometric definition of energy.
Our particular identification of $E$ is fixed by the choice of the affine parameter along the outgoing null geodesics. In particular, this brings out the significance of the \textit{radial} coordinate $r$ in spherically symmetric and stationary spacetimes; in either case, $r$ is the affine parameter along the outgoing null geodesics. 

To connect up with the previous discussion, let us consider again the spherically symmetric case with a compact horizon, in which $\lambda=r$ and $\chi=2$. We obtain $E=r_H/2$, with $r_H$ being the horizon radius, which matches with the standard expression for quasilocal energy for such spacetimes obtained previously. In general, for a compact, simply connected horizon 2-surface, $\chi=2$ (since any such manifold is homomorphic to a 2-sphere), and we have, $E = \lambda_H / 2$. Therefore, for spherically symmetric black holes, since $P_{\perp}=P_r$ is independent of the transverse coordinates $\left( \theta, \phi \right)$, we obtain
\begin{eqnarray}
T \delta S - \delta E = P_r \delta V
\end{eqnarray}
with $\delta S = 2 \pi r_H \delta r_H$, $\delta E = \delta r_H / 2$, $P_r = T^r_r(r_H)$, $\delta V = 4 \pi r_H^2 ~ \delta r_H$ and $T$ is the horizon temperature. We therefore recover the result in \eq{EHthermo}. 
 
 Exactly similar structure emerges for the near-horizon field equations of \LL\ gravity as well.  
In this case, the analysis proceeds along identical lines though the algebra is more complicated. To begin with, using the field equation $\mathcal{G}^a_b=(1/2)T^a_b$ on the horizon  for the $m$th order \LL\ theory in $D$ dimensions and manipulating the expressions, one can obtain \cite{tpdawoodgentds} the relation:
\begin{eqnarray}
2 \mathcal{G}^{\hatxi}_{\hatxi} \; \sqrt{\sigma} \; \delta \lambda &=& T \l( \frac{1}{8} \frac{m}{2^{m-1}} \r) \mathcal{E}^{B C} \; \delta_{\lambda} \sigma_{B C} \; \sqrt{\sigma} 
\nonumber \\
&-& L^{(D-2)}_m \sqrt{\sigma} \; \delta \lambda + \mathcal{O}[ (\lambda - \lambda_H)^{1/2} \; \delta \lambda] 
\label{eom-tds-1}
\end{eqnarray}
where $T$ is the horizon temperature,  $L^{(D)}_m$ is the \LL\ Lagrangian for the $m$th order \LL\ theory in $D$ dimensions and
\begin{eqnarray}
\mathcal{E}^B_A &=& \delta^{B A_1 \ldots B_{m-1}}_{A C_1 \ldots D_{m-1}} ~ ^{(D-2)}R^{C_1 D_1}_{~ A_1 B_1} \cdots ~ ^{(D-2)}R^{C_{m-1} D_{m-1}}_{~ A_{m-1} B_{m-1}} 
\end{eqnarray}
where the upper case Latin indices $A,B,....$ etc. run over the transverse coordinates.
 We can now prove that the factor multiplying $T$ in \eq{eom-tds-1} is directly related to the variation of the following quantity, with  the variation being evaluated at $\lambda=\lambda_H$:
\begin{eqnarray}
S = 4 \pi m \int \DM \Sigma \; L^{(D-2)}_{m-1} \label{wald-entropy-2} 
\end{eqnarray}
To do this, we  use the fact
 that the variation of  $S$ in  \eq{wald-entropy-2}  must give equations of motion for the $(m-1)^{\mathrm{th}}$ order \LL\ term in $(D-2)$ dimensions. (The variation would also produce surface terms, which would not contribute when evaluated at $\lambda=\lambda_H$ because the horizon is a compact surface with no boundary.) We therefore have:
\begin{eqnarray}
\delta_{\lambda} S = - 4 \pi m \int_\mathcal{H} \DM \Sigma \; \mathcal{E'}^{B C} ~ \delta_{\lambda} \sigma_{B C} 
\end{eqnarray}
where the variation is evaluated  on $\lambda=\lambda_H$. Noting that the Lagrangian is
\begin{eqnarray*}
L^{(D-2)}_{m-1} = \frac{1}{16 \pi} \frac{1}{2^{(m-1)}} \delta^{A_1 B_1 \ldots B_{m-1}}_{C_1 D_1 \ldots D_{m-1}} ~ \cdots ~ ^{(D-2)}R^{C_{m-1} D_{m-1}}_{~ A_{m-1} B_{m-1}} 
\end{eqnarray*}
and 
\begin{eqnarray}
\mathcal{G}^i_{j(m)} &=&  - \frac{1}{2} \frac{1}{16 \pi} \frac{1}{2^m} \delta^{i a_1 b_1 \ldots a_m b_m}_{j c_1 d_1 \ldots c_m d_m} R^{c_1 d_1}_{~ a_1 b_1} \cdots R^{c_m d_m}_{~ a_m b_m}
\nonumber \\
&=& \frac{1}{16 \pi} \frac{m}{2^m} \delta^{a_1 b_1 \ldots a_m b_m}_{j_{~} d_1 \ldots c_m d_m} R^{i d_1}_{~ a_1 b_1} \cdots R^{c_m d_m}_{~ a_m b_m} - \frac{1}{2} \delta^i_{j} L_m
\label{eom-lovelock}
\end{eqnarray}
we see that
\begin{eqnarray}
\mathcal{E'}^B_C = - \frac{1}{2} \frac{1}{16 \pi} \frac{1}{2^{(m-1)}} \mathcal{E}^B_C
\end{eqnarray}
Therefore, we obtain
\begin{eqnarray}
\delta_{\lambda} S = \frac{1}{8} \frac{m}{2^{(m-1)}} \int_\mathcal{H} \DM \Sigma \;\; \mathcal{E}^{B C} ~ \delta_{\lambda} \sigma_{B C}
\end{eqnarray}
\textit{which is precisely the integral of the factor multiplying $T$ in Eq.\;(\ref{eom-tds-1})}. As mentioned earlier, $S$ defined in Eq.\;(\ref{wald-entropy-2}) is a function of $\lambda$, and its derivative with respect to $\lambda$ is well defined and finite on the horizon. The expression for $S$, evaluated at $\lambda=\lambda_H$, 
\begin{eqnarray}
S\l( \lambda=\lambda_H \r) = 4 \pi m \int_\mathcal{H} \DM \Sigma \;\; L^{(D-2)}_{m-1} \label{wald-entropy-1} 
\end{eqnarray}
can be interpreted as the entropy of the horizon and it  matches with
the standard result in  \eq{app-ent-8} obtained by other methods. 
Multiplying Eq.~(\ref{eom-tds-1}) by ${\DM}^{(D-2)} y$, integrating over the horizon surface, and taking the $n \rightarrow 0$ limit, we now see that it can be written as
\begin{eqnarray}
T {\delta}_{\lambda} S - \int_\mathcal{H} \DM \Sigma \;\; L^{(D-2)}_{m} ~ {\delta} \lambda  &=& \int_\mathcal{H} \DM \Sigma \;\; T^{\hatxi}_{\hatxi} ~ {\delta} \lambda
= \int_\mathcal{H} \DM \Sigma \;\; T^{\hatn}_{\hatn} ~ {\delta} \lambda\nonumber \\
&=& \int_\mathcal{H} \DM \Sigma \;\; P_{\perp} ~ {\delta} \lambda
\label{firstlaweqn}
\end{eqnarray}
where we have used the field equations $\mathcal{G}^{\hatxi}_{\hatxi} = (1/2) T^{\hatxi}_{\hatxi}$ in the first equality, and the relation $\mathcal{G}^{\hatxi}_{\hatxi} \;|_\mathcal{H} = \mathcal{G}^{\hatn}_{\hatn} \;|_\mathcal{H}$ in the second equality. This equation now has the desired form of the first law of thermodynamics, when we identify: (i)  the $S$, defined by Eq.~(\ref{wald-entropy-1}), as the entropy of horizons in \LL\ gravity; indeed, \textit{exactly} the same expression for entropy has been obtained in the literature using the Wald entropy (e.g, ref.\cite{entropyLL}) and (ii)  the second term on the left hand side of \eq{firstlaweqn} as $\delta_{\lambda} E$; this leads to the definition of $E$ to be
\begin{eqnarray}
E = \int^{\lambda} \delta \lambda \int_\mathcal{H} \DM \Sigma \;\; L^{(D-2)}_{m}
\label{energy-ll}
\end{eqnarray}
where the $\lambda \rightarrow \lambda_H$ limit must be taken \textit{after} the integral is done. (Therefore, we need to know the detailed form of $L^{(D-2)}_{m}$ as a function of $\lambda$ to calculate this explicitly). For $D = 2 (m+1)$, the integral over $\mathcal{H}$  in \eq{energy-ll} is related to the Euler characteristic of the horizon and we get $E \propto \lambda_H$.  For $m=1, D=4$, this reduces to the expression obtained earlier in the case of Einstein gravity. 

One can also determine the scaling of $E$ for \textit{spherically symmetric} spacetimes for general \LL\ lagrangians, with horizon at $r=r_H$, and $\lambda=r$. In this case, $L^{(D-2)}_{m} \propto (1/{\lambda}^2)^{m}$ and $\sqrt{\sigma} \propto {\lambda}^{D-2}$. The integrand in \eq{energy-ll}  scales as ${\lambda}^{(D-2) - 2 m}$ giving $E \propto \lambda_H^{(D-2) - 2 m + 1}$. As mentioned above, for $D = 2 (m+1)$, $E \propto \lambda_H$. In fact, in the case of spherically symmetric spacetimes in \LL\ theory, the above expression can be  shown to be \textit{exactly equivalent} to the one derived by others (see \cite{KP}, and also Ref.~[14] therein). 
No general expression for energy in \LL\ theory exists in the literature, and \eq{energy-ll} could be thought of as first such definition which appears to be reasonable from physical point of view.

We thus conclude that for \textit{generic static spacetimes} in \LL\ gravity, the field equations can be written as a thermodynamic identity:
\begin{eqnarray}
T \delta_{\lambda} S - \delta_{\lambda} E = \overline{F} \delta \lambda  
\end{eqnarray} 
thereby showing that the thermodynamic relations hold for much more   general cases than Einstein's  theory.

\subsection{Structure of gravitational action functionals}

 Since the field equations of  a theory can be derived by varying the dynamical variables in a suitably defined action functional, it makes sense to examine the nature of the action functionals in order to learn more about the curious connection between horizon thermodynamics and gravitational dynamics discussed above.  
  We shall now  describe several peculiar features of action functionals in  Einstein's theory 
  as well as in more general
   \LL\ theories of gravity. Once again, these results have no explanation in 
  the conventional approach.

  Consider a theory based on a gravitational  Lagrangian $L_g$ which can be expressed in the form
  \begin{equation}
16\pi L_{g}\equiv Q_a^{\phantom{a}bcd}R^a_{\phantom{a}bcd}
= Q_{ab}^{cd}R^{ab}_{cd}
\label{lisrq}
\end{equation} 
where the tensor  $Q_a^{\phantom{a}bcd}$  has (i) all the symmetries of the curvature tensor and  (ii) has zero divergence on all the indices, $\nabla_a Q^{abcd}=0$ etc. 
With some simple algebraic manipulation, we can express \cite{ayan} any such action in the form
\begin{eqnarray}
\sqrt{-g}Q_{a}^{\phantom{a}bcd}R^{a}_{\phantom{a}bcd} &=&
2\sqrt{-g}Q_a^{\phantom{a}bcd}\Gamma^a_{dk}\Gamma^k_{bc} + 2\partial_c\left[\sqrt{-g}Q_a^{\phantom{a}bcd}\Gamma^a_{bd}\right]\nonumber\\
&\equiv&   \sqrt{-g}L_{\rm quad} + L_{\rm sur}
\label{gensq}
 \end{eqnarray} 
where we have separated out the expression into one term [$L_{\rm quad}$] which is quadratic in $\G abc$ (and hence quadratic in the first derivatives of the metric) and another term [$L_{\rm sur}$] which is a total divergence that could lead to a surface term on integration. 
Thus, to begin with, we have the result that a wide class of theories determined by the tensor $Q^{abcd}$ has an action which will naturally lead to a surface term.
We shall next show that: (i) there is a holographic relationship between the surface and bulk term and  that (ii) the surface term leads to the entropy of the horizon. 

\subsection{Holographic nature of surface term in Einstein-Hilbert action}

To set the stage, let us begin with Einstein's theory which has the Lagrangian  $L_g=(16\pi)^{-1} R$. This Lagrangian 
can be expressed in the form in \eq{lisrq}
with
\begin{equation}
 Q_a^{\phantom{a}bcd}=\frac{1}{2}(\delta^c_ag^{bd}-\delta^d_ag^{bc});\quad
 Q^{cd}_{ab}=\delta_{ab}^{cd}=\frac{1}{2}(\delta^c_a\delta_b^{d}-\delta^d_a\delta_b^{c})
 \label{altq}
\end{equation} 
In this particular case, the bulk term involving the quadratic part is given by
\begin{equation}
L_{\rm quad}=2Q_a^{\phantom{a}bcd}\Gamma^a_{dk}\Gamma^k_{bc}
=g^{ab} \left(\Gamma^i_{ja} \Gamma^j_{ib} -\Gamma^i_{ab} \Gamma^j_{ij}\right)
\label{quadpart}
\end{equation}
and  the surface term arises from 
\begin{equation}
L_{\rm sur}=2\partial_c\left[\sqrt{-g}Q_a^{\phantom{a}bcd}\Gamma^a_{bd}\right]
=2\partial_c\left[\sqrt{-g}Q_{ak}^{cd}g^{bk}\Gamma^a_{bd}\right]
\equiv\partial_c\left[\sqrt{-g}V^c\right]
\label{surpart}
\end{equation} 
where we have defined a four component object $V^c$ (which is \textit{not} a four-vector) by:
\begin{equation}
 V^c \equiv \left(g^{ik} \Gamma^c_{ik}-g^{ck} \Gamma^m_{km}\right) 
 = (g^{ia} g^{cj}-g^{aj} g^{ci})\partial_ig_{aj}
 =-\frac{1}{g} \partial_b (g g^{bc})
    \label{defpcone}
     \end{equation}
By direct computation one can verify that the bulk and the surface term are related by the equation 
\begin{equation}
    \sqrt{-g}L_{sur}=-\partial_a\left(g_{ij}
\frac{\partial \sqrt{-g}L_{bulk}}{\partial(\partial_ag_{ij})}\right)
\label{pecurrel}
\end{equation}
(We call this relation `holographic' for want of better terminology; it has no connection with the term holography used in string theory.)
This structure has a simple physical meaning that  can be understood as follows \cite{TPhol002,TPgravquantum}.
Given a Lagrangian $L_q(\dot q, q)$ in classical mechanics, say,  one can obtain the standard Euler-Lagrange equations
by varying $q$ in the action functional with the  condition $\delta q =0 $ at the boundary.
Consider now a different Lagrangian defined as 
\begin{equation}
L_p (\ddot q, \dot q, q) \equiv L_q(\dot q, q) - \frac{d}{dt} \left( q \frac{\partial L_q}{\partial \dot q}\right) 
\label{lp}
\end{equation} 
This Lagrangian, unlike $L_q$ contains $\ddot q$. If we  vary the action resulting from 
$L_p (\ddot q, \dot q, q)$ 
but --- instead of demanding $\delta q=0$ at the boundary --- demand that $\delta p=0$ at the boundary where $p(\dot q,q)\equiv (\partial L_q/\partial \dot q)$ is the momentum,
then we will 
get the same equations of motion as the one obtained from $L_q$. \textit{That is, even though 
$L_p$ contains the second derivatives of $q$, it leads to second order differential equations
for $q$ (rather than third order) if we fix $p$ at the boundary.}
 Lagrangians involving second derivatives
of dynamical variables but in a specific combination through the second term in $L_p (\ddot q, \dot q, q)$  are quite special. This idea generalizes trivially to field theory
and we see, comparing \eq{pecurrel} with \eq{lp}, that Einstein's theory has this special structure.
It is this holographic relationship which allows the surface terms to contain information about the bulk. 

\subsection{Horizon entropy and the surface term in Einstein-Hilbert action}

It is also easy to show that the surface term actually leads to the horizon entropy in the case of Einstein's theory. To do this, we will work with the Euclidean extension of the action in which the horizon is mapped to the origin. Close to the origin, in Rindler like coordinates, we have the metric of the form 
\begin{equation}
ds^2 =  \kappa^2 \xi^2 dt_E^2 + d\xi^2 + d\mathbf{x}_\perp^2
\label{surfrindnew}
\end{equation} 
To evaluate the surface integral arising from $L_{\rm sur}$ in \eq{surpart} on the  horizon, we shall compute it on the surface $\xi = \epsilon$ around the origin in the $\xi-t_E$ plane and then take the limit of $\epsilon \to 0$. 
So we need to integrate $\sqrt{h} n_c V^c$ where $V^c$ is defined by \eq{surpart},
$\sqrt{h}=\kappa \epsilon \sqrt{\sigma} $ with ${\sigma}$ being the determinant of the metric in the 
transverse coordinates and $n^i = \delta^i_\xi$ is the normal. In the integral, the range of $t_E$, being an angular coordinate, is 
 $(0,\beta =2\pi/\kappa)$. Using  $V^\xi= -2/\epsilon$,
 we find that the surface contribution to the action is 
 \begin{eqnarray}
  16\pi \mathcal{A}_{\rm sur} &=& \int_{\xi=\epsilon} d^3x\, \sqrt{h} n_c V^c = \int_0^{2\pi/\kappa} dt_E \int d^2 \mathbf{x}_\perp ( \kappa\epsilon \sqrt{\sigma}) \left(-\frac{2}{\epsilon}\right)\nonumber\\
   &=&-4\pi A_\perp
 \end{eqnarray} 
 Therefore,
 \begin{equation}
\mathcal{A}_{\rm sur} = - \frac{1}{4} A_\perp 
\label{onequarter}
\end{equation} 
where $A_\perp $ is the transverse area.
This result shows that the surface term in the action has a direct thermodynamic meaning
as the horizon entropy \cite{TPgrav06}.
The sign flips if we change the sign of $n_c$ and hence is not of real significance.
(But, with our choice for $n_c$, the sign   can be explained by the fact that the probability for a configuration
is related to Euclidean action by $\mathcal{P} \propto \exp(-\mathcal{A}_{\rm sur})$
while $\mathcal{P} \propto \exp(S)$ where $S$ is the entropy; hence
$S= - \mathcal{A}_{\rm sur}$.)
As we said before, this result has no explanation  in the conventional approach in
which the field equations know nothing about the surface term.

There is another curious aspect related to the surface term in Einstein-Hilbert action which is worth mentioning. In standard quantum field theory, the kinetic term for the field
$\phi$ will be quadratic in the derivatives of the field variable ($(\partial \phi)^2$) which will be integrated over the four volume to obtain the action. In natural units,
action is dimensionless and hence all fields will have the dimension of inverse length.
In the case of gravitational field, one might like to associate a second rank symmetric tensor field, $H_{ab}$,  to describe the graviton. In that case, the metric $g_{ab}$ will be interpreted as $g_{a b} = \eta_{ab}+l H_{ab}$ where $l$  is a constant with dimensions 
of length. [In normal units, $l^2 \propto (G\hbar/c^3)$.] Let us consider 
what happens to the Einstein-Hilbert action when we use this expansion and retain terms  up to the lowest non-vanishing order in the bulk and surface terms. 
Expanding $L_{\rm quad}$ and $L_{\rm sur}$ in Taylor series in $l$ and choosing   
 $l^2 = 16\pi G$, where $G$ is the Newtonian gravitational constant re-introduced for the sake of clarity, 
we find that the action functional becomes
\begin{equation}
\mathcal{A}\equiv  \frac{1}{16\pi G} \int d^4 x \sqrt{-g} R = \mathcal{A}_{\rm quad} + \mathcal{A}_{\rm sur}
\end{equation} 
where 
\begin{equation}
\mathcal{A}_{quad}=\frac{1}{4} \int d^4x\,   M^{abcijk}(\eta^{mn})\partial_a H_{bc}\partial_i H_{jk} 
+ \mathcal{O}(l)
\end{equation} 
and 
\begin{equation}
\mathcal{A}_{\rm sur}=
\frac{1}{4l}\int d^4x\,  \partial_a \partial_b[H^{ab}-\eta^{ab}H^i_i]+ \mathcal{O}(1)
\label{leading}
\end{equation}
with 
\begin{equation}
M^{abcijk}(\eta^{mn}) = \left[\eta^{ai} \eta^{bc}\eta^{jk}
 -\eta^{ai}\eta^{bj}\eta^{ck}
+2 \eta^{ak}\eta^{bj}\eta^{ci}
 -2\eta^{ak}\eta^{bc}\eta^{ij}\right]
 \label{formofmapp}
\end{equation}
This $\mathcal{A}_{\rm quad}$ matches exactly with the action for the spin-2 field
known as Fierz-Pauli action (see e.g. Ref. \cite{tpgraviton}). However, the surface term --- which is usually ignored in the discussions --- 
is \textit{non-analytic}
in the coupling constant. Hence
 one cannot provide an interpretation of black hole entropy 
(which, as we have seen, can be obtained from the surface term in the action)
in the linear, weak coupling limit of gravity. 
The integral we evaluated in the Euclidean sector around the origin to obtain
the result in \eq{onequarter} cannot even be defined usefully in the weak field limit
 because we used the fact that $g_{00}$ vanishes at the origin. When we take $g_{00}= \eta_{00}+h_{00}$ and treat $h_{00}$ as a  perturbation, it is obviously not possible 
to make $g_{00}$ vanish.

 In fact, the non-analytic behaviour of $\mathcal{A}_{\rm sur}$ on $l$  can be obtained from 
 fairly simple considerations 
  related to the algebraic structure 
  of  the curvature scalar.
 In terms of a spin-2 field, the final metric arises as $g_{ab}=\eta_{ab}+l\ H_{ab}$
where $l\propto \sqrt{G }$ has the dimension of length and $H_{ab}$ has the correct dimension of
(length)$^{-1}$ in natural units with $\hbar=c=1$. 
  Since the scalar curvature has the structure $R\simeq (\partial g)^2+\partial^2g$, substitution of $g_{ab}=\eta_{ab}+l\ H_{ab}$ gives to the lowest order:
\begin{equation}
L_{EH}\propto \frac{1}{l^2}R\simeq (\partial H)^2+\frac{1}{l}\partial^2 H
\end{equation}
Thus the full Einstein-Hilbert Lagrangian is non-analytic in $l$ because of the surface term.

\subsection{Holographic structure of action functional in \LL\ gravity}

We shall now consider the generalization of these results to \LL\ theories with an action described by a tensor $Q^{abcd}$.
In this case, $Q^{abcd}$  depends on the metric as well as curvature tensor but not on the derivatives of the curvature tensor.
The Lagrangian for the $m-$th order \LL\ theory in $D-$dimension is given by
\begin{equation}
L_{(m)}=\delta^{1357...2k-1}_{2468...2k}R^{24}_{13}R^{68}_{57}
....R^{2k-2\,2k}_{2k-3\,2k-1}; \qquad k=2m
\label{elmlll}
\end{equation} 
where $k=2m$ is an even number. 
The $L_{(m)}$ is clearly a homogeneous function of degree $m$ in
the curvature tensor $R^{ab}_{cd}$ so that it  can also  be expressed in the form:
\begin{equation}
L_{(m)}=\frac{1}{m}\left(\frac{\partial L_{(m)}}{\partial R^a_{\phantom{a}bcd} }\right)R^a_{\phantom{a}bcd}\equiv \frac{1}{m}P_a^{\phantom{a}bcd}R^a_{\phantom{a}bcd}.
\end{equation} 
where  $P_a^{\phantom{a}bcd}\equiv (\partial L_{(m)}/\partial R^a_{\phantom{a}bcd} )$
so that $P^{abcd}=mQ^{abcd}$.

The canonical momentum conjugate to the metric  has to be defined more carefully in this case because $L_{\rm bulk}$ will contain second derivatives of the metric (unlike in the case of Einstein's theory). Once this technical problem is taken care of, one can show by direct computation that the surface and bulk terms obey a holographic relation given by
\begin{equation}
[(D/2) - m]L_{sur}^{(m)} =-\partial_i \left[ g_{ab} \frac{\delta L_{bulk}^{(m)}}{\delta (\partial_i g_{ab})}
  +\partial_jg_{ab} \frac{\partial L_{bulk}^{(m)}}{\partial (\partial_i \partial_jg_{ab})}
  \right] 
\label{result1} 
\end{equation} 
where the \textit{Euler derivative} is defined as 
\begin{equation}
\frac{\delta K[\phi,\partial_i\phi,...]}{\delta\phi}=
\frac{\partial K[\phi,\partial_i\phi,...]}{\partial\phi}-
\partial_a \left[\frac{\partial K[\phi,\partial_i\phi,...]}{\partial(\partial_a\phi)}\right]+\cdots
\end{equation}
The proof involves straight forward combinatorics \cite{ayan}. This shows that a wide class of gravitational theories which have the surface term in the action functional  exhibits the holographic relationship between surface and bulk terms. 

\subsection{Horizon entropy from the surface term in the \LL\ action functional}

To generalize the result that the surface term in the action functional gives the entropy of the horizon, we need to compare the surface term with the entropy in the \LL\ theories 
which was obtained earlier  in \eq{SNoether} by using the Noether charge. We saw that, the Noether charge approach leads to the expression (see \eq{s}):
\begin{equation}
S_{Noether} = 2\beta m \mes{D-2}{t,r_H} Q^{cdr0}\partial_d g_{c0}; \qquad \beta = \frac{2\pi}{\kappa}
\label{s1}
\end{equation}  
We will now show that  the same  result can be  obtained by evaluating the surface term in the action on the horizon.
 For the \LL\ models, the bulk and the surface Lagrangians are given by 
 \begin{equation}
L_{\rm quad}=2Q_a^{\phantom{a}bcd}\Gamma^a_{dk}\Gamma^k_{bc};
\qquad
L_{\rm sur}=2\partial_c\left[\sqrt{-g}Q_a^{\phantom{a}bcd}\Gamma^a_{bd}\right]
\label{quadsur}
\end{equation}
  In the stationary case, the contribution of surface term on the horizon is given by
\begin{eqnarray}
S_{sur} &=& 2 \int d^D x \, \partial_c \left[ \sqrt{-g} Q^{abcd} \partial_b g_{ad}\right]\nonumber\\
&=& 2\int dt\mes{D-2}{r_H} Q^{abrd} \partial_b g_{ad}
\end{eqnarray} 
Once again, taking the integration over $t$ to be in the range $(0,\beta)$ and ignoring
transverse directions, we get
\begin{equation}
S_{sur} = 2\beta \mes{D-2}{r_H} Q^{abr0} \partial_b g_{a0}
\end{equation} 
Comparing with \eq{s1}, we find that 
\begin{equation}
S_{Noether} = mS_{sur}
\end{equation} 
The overall proportionality factor $m$ has a simple physical meaning. \eq{result1} tells us that the quantity $mL_{sur}$, rather than $L_{sur}$, which has the $``d(qp)"$ structure and it is this particular combination which plays the role of entropy, as to be expected. 

\textit{It must be stressed that the above results defy understanding in the conventional approach.} To begin with, it is not clear why the simplest generally covariant action in general relativity (and in \LL\ models)  contains a total divergence term which leads to a surface term.
  Further, in the conventional  approach, we \textit{ignore} the surface term completely
(or cancel it with a counter-term) and obtain the field equation
from the bulk term in the action. Any solution to the field equation obtained 
by this procedure is logically independent of the nature of the surface term.
But  when the \textit{surface term} (which was ignored) is evaluated at the horizon that arises
in any given solution, it does correctly give the entropy of the horizon!
This is possibly because of the specific relationship 
between the 
surface term and the bulk term
 given by \eq{pecurrel} and \eq{result1}. But these relations are
 again totally unexplained feature in the conventional
approach to gravitational dynamics.
Given that the surface term has the thermodynamic interpretation as the entropy of horizons,
 and is related holographically to the bulk term, we are again led to 
an indirect connection between spacetime dynamics and horizon thermodynamics.

\subsection{Gravitational action as the free energy of spacetime}\label{sec:gravactfree}

There is one more aspect of the gravitational action functional which adds strength to the thermodynamic interpretation. We can show that, in any static spacetime with a bifurcation horizon (so that the metric is periodic with a period $\beta = 2\pi/\kappa$ in the Euclidean sector), the action functional for gravity can be interpreted as the free energy
of spacetime. 

Since the spacetime is static, there exists a timelike Killing vector field $\xi^a$ with components $\xi^a = (1,\mathbf{0})$ in the natural coordinate system which exhibits the static nature of the spacetime. The conserved Noether current for the displacement $x^a \to x^a + \xi^a$ is given by \eq{current1} . We will work in the Euclidean sector and integrate this expression, taken on-shell with $2\mathcal{G}_{ab}=T_{ab}$, over a constant-$t$ hypersurface with the measure $d\Sigma_a = \delta_a^0 N\sqrt{h}\, d^{D-1} x$
where $g^{E}_{00} = N^2$ and $h$ is the determinant of the spatial metric. Multiplying by
the period $\beta$ of the imaginary time, we get
\begin{eqnarray}
\beta \int J^a d \Sigma_a 
&=& \beta \int T^a_b \xi^b d\Sigma_a  + \beta \int L \xi^a d \Sigma_a\nonumber\\
&=& \int  (\beta  N) T^a_b u_au^b \sqrt{h}\, d^{D-1}x  + \int_0^\beta dt_E \int L \sqrt{g}\, d^{D-1}x
\label{useja}
\end{eqnarray} 
where we have introduced the four velocity  $u^a = \xi^a/N = N^{-1}\delta^a_0$
of observers moving along the orbits of $\xi^a$ and the relation $d\Sigma_a = u_a\sqrt{h}\, d^{D-1} x$.  The term  in \eq{app-ent-5} involving the Lagrangian gives the Euclidean action
for the theory. In the term involving $T_{ab}$ we note that $\beta N \equiv \beta_{\rm loc}$ corresponds to the correct redshifted local temperature.
Hence we can define the (thermally averaged) energy $E$ as
\begin{equation}
\int  (\beta  N) T^a_b u_au^b \sqrt{h}\, d^{D-1}x =
\int  \beta_{\rm loc}   T^a_b u_au^b \sqrt{h}\, d^{D-1}x \equiv 
\beta E
\label{defE}
\end{equation} 
We thus get 
\begin{equation}
A= \beta \int J^a d \Sigma_a -\beta E
\end{equation} 
We have, however, seen earlier that the first term involving the Noether charge
gives the horizon entropy, which continues to hold true in the Euclidean sector.
Therefore, we find that 
\begin{equation}
A= S- \beta E = -\beta F 
\end{equation} 
where $F$ is the free energy.  (Usually, one defines the Euclidean action with an extra minus sign as $A=-A_E$ in which case the Euclidean action can be interpreted directly as the free energy.) The motivation for the definition of $E$ in \eq{defE}
now becomes clear since the entropy is related to the spatial integral of the energy density $\rho(\mathbf{x})$ with a weightage factor 
$\beta (\mathbf{x})$ when the temperature varies in space. One can also obtain from \eq{useja} the  relation:
\begin{equation}
S = \beta \int J^a d \Sigma_a =  \int \sqrt{-g}\, d^Dx (\rho+L ) 
\end{equation} 
where $\rho= T_{ab} u^a u^b$.  This equation gives the entropy in terms of matter energy
density and the Euclidean action. 
Alternatively, if we assume that the Euclidean action can be interpreted as 
the free energy, then these relations  provide an alternate justification
for interpreting the Noether charge as the entropy.
(Similar results have been obtained in 
Ref. \cite{mattdbh} by a more complicated procedure and with a different motivation.)

There is another result which one can obtain from the expression for the Noether current. Taking the $J^0$ component of \eq{current1} and writing $J^0 = \nabla_b J^{0b}$
we obtain 
\begin{equation}
L = \frac{1}{\sqrt{-g}} \partial_\alpha \left( \sqrt{-g}\, J^{0\alpha}\right) - 2 \mathcal{G}^0_0
\label{strucL}
\end{equation} 
Only spatial derivatives contribute in the first term on the right hand side when the 
spacetime is static. This relation shows that the action obtained by integrating $L\sqrt{-g}$ will generically have a surface term related to $J^{ab}$ (In Einstein gravity \eq{strucL} will read as $L=2R^0_0-2G^0_0$; our result generalizes the fact that $R^0_0$ can be expressed as a total divergence in static spacetimes \cite{TPbrazil05}.) This again illustrates, in a very general manner, why the surface terms in the action functional
lead to horizon entropy. 

\eq{strucL} can be expressed more formally by introducing a vector $l_k\equiv \nabla_k t=\partial_kt=(1,\mathbf{0})$ which is the unnormalised normal to $t=$ constant hypersurfaces. The unit normal is $\hat l_k=Nl_k=-u_k$ in the Lorentzian sector. Contracting $l_k$ with the expression for Noether charge, written for a Killing vector $\xi^a$, in the form
\begin{equation}
2\mathcal{G}^k_b\xi^b+L\xi^k=J^k=\nabla_aJ^{ka}
\end{equation} 
and using 
$
l_k\nabla_aJ^{ka}=\nabla_a(l_kJ^{ka})
$, we get
\begin{equation}
L=\nabla_a(l_kJ^{ka})-2\mathcal{G}^k_b\xi^bl_k=\frac{1}{\sqrt{-g}} \partial_\alpha \left( \sqrt{-g}\, l_kJ^{k\alpha}\right)-2\mathcal{G}^k_b\xi^bl_k
\end{equation} 
which is identical to \eq{strucL} but is sometimes more convenient for manipulation.

Finally, we show how several of the results obtained in ref.\cite{TPgravcqg04} giving a thermodynamic interpretation for Einstein-Hilbert action, can be generalized for \LL\ gravity. Using the \LL\ equations of motion, we can easily show that, for the $m$-th order \LL\ model in D-dimensions,
\begin{equation}
T=-(D-2m)L;\quad 2\mathcal{G}^a_b+L\delta^a_b=T^a_b-\frac{1}{(D-2m)}\delta^a_bT
\end{equation} 
Defining as before $\rho=T_{ab}u^au^b$ and 
\begin{equation}
\epsilon\equiv[T^a_b-\frac{1}{(D-2m)}\delta^a_bT]u_au^b
\end{equation}
 we have the relations 
\begin{equation}
(\epsilon-\rho)=-L; \quad  \nabla_a(l_kJ^{ka})=-\epsilon
\end{equation}  
Integrating over spacetime, treating the time integral as multiplication by $\beta$, and defining the entropy in terms of the Noether charge, we get
\begin{equation}
S-\beta E=-A=-\beta F
\end{equation}  
and $S=(\beta/2)M$ where $M$ is the generalization of the Komar mass, defined as the thermally weighted integral
\begin{equation}
\beta M=2\int \beta_{loc}\epsilon \sqrt{h}d^{D-1}x
\end{equation} 
All these results have been obtained previously in the context of general relativity in ref.\cite{TPgravcqg04} and the above analysis shows that they continue to remain valid for \LL\ models, strengthening the thermodynamic connection. In the context of general relativity,
using $S=A_H/4L_P^2$ where $A_H$ is the horizon area, the relation $S=(\beta/2)M$ can be written as $M=(1/2)nk_BT$ with $n=A_H/L_P^2$. This is just the law of equipartition of energy among the horizon degrees of freedom \cite{equipartition} if we think of horizon as divided into patches of size $L_P^2$. More generally, the definition of entropy given in  ref.\cite{TPgravcqg04} leads to the equipartition law
as an integral 
over the local acceleration temperature $T_{\rm loc}\equiv (N a^\mu n_\mu) /2\pi$ and is 
given by 
\begin{equation}
M  =  \frac{1}{2}k_B
 \int_{\partial\cal V}\frac{\sqrt{\sigma}\, d^2x}{L_P^2}\left\{\frac{N a^\mu n_\mu}{2\pi}
\right\}
\equiv \frac{1}{2} k_B \int_{\partial\cal V}dn\, T_{\rm loc}
\label{idn}
\end{equation} 
thereby identifying the number of degrees of freedom to be $dn=  \sqrt{\sigma}\, d^2x/L_P^2$ in an area element $\sqrt{\sigma} \, d^2 x$. (One can, alternatively, write down this relation by inspection and obtain the gravitational field equations as a consequence, provided one accepts the choice of various numerical factors.) This suggests that we  interpret the relation $S=(\beta/2)M$ as law of equipartition even in the more general context of \LL\ theories and identify $M/(\frac{1}{2}k_BT)=4S$ as the effective number of degrees of freedom.
(In the \LL\ models this will not be proportional to the area.)
The Noether current and its relation to entropy will play
a crucial role in our future discussions.

\subsection{Why are gravitational actions functionals holographic?}

We shall now reinterpret the above results by approaching them, from first principles, in  a manner which makes these  relations logically transparent \cite{TPhol002}.

We will start with the principle of equivalence and  draw from it three important consequences. First, it implies that, at least in the long wavelength limit, gravitational field has a geometrical description in terms of the metric of the spacetime \cite{mtw}. Second, the validity of laws of special relativity in the local inertial frames allows one to determine the influence of gravity on light rays. From this we conclude that, in any given spacetime, there will exist observers who will have 
access to only part of the spacetime. That is, horizons --- as perceived by a class of observers ---  will exist in any geometrical description of gravity that obeys principle of equivalence.
Finally, the principle of equivalence also demands (indirectly) a principle of `democracy of observers'. 
In flat spacetime, we can choose a special coordinate system with the global metric being $\eta_{ab}$; so  when the spacetime   metric is given to be $g_{ab}(t,\mathbf{x})$ 
we  can always attribute the part $(g_{ab}-\eta_{ab})$ to the choice of non-inertial coordinates. 
In a curved spacetime we cannot do this and hence it no longer makes sense to ask ``how much of $g_{ab}$" is due to our using non-inertial coordinates and ``how much" is due to genuine gravity.  
Different observers in different states of motion might use different coordinates leading to different sets of ten functions for $g_{ab}(t,\mathbf{x})$. Because we have no absolute reference metric, it follows that 
no coordinate system or  observer is special and the  
 laws of physics should not select out any special class of observers.

Therefore,  we need a  principle
 to handle the fact that different observers will have access to different regions of   the spacetime \cite{tpapoorva}.
  If a class of observers perceive a horizon, they should still be able to do physics using only the variables accessible to them without having to know what happens on the other side of the horizon.
  Classically, the light cone structure dictated by the horizon ensures that 
  the region inside the horizon cannot influence the outside. This is, however, not true quantum mechanically when, for example, we take into
  account the entanglement of fields across the horizon.
   Hence there should exist a mechanism which will encode
  the information in the region $\mathcal{V}$, which is inaccessible to a particular observer, at the boundary $\partial \mathcal{V}$ of that region.

One possible way of ensuring this is to have a suitable boundary term in the action principle which will contain  the necessary information for observers who perceive a horizon. Such a possibility leads to three immediate consequences. 
  First, if the theory is generally covariant, so that observers with horizons (like, for example, uniformly accelerated observers using a Rindler metric) need to be 
 accommodated in the theory, such a theory \textit{must} have an action functional that contains
  a surface term.
  (Since general covariance leads to the Noether current, which in turn shows that the action will have surface term --- see \eq{strucL} --- we already have a direct demonstration of this connection in all static spacetimes.)
  Second, if the surface term has to  encode the information which is blocked by the horizon,
   then there must exist a simple relation between the bulk term
   and surface term in the action and it cannot be  arbitrary.
   Third, 
   if the surface term encodes information which is blocked by the horizon, then
   it should actually lead to the entropy of the horizon. In other words, we should be able
   to compute the horizon entropy by evaluating the surface term.
  
  These three requirements are very strong constraints on the nature of action functional describing  a theory of gravity. 
  In fact these are sufficiently powerful for us to \textit{reconstruct} the action functional for this class of theories. We shall illustrate \cite{TPapspsci03,TPbrazil05} how this can be done in  the case of Einstein's theory.
  
To do this we shall start with the assumption that  the action in general relativity will have a surface term (obtained by integrating a local divergence) which is  holographically related 
 to the bulk term by \eq{pecurrel}. This specific form is needed to ensure that the
 {\it same}  equations
    of motion  are obtained from $A_{\rm bulk}$ or from another $A'$ (both, as yet, unknown) where:
    \begin{eqnarray}
    \label{aeh}
   A' &=& \int d^4x \sqrt{-g} L_{\rm bulk} - \int d^4x \partial_c \left[ g_{ab}
    \frac{\partial \sqrt{-g} L_{\rm bulk} }{ \partial(\partial_c g_{ab})}
    \right]   \nonumber\\
   &\equiv&  A_{\rm bulk} + \int d^4 x \partial_c (\sqrt{-g}V^c) 
    \end{eqnarray}   
    with $V^c $ constructed from  $g_{ab} $ and $\Gamma^i_{jk}$ with no further explicit dependence on $\sqrt{-g}$, which has been factored out. Further, $V^c$ must be linear
    in the $\Gamma$'s since the original Lagrangian $L_{\rm bulk}$ was quadratic in the first derivatives
    of the metric. 
    Since $\Gamma$s vanish in the local inertial frame and the metric reduces to
    the Lorentzian form, the action $A_{\rm bulk}$ cannot be generally covariant. Our aim is to determine the surface term using the known expression for horizon entropy and then determine the bulk term which is consistent with the holographic relation. At this stage we have no assurance that the resulting action $A'$ will be generally covariant but it is an important consistency check on the idea.

    To obtain a quantity $V^c$, which is linear in $\Gamma$s
     and having  a single index $c$, from $g_{ab} $ and $\Gamma^i_{jk}$,
     we must contract on two of the indices on $\Gamma$
    using the metric tensor. 
    (Note that we require $A_{\rm bulk}$, $A'$ etc. to be Lorentz scalars and $P^c, V^c$ etc.
    to be vectors under Lorentz transformation.)
    Hence the most general choice for $V^c$ is the linear combination
      \begin{equation}
    V^c =  \left(a_1 g^{ck} \Gamma^m_{km} +a_2 g^{ik} \Gamma^c_{ik}\right) 
    \label{defvc}
     \end{equation}
     where $a_1$ and $a_2$ are two constants. 
      Using the identities $\Gamma^m_{km} =\partial_k
     (\ln \sqrt{-g})$, \ $\sqrt{-g}g^{ik}\Gamma^c_{ik} = -\partial_b(\sqrt{-g}g^{bc})$,
     we can rewrite the expression for $P^c \equiv \sqrt{-g}V^c$ as 
     \begin{equation}
    P^c =\sqrt{-g}V^c=
    c_1g^{cb} \partial_b \sqrt{-g} +c_2 \sqrt{-g} \partial_b g^{bc}
    \label{defpc}
    \end{equation}
    where $c_1\equiv a_1 - a_2,\  c_2\equiv -a_2$ are two other unknown constants.
    We now fix these coefficients by using 
     our demand that $A_{\rm sur}$ should lead to the entropy of the horizon, thereby determining
      the surface term and --- by integrating --- the Lagrangian $L_{\rm bulk}$.
      We will use a Euclidean Rindler frame for this purpose.

      The form of Rindler metric that is sufficiently general for our purpose can be taken to be
  \begin{eqnarray}
    \label{rindmetric}
  ds^2 &=& 2\kappa l\, dt_E^2 + \frac{dl^2}{2\kappa l} + (dy^2+dz^2) \\
  &=& 2\kappa l(x) \, dt_E^2 + \frac{l^{'2}dx^2 }{2\kappa l(x)} + (dy^2+dz^2)\nonumber
  \end{eqnarray}
  where $l(x)$ is an arbitrary function and $l' \equiv (dl/dx)$. 
  The metric in the first line is same as the one in \eq{standardhorizon}; the second line introduces an arbitrary coordinate $x$ through the function $l(x)$. The horizon is at $l(x)=0$ and $\kappa$ is the surface gravity for all $l(x)$, giving the horizon temperature to be $\kappa/2\pi$.
  Evaluating the surface term $P^c$ in Eq.~(\ref{defpc})
  for this metric, we get
  the only nonzero component to be
  \begin{equation}
  P^x = 2\kappa \left[  c_2 +  \frac{ll''}{l^{'2}} [c_1- 2 c_2]\right]
  \end{equation}
       so that the surface term in the action in Eq.~(\ref{aeh})
    becomes 
    \begin{equation}
    A_{\rm sur}=\beta P^x \int d^2x_\perp =\beta P^x \mathcal{A}_\perp
    = 4\pi  \mathcal{A}_\perp\left[  c_2 +  \frac{ll''}{l^{'2}} [c_1- 2 c_2]\right]
    \label{detcone}
    \end{equation}   
    where $\mathcal{A}_\perp$ is the transverse area of the $(y-z)$ plane and the time integration is limited to the range $(0, \beta)$ with $\beta = 2\pi/\kappa$.
     Demanding that $A_{\rm sur}$ should give the horizon entropy, which we take
     to be proportional to the transverse area,
     as  $-(\mathcal{A}_\perp / 4\mathcal{A}_p)$ where $\mathcal{A}_p$ is a constant, we get the condition:
     \begin{equation}
      \left[  c_2 +  \frac{ll''}{l^{'2}} [c_1- 2 c_2]\right]= - \frac{1}{16\pi \mathcal{A}_P}
     \label{eqnxxx}
     \end{equation}
     (The minus sign arises because we are working with Euclidean signature as in, for example, \eq{onequarter}.)
    We demand that this equation should hold,  on the horizon, for all functions $l(x)$.
    If we take $l(x)=x^n$, then $ll''/l'^2 = (n-1)/n$. Taking $n=1 $ makes the second term on the left hand side of \eq{eqnxxx} vanish giving $c_2=-(16\pi \mathcal{A}_P)^{-1}$.
    For other values of $n$, we now require the second term to vanish identically which is possible only if  $c_1 = 2 c_2$. 
    (One might have thought that, since the horizon is at $l=0$, the second term
    on the left hand side of \eq{eqnxxx} vanishes identically on the horizon and 
    we can only determine $c_2$. The above analysis shows that, for $l(x) = x^n$
    with $n>0$, the horizon is still at $l=x=0$ but $ll''/l'^2 = (n-1)/n$ does not
    vanish on the horizon. It is this fact which allows us to determine $c_1$ and 
    $c_2$. In fact we only need to use $n=1$ and $n=2$ corresponding to the two standard forms of the Rindler metric to fix the two constants.)
    This completely determines $c_1$ and $c_2$ and hence the surface term. Continuing
    back to Lorentzian sector, we get
    \begin{eqnarray}
        \label{pcfix}
    P^c &=&  - \frac{1}{16\pi \mathcal{A}_P} \left( 2g^{cb} \partial_b \sqrt{-g} + \sqrt{-g} \partial_b g^{bc}\right) 
     \nonumber\\
    &=& \frac{1}{16\pi \mathcal{A}_P}\frac{1}{\sqrt{-g}} \partial_b(gg^{bc})
    \end{eqnarray}    
    This is precisely the surface term in Einstein-Hilbert action as can be seen by comparing with \eq{defpcone} and recalling $P^c = \sqrt{-g}\, V^c$.
        Given the form of $P^c$ we need to solve the 
   equation   
    \begin{equation}
 \left(\frac{\partial \sqrt{-g}L_{\rm bulk}}{\partial g_{ab,c}}g_{ab}\right)=
 P^c=  \frac{1}{ 16\pi \mathcal{A}_P}\frac{1}{\sqrt{-g}} \partial_b(gg^{bc})
 \label{dseq}
\end{equation}
to obtain 
the first order Lagrangian density.
    It is straightforward to show that this equation is satisfied by the Lagrangian
\begin{equation}
\sqrt{-g}L_{\rm bulk}  = 
 \frac{1}{ 16\pi \mathcal{A}_P} \left(\sqrt{-g} \, g^{ik} \left(\Gamma^m_{i\ell}\Gamma^\ell_{km} -
\Gamma^\ell_{ik} \Gamma^m_{\ell m}\right)\right).
\label{ds}
\end{equation}  
which we already know from \eq{pecurrel}. 
(The solution to Eq.~(\ref{dseq}) obtained in Eq.~(\ref{ds}) is not unique. However, self consistency requires that the final equations of motion for gravity must admit the line element in  Eq.~(\ref{rindmetric}) as a solution.  It can be shown, by fairly detailed algebra, that this condition makes the Lagrangian in Eq.~(\ref{ds}) to be the only solution.) 
    Given the two pieces, the final second order Lagrangian follows from our Eq.~(\ref{aeh})
    and is, of course, the standard Einstein-Hilbert Lagrangian.
    In this approach, 
     our full second order Lagrangian  {\it turns out} to be the standard 
Einstein-Hilbert Lagrangian.
Our result has been obtained,
by relating the surface term in the action  to the entropy per unit area; i.e., the surface terms dictates the form of the  Lagrangian in the bulk through the holographic relation.

The same procedure works for \LL\ models though the mathematics is much more tedious.
(This approach has also uncovered several other issues related to entropy, quasi-normal modes etc. and even a possibility of entropy being quantized \cite{entropyideas} but we will not discuss these aspects.)
Also, in the case of \LL\ gravity, the expression for entropy has no simple physical motivation unlike in Einstein's theory. Hence, while this does illustrate the power of the holographic action principle it does not allow one to make significant further progress. We shall see later (see Sec \ref{sec:originofll}) that the thermodynamic interpretation of the field equations actually arises from different approach in which we do \textit{not} consider the metric as a fundamental variable.

\subsection{Summary}

The various results obtained in the previous sections can be briefly summarized as follows. 

(a) It is possible to associate the notion of temperature with any bifurcation horizon in a fairly general manner by, for example, using the periodicity in the imaginary time.
Association of entropy with  a generic horizon is conceptually more involved but it seems quite unnatural to associate temperature with \textit{all} horizons and entropy with \textit{only a subset} of them. 
 The association of thermodynamic variables to  a horizon is not a special feature
of Einstein's theory and extends to a much wider class of theories of gravity. 

(b) There are strong hints which suggests that horizon thermodynamics has a deep connection with gravitational dynamics, which is not apparent in the conventional approach to gravity. 
In particular, we have seen that:

\begin{itemize}

\item The field equations of gravity in a very wide class of theories reduce to the thermodynamic identity $TdS = dE + P dV$ on the horizon. This result has no explanation
in the conventional approach. 

\item The action functional in a wide class of theories of gravity contains both a bulk term and a surface term. There is a specific relationship between these two terms which allows these action functionals to be interpreted as a momentum space action functionals.
It is not clear why this peculiar relation exists between the bulk and surface terms.

\item The surface term of the action functional, when evaluated on the horizon in a solution, gives the entropy. This is quite mysterious since the field equations are  obtained by varying the bulk term after the surface term has been ignored (or cancelled out by a counter-term). Therefore, the field equations and their solutions are completely independent of the surface term. We do not expect a specific property of the solution (for example, the horizon entropy) to be obtainable from the surface term which played no role in the field equations.

\item The Euclidean action in any static spacetime can be interpreted as the free energy
in a wide class of theories of gravity showing that the minimization of the action  can be related to the minimization of free energy.

\end{itemize}

In the next two sections of this review, we shall provide an alternate perspective  
on gravity which will help us to understand these features better and in a fairly unified manner.

\section{The emergent spacetime}\label{sec:emergentspacetime}

In this section and the next, we shall present an alternative perspective on the nature of gravity motivated by the results described in the previous section. This approach, as we shall see, provides a natural setting for many of the results obtained in the earlier sections which are somewhat mysterious in the context of the conventional approach. The key feature of the new perspective is that it treats gravity as an emergent phenomena and describes its dynamics in the thermodynamic limit. We will, therefore, begin by making clear what is meant by emergent phenomena in this context and establishing its connection with thermodynamics.

As we have already argued in Sec. \ref{sec:intro}, the fact that spacetimes can be hot strongly indicates the existence of internal degrees of freedom for the spacetime.
This fact probably would be accepted by most people working in quantum gravity since almost all these models introduces extra structures at microscopic (Planck) scales in the spacetime. The natural picture which then emerges is that there are some ``atoms of spacetime'' at the microscopic level and the description of spacetime in terms of variables like metric, curvature etc. is a continuum, long wavelength, approximation. This is analogous to description of a gas or a fluid in terms of dynamical variables like density $\rho$, velocity $\mathbf{v}$ etc. in the continuum limit, none of which have any relevance in the microscopic description. 

The new ingredient we will introduce is based on the fact that, while we do not have definite knowledge about the \textit{statistical mechanics} of atoms of spacetime, we should be able to develop the \textit{thermodynamic limit} of the theory taking clues from horizon thermodynamics.

As emphasized in Sec.~\ref{sec:intro}, such an approach has two major advantages.
First, the thermodynamic description has a universal validity which is fairly independent of the actual nature of the microscopic degrees of freedom. 
Second, the entropy of the system arises due to our ignoring the microscopic degrees of freedom. Turning this around, one can expect the form of entropy functional to encode the essential aspects of microscopic degrees of freedom, even if we do not know what they are. If we can arrive at the appropriate form of entropy functional, in terms of some effective degrees of freedom using our knowledge of horizon thermodynamics, then we can expect it to provide the correct description.\footnote{Incidentally, this is why thermodynamics needed no modification due to either relativity or quantum theory. An equation like $TdS=dE+PdV$ will have universal applicability as long as effects of relativity or quantum theory are incorporated in the definition of $S(E,V)$ appropriately.}
As we know, thermodynamics was developed
and used effectively decades before we  understood  the molecular structure of matter or its statistical mechanics. 

Similarly, even without knowing the microstructure of spacetime or the full quantum theory of gravity, we should be able to make significant progress with the thermodynamic description of spacetime. 
The horizon thermodynamics
provides \cite{tworeviews} valuable insights about the nature of gravity totally independent of what ``the atoms of spacetime'' may be. 
This is what we will attempt to do. (There has been several other attempts in literature to implement the idea that gravity is an emergent phenomenon, which we shall not discuss. They do not: (a) address issues we have raised regarding  the action functionals and (b) cannot handle \LL\ models effectively; for a small sample of papers on other approaches, which contain additional references, see Ref.\cite{others}.)

\subsection{Thermodynamic interpretation of field equations of gravity}\label{sec:interpretthermo}

Consider the action functional in \eq{genAct} which, on variation, leads to the field equations 
\begin{equation}
2\mathcal{G}_{ab} -   T_{ab}=0
\label{eabminustab}
\end{equation} 
where the explicit form of $\mathcal{G}_{ab}$ is given by \eq{genEab}. 
As mentioned earlier,  this 
equation --- which equates   a geometrical quantity to matter variables ---
does not have any simple physical interpretation. 
The lack of an elegant principle to determine the \textit{dynamics} of gravity
is in sharp contrast with the issue of determining the \textit{kinematics}
of gravity which  can be tackled through the principle of equivalence by demanding that all freely falling observers  must find that the equations of motion for matter degrees of matter must reduce to their special relativistic form.

Our first aim will be to remedy this situation and provide a physical interpretation to 
\eq{eabminustab}. We will do this in a manner very similar in spirit to using freely falling observers to determine the kinematics of gravity. At every event in spacetime, we will introduce uniformly accelerating local Rindler observers and use the horizon thermodynamics
perceived by these Rindler observers to constrain the background geometry.
 We shall begin by making the notion of local Rindler observers and their coordinate system well defined.

 Let us choose any event $\mathcal{P}$ and introduce a local inertial frame (LIF) around it with Riemann normal coordinates $X^a=(T,\mathbf {X})$ such that $\mathcal{P}$ has the coordinates $X^a=0$ in the LIF.  
Let $k^a$  be
 a future directed null vector at $\mathcal{P}$ and we align the coordinates of LIF
 such that it lies in the $X-T$ plane at $\mathcal{P}$. We next  transform from the LIF to a local Rindler frame (LRF) coordinates $x^a$ by accelerating along the X-axis with an acceleration $\kappa$ by the usual transformation. The metric near the origin now reduces to the form  
 \begin{eqnarray}
 ds^2 &=& -dT^2 + dX^2 + d\mathbf{x}_\perp^2=-\kappa^2 x^2 dt^2 + dx^2 + dL\mathbf{x}_\perp^2\nonumber\\
&=& - 2 \kappa l \ dt^2 + \frac{dl^2}{2\kappa l}  + d\mathbf{x}_\perp^2
\label{surfrind}
\end{eqnarray} 
where  ($t,l, \mathbf {x}_\perp$) and ($t,x, \mathbf {x}_\perp$) 
are the coordinates of LRF.  
 Let $\xi^a$ be the approximate Killing vector corresponding to translation in the Rindler time such
that the vanishing of $\xi^a\xi_a \equiv -N^2$ characterizes the location of the 
local horizon $\mathcal{H}$ in LRF. Usually, we shall do all the computation 
on a timelike surface infinitesimally away from $\mathcal{H}$
with $N=$ constant,  called a  ``stretched horizon''. 
 Let the timelike unit normal to the stretched horizon
be  $r_a$. 

 This LRF (with  metric in \eq{surfrind}) and its local horizon $\mathcal{H}$ will exist within a region of size $L\ll\mathcal{R}^{-1/2}$ 
 (where $\mathcal{R}$ is a typical component of curvature tensor of the background spacetime) as long as $\kappa^{-1}\ll\mathcal{R}^{-1/2}$. This condition can always be satisfied by taking a sufficiently large $\kappa$ (see \fig{fig:ray1}). This procedure introduces a class of uniformly accelerated observers
 who will perceive the null surface $T=\pm X$ as the local Rindler horizon $\mathcal{H}$.

 \begin{figure}
\begin{center}
\scalebox{0.25}{\input{ray1.pstex_t}}\qquad\scalebox{0.25}{\input{ray2.pstex_t}}
\end{center}
\caption{The left frame illustrates schematically the light rays near an event
$\mathcal{P}$ in the $\bar t - \bar x$ plane of an arbitrary spacetime. The right frame
shows the same neighbourhood of $\mathcal{P}$ in the locally inertial frame at $\mathcal{P}$
in Riemann normal coordinates $(T,X)$. The light rays now become 45 degree lines and the trajectory of the local Rindler observer becomes a hyperbola very close to $T=\pm X$ lines
which act as a local horizon to the Rindler observer.}
\label{fig:ray1}
\begin{center}
\scalebox{0.25}{\input{ray3.pstex_t}}
\end{center}
\caption{The region around $\mathcal{P} $ shown in \fig{fig:ray1} is represented in the 
Euclidean sector obtained by analytically continuing to imaginary values of $T$ by $T_E = iT$. The horizons $T=\pm X$ collapse to the origin and the hyperbolic trajectory of the Rindler 
observer becomes a circle of radius $\kappa^{-1}$ around the origin. The Rindler coordinates
$(t,x)$ become --- on analytic continuation to $t_E = it$ --- the polar coordinates
$(r=x, \theta = \kappa t_E$) near the origin.}
\label{fig:ray3}
\end{figure}

  Essentially,  the introduction of the LRF uses  the fact that we have two length scales in the problem at any event. First is the length scale $\mathcal{R}^{-1/2}$ associated with the curvature components of the background metric over which we have no control while the second is the length scale $\kappa^{-1}$ associated with the accelerated trajectory which we can choose. Hence  we can always ensure that  $\kappa^{-1}\ll\mathcal{R}^{-1/2}$. In fact,  this
  is clearly seen in the  Euclidean sector in which the horizon maps to the origin (see \fig{fig:ray3}). The locally flat frame in the Euclidean sector will exist in a region of radius $\mathcal{R}^{-1/2}$ while the 
 trajectory of a uniformly accelerated observer will be a circle of radius $\kappa^{-1}$ and hence one  can always keep the latter inside the former. The metric in \eq{surfrind} is just the metric of the locally flat region in polar coordinates.

 More generally, one  can choose a trajectory $x^i(\tau)$ such that its acceleration $a^j=u^i\nabla_i u^j$ (where $u^i$ is the time-like four velocity) satisfies the condition
 $a^ja_j =\kappa^2$. In a suitably chosen LIF this trajectory will reduce to the standard hyperbola of a uniformly accelerated observer.

 Our construction also defines local Rindler horizons around any event.
 Further, the local temperature   on the stretched horizon will be $\kappa/2\pi N$ so that 
$\beta_{\rm loc} = \beta N$ with $\beta \equiv \kappa/2\pi$.
Note that in the Euclidean sector the Rindler observer's trajectory is a circle of radius $\kappa^{-1}$ which can be made arbitrarily close to the origin. Suppose the observer's trajectory has the usual form $X=\kappa^{-1}\cosh \kappa t; T=\kappa^{-1}\sinh \kappa t$ which is maintained for a time interval  of the order of $ 2\pi/\kappa$. Then, the trajectory will complete a full circle \textit{in the Euclidean sector} irrespective of what happens later. When we work in the limit of $\kappa\to\infty$, our construction becomes arbitrarily local in both space \textit{and} time \cite{loctemp,dawood-paddy-varg}.

 As stressed earlier in Sec.~\ref{sec:obsdependence}, the local Rindler observers will perceive the thermodynamics of matter around them very differently from the freely falling observers. In particular, they will attribute a loss of entropy $\delta S = (2\pi/\kappa) \delta E$ (see \eq{delS}) when an amount of energy $\delta E$ gets close to the horizon
 (within a few Planck lengths, say). In the Rindler frame the appropriate energy-momentum  density is $T^a_b\xi^b$. (It is the integral of $T^a_b\xi^b d\Sigma_a$ that gives the Rindler Hamiltonian $H_R$, which leads to evolution in Rindler time $t$ and appears in the thermal density matrix $\rho=\exp-\beta H_R$.) 
 A local Rindler observer, moving along the orbits of the Killing vector field $\xi^a$ with four velocity $u^a = \xi^a/N$, will associate an energy density $u^a(T_{ab}\xi^b)$ and an energy
  $\delta E =u^a(T_{ab}\xi^b) dV_{\rm prop}$ with a proper volume $dV_{\rm prop}$.
  If this energy gets transfered across the horizon, the corresponding entropy transfer will be $\delta S_{\rm matter} = \beta_{\rm loc}\delta E$ where $\beta_{\rm loc} = \beta N = 
(2\pi/\kappa)N$ is the local  (redshifted) temperature of the horizon
 and $N$ is the lapse function. 
 Since $\beta_{\rm loc} u^a = (\beta N)(\xi^a/N) = \beta \xi^a$, we find that 
\begin{equation}
 \delta S_{\rm matter} = \beta \xi^a \xi^b T_{ab}\ dV_{\rm prop}
 \label{defsmatter}
\end{equation} 
Consider now the gravitational entropy associated with the local horizon. From the 
discussion of Noether charge as horizon entropy in Section \ref{sec:noetherent} [see \eq{noetherint}], we know that $\beta_{\rm loc} J^a$, 
associated with the Killing vector $\xi^a$,
can be thought of 
as local entropy current. Therefore,  $\delta S = \beta_{\rm loc} u_a J^a dV_{\rm prop}$ can be interpreted
as the gravitational entropy associated with a volume $dV_{\rm prop}$ as measured
by an observer with four-velocity $u^a$. (The conservation of  $J^a$ ensures that  there is no irreversible entropy production in the spacetime.)
   Since $\xi^a$ is a Killing vector locally, satisfying \eq{cond1} it follows that $\delta_\xi v=0$ giving the current to be
$
J^a = \left( L \xi^a + 2 \mathcal{G}^{a b} \xi_b \right) $.
 For observers moving along the orbits of the Killing vector $\xi^a$
with $u^a = \xi^a/N$  we get
\begin{equation}
\delta S_{\rm grav} = \beta_{\rm loc} N u_a J^a dV_{\rm prop} =\beta \xi_a J^a dV_{\rm prop} =\beta  [\xi_j \xi_a (2\mathcal{G}^{aj}) +  L  (\xi_j \xi^j)]\, dV_{\rm prop}
\end{equation} 
As one approaches the horizon, $\xi^a\xi_a\to 0$ making the second term vanish and we find that
\begin{equation}
\delta S_{\rm grav} = \beta  [\xi^j \xi^a (2\mathcal{G}_{aj})] \, dV_{\rm prop}
\end{equation}
In the same limit $\xi^j$ will become proportional to the original null vector $k^j$ we started with keeping everything finite.   
We now see that the condition $\delta S_{\rm grav} = \delta S_{\rm matter}$ leads
to the result
\begin{equation}
 [2\mathcal{G}_{ab} - T_{ab}]k^ak^b =0
 \label{myeqn}
\end{equation} 
Since the original null vector $k_a$ was arbitrary, this equation should hold for all
null vectors for all events in the spacetime.
This is equivalent to $2\mathcal{G}^{ab} - T^{ab}=\lambda g^{ab}$ with some constant $\lambda$.
(Because of the conditions $\nabla_a\mathcal{G}^{ab}=0,\ \nabla_a T^{ab}=0$,
it follows that 
$\lambda$ must be a constant.)

 We have thus succeeded in providing a purely thermodynamical interpretation of the 
 field equations of any diffeomorphism invariant theory of gravity. Note that the equations 
 \eq{myeqn} has an extra symmetry
 which standard gravitational field equations do not have: This equation is  invariant
 under the shift $T^{ab}\to T^{ab}+\mu g^{ab}$ with some constant $\mu$. This symmetry has important implications for cosmological constant problem which we will discuss in Section \ref{sec:cc}. 
 
 It is clear that the properties of LRF are relevant conceptually to define the intermediate notions (local Killing vector, horizon temperature ....) but the essential result is independent of these notions. Just as we introduce local inertial frame to decide how gravity couples to matter, we use local Rindler frames to interpret the physical content of the field equations.

 In Section \ref{sec:noetherent} we mentioned that $J^a$ is not unique and one can
 add to it the divergence of any anti-symmetric tensor (see the discussion after
 \eq{variations}). In providing the thermodynamic interpretation to the field equations,
 we have ignored this ambiguity and used the expression in \eq{current}. 
 There are several reasons why the ambiguity is irrelevant for our purpose. First,
 in a truly thermodynamic approach, one specifies the system by specifying a thermodynamic potential, say, the entropy functional. In a local description, this translates into 
 specifying the entropy current which determines the theory. So it is perfectly acceptable to make a specific choice for $J^a$ consistent with the symmetries of the problem. Second, we shall often be interested in theories in which the equations of motion are no higher
 than second order and has the form in \eq{genEab1}. In these 
 \LL\ models it is not natural to add any extra term to the Noether current such that
 it is linear in $\xi^a$ as we approach the horizon with a coefficient determined entirely from metric and curvature. Finally, we shall obtain  in Section \ref{sec:gravitystory}
 the field equation from maximizing an entropy functional when this ambiguity will not 
 arise. 
 
 It may be noted that our result only required the part of $J^a$ given by $2\mathcal{G}^a_b\xi^b$.
 Hence, one can obtain the same results by postulating the entropy current to be $J^a \equiv 2\mathcal{G}^a_b\xi^b$ which is also conserved off-shell when $\xi^a$ is a Killing vector.
  In fact one can give $2\mathcal{G}^a_b\xi^b$ an interesting interpretation. Suppose there are some microscopic degrees of freedom in spacetime, just as there are atoms in a solid. If one considers   an elastic deformation $x^\alpha\to x^\alpha+\xi^\alpha(x)$ of the solid, the physics can be formulated in terms of the displacement field $ \xi^\alpha(x)$ and one can ask how thermodynamic potentials like entropy change under such displacement. Similarly, in the case of spacetime, one could think of
\begin{equation}
\delta S_{grav}=\beta_{loc}(2\mathcal{G}^a_b)u_a\delta x^b
\label{gradentropy}
\end{equation} 
 as the change in the gravitational entropy density under the `deformation' of the spacetime $x^a\to x^a+\delta x^a$ as measured by the Rindler observer with velocity $u^a$. (One can show that this interpretation is consistent with all that we know about horizon thermodynamics.) So, one can interpret the left hand side of gravitational field equation $(2\mathcal{G}^a_b)$ as giving the response of the spacetime entropy to the deformations. This
 matches with the previous results  because $ \beta_{loc} u_a=\beta\xi_a$ implies that  the entropy density will be proportional to $2\mathcal{G}_{ab}k^ak^b$ on the horizon. We will see later that this interpretation remains valid in a very general context.
 
 Once it is understood that the real physical meaning of the field equations
 lies in writing them in the form of \eq{myeqn}, it is possible to re-interpret
 these equations in several alternative ways all of which have the same physical content.
 We shall mention two of them. 
 
 Consider an observer who sees 
 some matter energy flux crossing the  horizon.   Let $r_a$ be the spacelike unit normal to the stretched horizon $\Sigma$, pointing in the direction of increasing $ N$.   The energy flux through a
 patch of stretched horizon  will be $T_{ab}\xi^ar^b$ and the associated entropy flux will be $\beta_{loc} T_{ab}\xi^ar^b$.
 To maintain the second law of thermodynamics, this entropy flux must match  the entropy change of the locally perceived  horizon.
 The gravitational entropy current is given by  $\beta_{loc}J^a$, such that
 $\beta_{loc}(r_aJ^a)$  gives the corresponding  gravitational entropy flux.  So we require
  \begin{equation}
\beta_{loc} r_a J^a =\beta_{loc} T^{a b} r_a \xi_b
\label{crucial1}
\end{equation}  
 to hold at 
 all events where  $J^a$ is the conserved Noether current corresponding to $\xi^a$.
 The product $r_a J^a$ for the vector $r^a$, which satisfies $\xi^ar_a=0$ on the stretched horizon is
$
r_a J^a = 2 \mathcal{G}^{a b} r_a \xi_b $.
 Hence we get
\begin{equation}
\beta_{loc} r_a J^a =2 \mathcal{G}^{a b} r_a \xi_b=\beta_{loc} T^{a b} r_a \xi_b
\label{final}
\end{equation}
As $ N\to0$ and the stretched horizon approaches the local horizon and $ N r^i$ approaches $\xi^i$ (which in turn is proportional to $k^i$) so that $\beta_{loc}r_a=\beta N r_a\to\beta\xi_a$. So, 
 as we approach the horizon \eq{final} reduces to \eq{myeqn}. 
 
There is another way of interpreting this result which will be useful for further generalizations. Instead of allowing  matter to flow  across the horizon, one can equally well
  consider a  virtual, infinitesimal (Planck scale), displacement of the $\mathcal{H}$ normal to itself
engulfing some matter. We only need to consider infinitesimal displacements because the entropy of the matter is not `lost' until it crosses the horizon; that is, until when the matter is at an infinitesimal distance (a few Planck lengths) from the horizon. 
 Some entropy will be again lost to the  outside observers unless  displacing a piece of local Rindler horizon  costs some entropy. 
 
 We can verify this as follows:
An infinitesimal displacement of a local patch of the stretched horizon in the direction of $r_a$, by an infinitesimal proper distance $\epsilon$, will change the proper volume by $dV_{prop}=\epsilon\sqrt{\sigma}d^{D-2}x$ where $\sigma_{ab}$ is the metric in the transverse space.
 The flux of energy through the surface will be  $T^a_b \xi^b r_a$ and the corresponding  entropy flux
 can be obtained by multiplying the energy flux by $\beta_{\rm loc}$.  Hence
 the `loss' of matter entropy to the outside observer because the virtual displacement of the horizon has engulfed some matter is 
$\delta S_m=\beta_{\rm loc}\delta E=\beta_{\rm loc} T^{aj}\xi_a r_j dV_{prop}$. 
To find the change in the gravitational entropy, we again use the Noether current $J^a$ corresponding
to the local Killing vector $\xi^a$. 
Multiplying by $r^a$ and 
 $\beta_{\rm loc} = \beta N$, we get
\begin{equation}
\beta_{\rm loc} r_a J^a  = \beta_{\rm loc}\xi_a r_a T^{ab}  + \beta N ( r_a \xi^a) L
\end{equation}  
As the stretched horizon approaches the true horizon, we know that  $N r^a \to \xi^a$
 and $\beta \xi^a \xi_a L \to 0$ making the last term vanish. So
\begin{equation}
\delta S_{\rm grav} \equiv  \beta \xi_a J^a dV_{prop} = \beta T^{aj}\xi_a \xi_j dV_{prop}
=\delta S_m
\end{equation} 
showing the validity of local entropy balance for any $\beta$. 
In this limit, $\xi^i$ also goes to $\kappa \lambda k^i$ where $\lambda $ is the affine parameter associated with the null vector $k^a$ we started with  and all the reference to LRF goes away.

 \section{Gravity: The inside story}\label{sec:gravitystory}
 
 \subsection{An Entropy maximization principle for gravitational field equations}
 
 The last interpretation given above is similar to switching from a passive point of view to an active point of view. 
 Instead of allowing matter to fall into the horizon, we are making a  displacement of the horizon surface to engulf the matter when it is infinitesimally close to the horizon. But in the process, we have introduced the notion of virtual displacement of horizons and, for the theory to be consistent, this displacement
of these surface degrees of freedom should cost  some entropy. 
This allows one to associate an entropy functional with the normal displacement of any horizon. 

An analogy may be helpful in this context.
If  gravity is an emergent, long wavelength, phenomenon like elasticity  then the diffeomorphism $x^a\to x^a+\xi^a$ is analogous to  the elastic deformations of the ``spacetime solid"
\cite{elasticgravity}. It then makes sense to demand that the entropy density should be a functional of $\xi^a$ and its derivatives $\nabla_b \xi^a$. By constraining 
the functional form of this entropy density, we should be able to obtain the field equations of gravity by a maximization principle.
Recall that thermodynamics relies entirely on the form of the entropy functional to make predictions.
Hence, if we can determine  the  form of entropy functional for gravity  ($S_{grav}$) in terms of the normal to the null surface, then it seems
 natural to demand that  the dynamics should follow from the extremum prescription
$\delta[S_{grav}+S_{matter}]=0$ for \textit{all null surfaces in the spacetime} where $S_{matter}$ is the relevant  matter entropy.

The form of $S_{matter}$ and $S_{\rm grav}$ can be determined as follows.
Let us begin with $S_{matter}$ which  is easy to ascertain from the previous discussion.
If $T_{ab}$ is the matter energy-momentum tensor in a general $D(\ge 4)$ dimensional spacetime then an expression for matter entropy \textit{relevant for our purpose} can be taken to be 
\begin{equation}
S_{\rm matt}=\int_\Cal{V}{d^Dx\sqrt{-g}}
      T_{ab}n^an^b
      \label{Smatt}
\end{equation} 
where $n^a$ is a null vector field.  From our \eq{defsmatter} we see that the entropy density associated with proper 3-volume is $\beta(T_{ab}\xi^a\xi^b)dV_{prop}$ where --- on the horizon --- the vector $\xi^a$ becomes proportional to a null vector $n^a$. 
If we now use the Rindler coordinates in \eq{standardhorizon} in which $\sqrt{-g}=1$ and 
interpret the factor $\beta$ as arising from an integration of $dt$ in the range $(0,\beta)$ we find that the entropy density associated with a proper four volume is $(T_{ab}n^an^b)$. This suggests treating \eq{Smatt} as the matter entropy. 
 For example, if $T_{ab}$ is due to an ideal fluid at rest in  the LIF then $T_{ab}n^an^b$ will contribute $(\rho+P)$, which --- by Gibbs-Duhem relation --- is just $T_{local}s$ where $s$ is the entropy density and $T_{local}^{-1}=\beta N$ is the properly redshifted temperature. Then
 \begin{eqnarray}
 \label{intlim}
\int dS&=&\int\sqrt{h}d^{D-1}x s=\int\sqrt{h}d^{D-1}x\beta_{\rm loc}(\rho+P)=\int\sqrt{h}Nd^{D-1}x \beta (\rho+P)\nonumber\\
&=&\int_0^\beta dt \int d^{D-1}x\g T^{ab}n_an_b
\end{eqnarray} 
which matches with \eq{Smatt} in the appropriate limit.

It should be stressed that this argument works for any matter source, not necessarily the ones with which we conventionally associate an entropy. What is really relevant is only the \textit{energy} flux close to the horizon from which one can obtain an entropy flux.
We \textit{do} have the notion of \textit{energy} flux across a surface with normal $r^a$ being $T_{ab}\xi^br^a$ which holds for \textit{any} source $T^{ab}$. Given some energy flux $\delta E$ in the Rindler frame, there is an associated entropy flux loss $\delta S=\beta\delta E$ as given by \eq{delS}. (One might think, at first sight, that an ordered field, say, a scalar field, has no temperature or entropy but a Rindler observer will say something different. For any state, she will have a corresponding density matrix $\rho$ and an entropy $-Tr(\rho\ln\rho)$; after all, she will attribute entropy even to vacuum state.) It is \textit{this} entropy which is given by \eq{delS} and \eq{Smatt}. 
The only non-trivial feature in \eq{Smatt} is the integration range for time which is limited to $(0,\beta)$. This is done by considering the integrals in the Euclidean sector and rotating back to the Lorentzian sector but the same result can be obtained working entirely in the Euclidean sector. (There is an ambiguity in the overall scaling of $n^a$ since if $n^a$ is null so is $f(x)n^a$ for all $f(x)$; we will comment on this ambiguity, which anyway turns out to be irrelevant, later on.)

Next, let us consider the expression for $S_{\rm grav}$. We will first describe the simplest possible choice and  then consider a more general expression. The simplest choice is to  postulate $S_{grav}$ to be a quadratic expression \cite{aseementropy} in the derivatives of the normal:  
 \begin{equation}
S_{grav}= - 4\int_\Cal{V}{d^Dx\sqrt{-g}}
    P_{ab}^{\ph{a}\ph{b}cd} \D_cn^a\D_dn^b 
    \label{Sgrav}
\end{equation}  
where the explicit form of $P_{ab}^{\ph{a}\ph{b}cd}$ is ascertained below. The expression for the total entropy  now becomes:
\begin{equation}
S[n^a]=-\int_\Cal{V}{d^Dx\sqrt{-g}}
    \left(4P_{ab}^{\ph{a}\ph{b}cd} \D_cn^a\D_dn^b - 
    T_{ab}n^an^b\right) \,,
\label{ent-func-2}
\end{equation}

We should be able to determine the field equations of gravity by extremizing this entropy functional. However, there is one 
 crucial conceptual difference 
between the extremum principle introduced here and the conventional one. Usually, given a set of dynamical variables $n_a$ and a functional $S[n_a]$, the extremum principle will give a set of equations for the dynamical variable $n_a$. Here the situation is completely different. We expect the variational principle to hold for   \textit{all} null vectors $n^a$ thereby  leading  to a condition on  the \textit{background
metric.} Obviously, the functional in \eq{ent-func-2} must be rather special to accomplish this and one \textit{needs} to impose  restrictions on  $P_{ab}^{\ph{a}\ph{b}cd}$ --- and $T_{ab}$ though that condition turns out to be trivial --- to achieve this.
(Of course, one can specify any null vector $n^a(x)$ by giving its components $f^A(x)\equiv n^ae_a^A$ with respect to fixed set of basis vectors $e_a^A$ with $e_A^be_b^B=\delta^B_A$ etc so that $n^a=f^Ae_A^a$. So the class of all null vectors can be mapped to the scalar functions $f^A$ with the condition $f_Af^A=0$.)

It turns out --- as we shall see below --- that two conditions are sufficient
to ensure this.
First, the tensor $P_{abcd}$ should
have the same algebraic symmetries as the Riemann tensor $R_{abcd}$
of the $D$-dimensional spacetime. 
This condition can be ensured if we define $P_a^{\phantom{a}bcd}$ as
\begin{equation}
P_a^{\phantom{a}bcd} = \frac{\partial L}{\partial R^a_{\phantom{a}bcd}}
\label{condd1}
\end{equation} 
where $L = L(R^a_{\phantom{a}bcd}, g^{ik})$ is some scalar. 
The motivation for this choice
arises from the fact that this approach leads to the same field equations as the one with $L$ as gravitational Lagrangian in the conventional approach (which explains the choice of the symbol $L$).
Second, we will postulate the condition:
\begin{equation}
\D_{a}P^{abcd}=0.
\label{ent-func-1}
\end{equation}
as well as $\D_{a}T^{ab}=0$ which is anyway satisfied by any matter energy-momentum tensor.

One possible motivation for this condition in \eq{ent-func-1} arises from the
 following fact: It will ensure that the field equations do not contain any derivative of the metric which is of higher order than second. Another possible interpretation arises from the 
 analogy introduced earlier.
If we think of $n^a$ as analogous to deformation field in elasticity, then, in theory of elasticity \cite{landau7} one usually postulates the form of the thermodynamic potentials which are quadratic in first derivatives of $n_a$. The coefficients of this term will be the elastic constants. Here the coefficients are $P^{abcd}$
and the condition in  \eq{ent-func-1}
may be interpreted 
as saying the `elastic constants of spacetime solid' are actually `constants'. 
This is, however, not a crucial condition and in fact we will see below how this condition in \eq{ent-func-1} can be relaxed. 

\subsection{The field equations}
 
Varying the normal vector field $ n^a$  in \eq{ent-func-2} after adding a
Lagrange multiplier function $\lambda(x)$ for imposing the   condition
$ n_a\delta  n^a=0$, we get 
\begin{eqnarray}
-\delta S &=& 2\int_\Cal{V} d^Dx\sqrt{-g}
  \left[4P_{ab}^{\ph{a}\ph{b}cd}\D_c n^a\left(\D_d\delta n^b\right)
  - T_{ab} n^a\delta n^b\right. \nonumber\\
   &&\left. \qquad  - \lambda(x) g_{ab} n^a\delta n^b\right]
 \label{ent-func-3}
\end{eqnarray}
where we have used the symmetries of $P_{ab}^{\ph{a}\ph{b}cd}$ and
$T_{ab}$.  (We note, for future reference, that the Lagrange multiplier in the calculation only imposes the constancy of $n_in^i$ under variation and  does not require  $n_i$ to be null vector.) An integration by parts and the
condition $\D_dP_{ab}^{\ph{a}\ph{b}cd}=0$, leads to 
\begin{eqnarray}
-\delta
S&=& 2\int_\Cal{V}{d^Dx\sqrt{-g}\left[-4P_{ab}^{\ph{a}\ph{b}cd}
  \left(\D_d\D_c n^a\right) - ( T_{ab}+ \lambda g_{ab}) n^a\right]\delta n^b}\nonumber\\
  &&+8\int_{\dV}{d^{D-1}x\sqrt{h}\left[k_d
  P_{ab}^{\ph{a}\ph{b}cd}\left(\D_c n^a\right)\right]\delta n^b}
\,,
\label{ent-func-4}
\end{eqnarray}
where $k^a$ is the $D$-vector field normal to the boundary \dV\ and
$h$ is the determinant of the induced metric on \dV.  As usual, in order for
the variational principle to be well defined, we require that the
variation $\delta n^a$ of the  vector field should vanish on the
boundary. The second term in \eq{ent-func-4} therefore vanishes, and
the condition that $S[ n^a]$ be an extremum for arbitrary variations of
$ n^a$ then becomes  
\begin{equation}
2P_{ab}^{\ph{a}\ph{b}cd}\left(\D_c\D_d-\D_d\D_c\right) n^a
-( T_{ab}+\lambda g_{ab}) n^a = 0\,,
\label{ent-func-5}
\end{equation}
where we used the antisymmetry of $P_{ab}^{\ph{a}\ph{b}cd}$ in its
upper two indices to write the first term. Using the definition of the
Riemann tensor in terms of the commutator of covariant derivatives and writing
$\mathcal{R}^a_b=P_b^{\ph{b}ijk}R^a_{\ph{a}ijk}$ 
the above expression reduces to
\begin{equation}
\left(2\mathcal{R}^a_b -  T{}^a_b+\lambda \delta^a_b\right) n_a
=\left( 2 \mathcal{G}^a_b - T^a_b + (L+\lambda) \delta^a_b\right) n_a
=0\,, 
\label{ent-func-6}
\end{equation}
where we have used the definition of $\mathcal{G}^a_b$ in \eq{genEab1}.
We see that the equations of motion \emph{do not contain}
derivatives with respect to $n^a$ which is, of course, the crucial point. This peculiar feature arose because
of the symmetry requirements we imposed on the tensor
$P_{ab}^{\ph{a}\ph{b}cd}$. 
(Multiplying by $n^a$ and noting $n^an_a=0$ we see that \eq{ent-func-6} and \eq{myeqn} are identical.)
We  need the condition in
\eq{ent-func-6} holds for \emph{arbitrary}  vector fields
$ n^a$.
One can easily show \cite{aseementropy} using $\nabla_a \mathcal{G}^a_b =0=\nabla_aT^a_b$ that this requires $\lambda +L = $ constant leading to the field equation
\begin{equation}
\mathcal{G}^a_b = \left[ \mathcal{R}^a_b-\frac{1}{2}\delta^a_b L \right]=
  \frac{1}{2}T{}_b^a +\Lambda\delta^a_b   
\label{ent-func-71}
\end{equation}
where $\Lambda$ is a constant.
Comparison of \eq{genEab1} (or \eq{genEab}) with \eq{ent-func-71} shows that these are  precisely the field equations for  gravity  in a theory with Lagrangian $L$ when \eq{ent-func-2} is satisfied.
One crucial difference between the two equations is the introduction of the cosmological constant $\Lambda$ as an integration constant in \eq{ent-func-71}; we will discuss this  later in Section~\ref{sec:cc}.

We mentioned earlier that the expression in \eq{ent-func-2} depends on the overall scaling of $n^a$ which is arbitrary, since $f(x)n^a$ is a null vector if $n^a$ is null. But since the arbitrary variation of $n^a$ with the constraint $n_an^a=0$ includes scaling variations of the type $\delta n^a=\epsilon(x) n^a$, it is clear that this ambiguity is irrelevant for determining the equations of motion.

To summarize, we have proved the following. 
 Suppose we start with a total Lagrangian $L(R_{abcd},g_{ab})+L_{matt}$, define a $P^{abcd}$ by \eq{condd1} ensuring it satisfies
\eq{ent-func-1}. 
Varying the metric with this action will lead to certain field equations. We have now shown that we will get the \textit{same} field equations (but with a cosmological constant) if we start with the expression in \eq{ent-func-2}, maximize it with respect to $n^a$ and demand that it holds for all $n^a$. 

This result might appear a little mysterious at first sight, but the following alternative description will make clear why this works.
Note that, using the constraints on $P^{abcd}$ we can prove the identity
\begin{eqnarray}
\label{details1}
4P_{ab}^{\ph{a}\ph{b}cd} \D_cn^a\D_dn^b&=&
4\D_c[P_{ab}^{\ph{a}\ph{b}cd} n^a\D_dn^b]-4n^aP_{ab}^{\ph{a}\ph{b}cd} \D_c\D_dn^b\nonumber\\
&=&4\D_c[P_{ab}^{\ph{a}\ph{b}cd} n^a\D_dn^b]-2n^aP_{ab}^{\ph{a}\ph{b}cd} \D_{[c}\D_{d]}n^b\nonumber\\
&=&4\D_c[P_{ab}^{\ph{a}\ph{b}cd} n^a\D_dn^b]-2n^aP_{ab}^{\ph{a}\ph{b}cd} R^b_{\phantom{b}icd}n^i\nonumber\\
&=&4\D_c[P_{ab}^{\ph{a}\ph{b}cd} n^a\D_dn^b]+2n^a\mathcal{G}_{ai}n^i
\end{eqnarray} 
where the first line uses \eq{ent-func-1}, the second line uses the antisymmetry of $P_{ab}^{\ph{a}\ph{b}cd}$ in c and d, the third line uses the standard identity for commutator of covariant derivatives and the last line is based on 
\eq{genEab} when $n_an^a=0$ and \eq{ent-func-1} hold. Using this in the expression for
 $S$ in \eq{ent-func-2} and integrating the four-divergence term, we can write
 the entropy functional as
\begin{eqnarray}
S[n^a]&=&-\int_{\partial\Cal{V}}{d^{D-1}x k_c\sqrt{h}}
\left(4P_{ab}^{\ph{a}\ph{b}cd} n^a\D_dn^b\right)\nonumber\\
&& \qquad \quad -\int_\Cal{V}{d^Dx\sqrt{-g}}\left[(2\mathcal{G}_{ab}-T_{ab})n^an^b\right]
\label{thetrick}
\end{eqnarray}
So, when we consider variations ignoring the surface term we are  effectively varying $(2\mathcal{G}_{ab}-T_{ab})n^an^b$ with respect to $n_a$ and demanding that it holds for all $n_a$. This is the reason why  we get  $(2\mathcal{G}_{ab}=T_{ab})$ except for a cosmological constant.
There is an ambiguity of adding a term of the form $\lambda(x)g_{ab}$ in the integrand of the second term in \eq{thetrick} leading to the final equation
$(2\mathcal{G}_{ab}=T_{ab}+\lambda(x)g_{ab})$ but the Bianchi identity $\nabla_a\mathcal{G}^{ab}=0$ along with
$\nabla_aT^{ab}=0$ will make $\lambda(x)$ actually a constant. 
(We see from \eq{details1}
that, in the case of Einstein's theory, we have a bulk Lagrangian $n^a(\nabla_{[a}\nabla_{b]})n^b$ for a vector field $n^a$
--- plus for  a surface term which does not contribute to variation. 
In flat spacetime, in which covariant derivatives become partial derivatives, the bulk lagrangian becomes vacuous; i.e., there is  no bulk dynamics in $n^a$, in the usual sense. Nevertheless, they do play a crucial role.) 

The expression in \eq{thetrick} also connects up with our previous use of $2\mathcal{G}_{ab}n^an^b$ as  gravitational entropy density. The gravitational part of the entropy in \eq{thetrick} can  be written as
\begin{eqnarray}
\label{thetrick1}
S_{grav}[n^a]&=&-\int_\Cal{V}{d^Dx\sqrt{-g}}4P_{ab}^{\ph{a}\ph{b}cd} \D_cn^a\D_dn^b\\
&=&-\int_{\partial\Cal{V}}{d^{D-1}x k_c\sqrt{h}}
(4P_{ab}^{\ph{a}\ph{b}cd} n^a\D_dn^b)
-\int_\Cal{V}{d^Dx\sqrt{-g}}(2\mathcal{G}_{ab}n^an^b)\nonumber
\end{eqnarray}
with one  bulk contribution (proportional to  $2\mathcal{G}_{ab}n^an^b$) and a surface contribution. When equations of motion hold, the bulk also get a contribution from matter which cancels it out leaving the entropy of a region $\Cal{V}$ to reside in its boundary  $\partial\Cal{V}$.

It is now clear how we can find  an $S$ for any theory, even if \eq{condd1} does not hold. 
This can be achieved by starting from  the expression  $(2\mathcal{G}_{ab}-T_{ab})n^an^b$ as the entropy density,  using \eq{genEab} for $\mathcal{G}_{ab}$ and integrating by parts. In this case, we get
for $S_{\rm grav}$ the expression:
\begin{eqnarray}
S_{\rm grav} &=& - 4 \int_V d^Dx\, \sqrt{-g}\, \left[ P^{abcd} \nabla_c n_a \, \nabla_d n_b + (\nabla_d P^{abcd})n_b \nabla_cn_a \right.\nonumber\\
&&\qquad \qquad \left. + (\nabla_c \nabla_dP^{abcd}) n_a n_b\right]
\label{generalS}
\end{eqnarray} 
Varying this with respect to $n^a$ will then lead to the correct equations of motion and --- incidentally --- the same surface term.

While one could indeed work with the more general expression in \eq{generalS}, there are four reasons to prefer the imposition of the condition in  \eq{condd1}.
First, as we shall see   below,  with that condition we can actually determine the form of $L$; it turns out that in D=4, it uniquely selects Einstein's theory, which  is probably a nice feature. In higher dimensions, it picks out a very geometrical extension of Einstein's theory in the form of \LL\ theories.
Second,
it is difficult to imagine why the terms in \eq{generalS} should occur with very specific coefficients. In fact, it is not clear  why we cannot have derivatives of $R_{abcd}$ in $L$, if the derivatives of $P_{abcd}$ can occur in the expression for entropy. 
Third, it is clear from \eq{genEab} that when $L$ depends on the curvature tensor and
the metric, $\mathcal{G}_{ab}$ can depend up to the fourth derivative of the metric if \eq{condd1}
is not satisfied. But when we impose \eq{condd1} then we are led to field equations
which have, at most, second derivatives of the metric tensor which is again a desirable feature.
Finally, if we take the idea of elastic constants being constants, then one is led to \eq{condd1}. 
None of these rigorously exclude the possibility in \eq{generalS} and in fact
this model has been
 explored  recently \cite{sfwu}.
 
Our variational principle  extremises the \textit{total} entropy of matter and gravity when $n_a$ is a null vector. It is, however, possible to provide an alternative interpretation of our variational principle (along the lines of ref.\cite{TPgravcqg04}), which is of interest  when we are dealing with static 
spacetimes with a horizon.  Such spacetimes are described by the line element
\begin{equation}
ds^2=-N^2(\mathbf{x})dt^2+\gamma_{\mu\nu}(\mathbf{x}) dx^\mu dx^\nu
\label{dssquare}
\end{equation} 
If  $n^i=\xi^i/N$ denotes the four velocity of static observers with $x^\alpha$= constant, where $\xi^i$ is a timelike Killing vector such that $\xi^i\xi_i\equiv N^2(\mathbf{x})=0$ is the location of the horizon $\mathcal{H}$, then the matter energy is given by the integral of $dU=T_{ab}\xi^an^b\sqrt{\gamma}d^{D-1}x
=T_{ab}n^an^b\sqrt{-g}d^{D-1}x$. 
In this case, our variational principle can be thought of  as exremising just the \textit{gravitational} entropy in \eq{Sgrav} subject to two constraints: (i) $\delta(n_in^i)=0$  where $n^i$ is now the  velocity vector of static observers with $n_in^i=-1$ and (ii) the total matter energy $U$ is constant.
Implementing the constancy of $U$ under variation by a lagrange multiplier $\beta$ and
extremising $S-\beta U$, we can as usual identify $\beta$ with the range of time integration by analytic continuation from the Euclidean sector so that $\beta U$ becomes an integral over
$T_{ab}n^an^b\sqrt{-g}d^Dx$.
Also note that, when $n^i$ is a non-null vector, the identity in \eq{details1} becomes
\begin{equation}
\label{details2}
4P_{ab}^{\ph{a}\ph{b}cd} \D_cn^a\D_dn^b=4\D_c[P_{ab}^{\ph{a}\ph{b}cd} n^a\D_dn^b]+2\mathcal{R}_{ai}n^an^i
\end{equation}
which
allows us to work with an alternative definition of $S$ given by
\begin{equation}
S[n^a]\propto\int_\Cal{V}{d^Dx\sqrt{-g}}
    \left( 2\mathcal{R}_{ai}n^an^i\right) 
    =\beta\int{d^{D-1}x\sqrt{h}}
    \left( J_an^a\right)
\label{altS}
\end{equation}
where the second equality arises on replacing the time integration by multiplication by $\beta$ and using $\sqrt{-g}=N\sqrt{h},n^i=\xi^i/N$ along with the expression for Noether current in \eq{current1}. This result reinforces the idea that this expression is gravitational entropy.
In the context of Einstein's theory (with $\mathcal{R}_{ai}=R_{ai}$) this reduces to the expression used in ref.\cite{TPgravcqg04}. More details regarding this approach can be found in ref.\cite{equipartition,TPgravcqg04}.

Having determined the gravitational field equations, we will make a brief comment on the matter sector, before proceeding further. 
 In the conventional action principle, one will have a functional which depends on the gravitational degrees of freedom through the metric and on the matter degrees of freedom through the matter variables and we will vary both to get the equations of motion for gravity and matter. In maximizing the entropy we have only varied $n_a$.
 However,
 at the classical level, the  equations of motion for matter are already contained in the condition $\nabla_a T^{ab} =0$ which we have  imposed.  One can do quantum field theory in a curved spacetime  using these field equations in the Heisenberg picture. 
 Only in the context of   path integral quantization of the matter fields, one needs
 to exercise some care. In this case, we should  vary $n^a$ first  and get the classical equations for gravity
 because the expression in \eq{ent-func-2} is designed as an entropy functional.  
 But once we have obtained 
 the field equations for gravity, we can perform the usual variation of matter Lagrangian in a given curved spacetime and get the standard equations \cite{aseementropy}.

So far we have not fixed $P^{abcd}$ and so we have not fixed the theory. 
It is, however, possible to determine the form of $P^{abcd}$ using \eq{condd1}
which we shall now describe.

\subsection{The origin of \LL\ models}\label{sec:originofll}

In a
complete theory, the explicit form of $P^{abcd}$ will be determined by the
long wavelength limit of the microscopic theory just as the elastic
constants can --- in principle --- be determined from the microscopic
theory of the lattice. 
In the absence of such a theory, we need to determine $P^{abcd}$ by general considerations which is possible when $P^{abcd}$ satisfies \eq{ent-func-1}.
Since this condition is identically satisfied by \LL\ models which are known to be unique, our problem can be completely solved by taking the  $P^{abcd}$  as a series in the
 powers of  derivatives of
the metric as:
\begin{equation}
P^{abcd} (g_{ij},R_{ijkl}) = c_1\,\stackrel{(1)}{P}{}^{abcd} (g_{ij}) +
c_2\, \stackrel{(2)}{P}{}^{abcd} (g_{ij},R_{ijkl})  
+ \cdots \,,
\label{derexp}
\end{equation} 
where $c_1, c_2, \cdots$ are coupling constants with the $m$ th order term derived from the \LL\ Lagrangian:
\begin{equation}
\stackrel{(m)}{P}{}_{ab}^{cd}\propto
\AltC{c}{d}{a_3}{a_{2m}}{a}{b}{b_3}{b_{2m}}
\Riem{b_3}{b_4}{a_3}{a_4} \cdots
\Riem{b_{2m-1}}{b_{2m}}{a_{2m-1}}{a_{2m}} 
 =
\frac{\partial\LDm}{\partial R^{ab}_{{cd}}}\,. 
\label{zerothree}
\end{equation}
where $\AltC{c}{d}{a_3}{a_{2m}}{a}{b}{b_3}{b_{2m}}$ is the alternating tensor.
 The lowest order
term depends only on the metric with no derivatives. The next
term depends (in addition to metric) linearly on curvature tensor and the next one will be quadratic in curvature etc.
The lowest order term in \eq{derexp} (which  leads to Einstein's theory) is
\begin{equation}
\stackrel{(1)}{P}{}^{ab}_{cd}=\frac{1}{16\pi}
\frac{1}{2} \delta^{ab}_{cd} =\frac{1}{32\pi}
(\delta^a_c \delta^b_d-\delta^a_d \delta^b_c)
  \,.
\label{pforeh}
\end{equation}
so that when we use \eq{pforeh} for $P_{b}^{\ph{b}ijk}$, 
   \eq{ent-func-71} 
 reduces to Einstein's equations.
 The corresponding gravitational entropy functional
is:
 \begin{equation}
S_{\rm GR}[n^a]=\int_\Cal{V}\frac{d^Dx}{8\pi}
   \left(\D_an^b\D_bn^a - (\D_cn^c)^2 \right)
\end{equation}
That is, we can obtain the field equations in general relativity by varying
the vector fields $n^a$ in the above functional and demanding that the resulting equations hold for all null vector fields. 
 Interestingly, the integrand in $S_{GR}$ has the $Tr(K^2)-(Tr K)^2$ structure. If we think of the $D=4$ spacetime being embedded in a sufficiently large k-dimensional \textit{flat} spacetime we can obtain the same structure using the Gauss-Codazzi equations relating the (zero) curvature of k-dimensional space with the curvature of spacetime.
 As mentioned earlier, one can express any vector field $n^a$ in terms of a set of basis vector fields $e^a_A$. Therefore, one can equivalently think of the functional $S_{\rm GR}$ as given by 
 \begin{equation}
S_{\rm GR}[n^a_A]=\int_\Cal{V}\frac{d^Dx}{8\pi}
   \left(\D_an^b_I\D_bn^a_J - \D_cn^c_I\D_an^a_J \right)P^{IJ}
\end{equation}
where $P^{IJ}$ is a suitable projection operator. It is not clear whether the embedding approach leads to any better understanding of the formalism; in particular, it does not seem to generalize in a natural fashion to \LL\ models.

 The next order term (which arises from  the  Gauss-Bonnet Lagrangian) in \eq{twotw} is:
\begin{eqnarray}
\stackrel{(2)}{P}{}^{ab}_{cd}&=& \frac{1}{16\pi}
\frac{1}{2} \delta^{ab\,a_3a_4}_{cd\,b_3\,b_4}
R^{b_3b_4}_{a_3a_4} \nonumber\\
&=&\frac{1}{8\pi} \left(R^{ab}_{cd} -
         G^a_c\delta^b_d+ G^b_c \delta^a_d +  R^a_d \delta^b_c -
         R^b_d \delta^a_c\right) 
\label{pingone}
\end{eqnarray} 
and similarly for all the  higher orders terms. None of them can contribute in $D=4$ so we get Einstein's theory as the unique choice if we assume $D=4$. If we assume that $P^{abcd}$ is to be built \textit{only} from the metric, then this choice is unique in all $D$. 

\subsection{On-shell value of entropy functional}

The analysis so far used a variational principle based on the functional in 
\eq{ent-func-2}.  While the matter term in this functional has a natural interpretation
in terms of entropy transfered to the horizon, the interpretation of the gravitational part needs to be made explicit. The interpretation of $S_{\rm grav}$ as entropy  arises from the following two facts. First, we see from the identity \eq{details1} that this term differs from $2\mathcal{G}_{ij} n^i n^j$ by a total divergence. On the other hand, we have seen earlier that the term $2\mathcal{G}_{ij} n^i n^j$
can be related to the gravitational entropy of the horizon through the Noether current.
In fact, 
 \eq{thetrick} shows that when the equations of motion holds the total entropy of a bulk region is entirely on its boundary. Further
if we evaluate this boundary term
\begin{equation}
-S|_{\rm on-shell}=4\int_{\dV}{d^{D-1}xk_a\sqrt{h}\,\left(P^{abcd}n_c\D_bn_d\right)}
\label{on-shell-2}
\end{equation} 
(where we have manipulated a few indices using the symmetries of
$P^{abcd}$) 
in the case of a \textit{stationary} horizon which can be locally approximated as Rindler spacetime, one gets exactly the Wald entropy of the horizon \cite{aseementropy}.

 To prove  this, we will  use a limiting procedure and provide
a physically motivated choice of $n^a$ based on  the local Rindler frame. Making such a choice is necessary for two reasons. First, we do not expect the value of on-shell $S$ to have any direct physical meaning for a solution which does not have a horizon. So some choices have to be made. Second, we had already mentioned that the expression for $S$ is not invariant under the scaling $n^a\to f(x)n^a$. While this is irrelevant for obtaining the field equations, it does change the value of on-shell $S$. So we also need to have a prescription for normalization. We expect, however, to find sensible results when we evaluate this expression on a local Rindler approximation to the horizon which is what we shall do.

As usual, we shall introduce the LIF and LRF around an event and take  the normal to the stretched horizon (at $N=\epsilon$) to be $r_a$. In the coordinates used in \eq{surfrind}, we have the components:
\begin{equation} 
n_a=(0,1,0,0,...) ~~;~~ n^a=(0,1,0,0,...)
~~;~~ \sqrt{h}=\kappa \epsilon\sqrt{\sigma}\,,
\label{app-ent-1}
\end{equation}
where $\sigma$ is the metric determinant of the transverse  surface. This vector field $n^a$ is a natural choice for
evaluation of \eq{on-shell-2} if we evaluate the  integral on a surface with
$N=\epsilon=\,$constant, and take the limit $\epsilon\to 0$ at the end of the calculation. 
  In the integrand of \eq{on-shell-2} for the entropy functional, 
   we use
   $d^{D-1}x=dtd^{D-2}x_\perp$, and $\D_bn_d =
-\Gamma^a_{\ph{a}bd}n_a = -\Gamma^x_{\ph{a}bd}$, of which
only $\Gamma^x_{\ph{a}00}=\kappa^2\epsilon$ is nonzero. 
 The integrand for the $m-$th order term  in \eq{on-shell-2} can be
evaluated as follows:  
\begin{eqnarray}
\sqrt{h}\,k_a\left(4P^{abcd}n_c\D_bn_d\right) &=&
\kappa\epsilon\sqrt{\sigma}\left(4P^{x bx d}\D_bn_d\right)
=\kappa\epsilon\sqrt{\sigma}\left(
-4P^{x 0x 0}\Gamma^x_{\ph{a}00}\right) \nonumber\\
&=&\kappa^3\epsilon^2\sqrt{\sigma}\left(-4P^{x 0x 0}\right)
= \kappa^3\epsilon^2\sqrt{\sigma}\left(-4 m g^{00}g^{xx}
Q_{x 0}^{x 0}\right) \nonumber\\
&=& \kappa\sqrt{\sigma}\left(4 m Q_{x 0}^{x 0}\right)\,.
\label{app-ent-3}
\end{eqnarray}
where $Q^{abcd} = (1/m) P^{abcd}$.
Rest of the calculation proceeds exactly as from \eq{app-ent-4} to \eq{app-ent-8}  and we  find that \eq{on-shell-2} gives the horizon entropy.
This is a clear reason why we can think of $S$ as entropy.

 \subsection{Cosmological constant and gravity}\label{sec:cc}
 
 The approach outlined above has important implications for the \cc\ problem  \cite{ccreview} which we shall now briefly mention.
In the conventional approach, 
 we start with an action principle which depends on matter degrees of freedom and the metric 
 and vary (i) the matter degrees of freedom to obtain the equations of motion for
matter and (ii) the metric $g^{ab}$ to obtain the field equations of gravity.
The equations of motion for \textit{matter} remain invariant if one adds a constant,
say, $-\rho_0$ to the matter Lagrangian.
However, gravity breaks
this symmetry which the matter sector has and $\rho_0$ appears as a \cc\ term in the field equations of gravity. If we interpret the evidence for dark energy in the 
universe (see ref. \cite{sn}; for a critical look at data, see ref. \cite{tptirthsn1} and references therein)
as due to the cosmological constant, then its value has to be
fine-tuned  to satisfy the observational constraints.
It is not clear why a particular parameter in the low energy  sector has to be fine-tuned in such a manner. 

In the alternative perspective described here,
the functional in \eq{ent-func-2} is clearly invariant under the shift $L_m \to L_m - \rho_0$ or equivalently, $T_{ab} \to T_{ab} + \rho_0 g_{ab}$, 
since it only introduces a term $-\rho_0 n_a n^a =0$ for any null vector $n_a$.
In other words, one \textit{cannot} introduce the cosmological constant
as  a low energy parameter in the action in this approach. We saw, however, 
that the cosmological constant can re-appear as  \textit{an integration constant} when the equations 
are solved. The integration constants which appear in a particular solution  have a completely different conceptual status compared to the parameters which appear in the action describing
the theory. It is much less troublesome to choose a fine-tuned value for a particular integration constant in the theory if observations require us to do so.
From this point of view, the cosmological constant problem is considerably less severe
when we view gravity from the alternative perspective. 

This extra symmetry  under the shift 
$T_{ab} \to T_{ab} + \rho_0 g_{ab}$ arises because we are not treating metric  as a 
dynamical variable in an action principle.\footnote{It is sometimes claimed that a spin-2 graviton in the linear limit \textit{has to} couple to $T_{ab}$ in a universal manner, in which case, one will have the graviton coupling to the cosmological constant. In our approach, the linearized  field equations for the spin-2 graviton field $h_{ab}=g_{ab}-\eta_{ab}$, in a suitable gauge, will be $(\square h_{ab} -T_{ab})n^an^b=0$ for all null vectors $n^a$. This equation is still invariant under $T_{ab} \to T_{ab} + \rho_0 g_{ab}$ showing that the graviton does \textit{not} couple to cosmological constant.}
In fact one can state a stronger result \cite{TPdarkenergy1,TPdarkenergy2}.
Consider any model of gravity satisfying the following three conditions: (1)
The metric  is varied in a local action to obtain the equations of motion. (2) We demand full general covariance of the equations of motion. (3) The equations of motion for matter sector are invariant under the addition of a constant to the matter Lagrangian.  Then, we can prove a `no-go' theorem that the \cc\ problem cannot be solved in such model. That is, we cannot solve \cc\ problem unless we drop one of these three demands. Of these, we do not want to sacrifice general covariance encoded in (2); neither do we have a handle on low energy matter Lagrangian so we cannot avoid (3). So the only hope we have is to introduce an approach in which gravitational field equations are obtained by varying some degrees of freedom other than $g_{ab}$ in a maximization principle.
This suggests that the so called cosmological constant problem has its roots in our misunderstanding of the nature of gravity.
 
 Our approach   is not yet developed far enough to predict the value of the \cc. But providing a mechanism in which the \textit{bulk cosmological constant
decouples from gravity} is a major step forward.  It was always thought
that  some unknown symmetry should make the \cc\ (almost) vanish and weak (quantum gravitational) effects which break this symmetry could lead to its small value. \textit{Our approach provides a model which has such symmetry.} The small value of the observed \cc\ has to arise from non-perturbative quantum gravitational effects at the next order, for which we do not yet have a fully satisfactory model. (See, however, Ref. \cite{TPcccqg1,TPcccqg2}.)

\begin{figure}
	\begin{center}
	\includegraphics[scale=0.7]{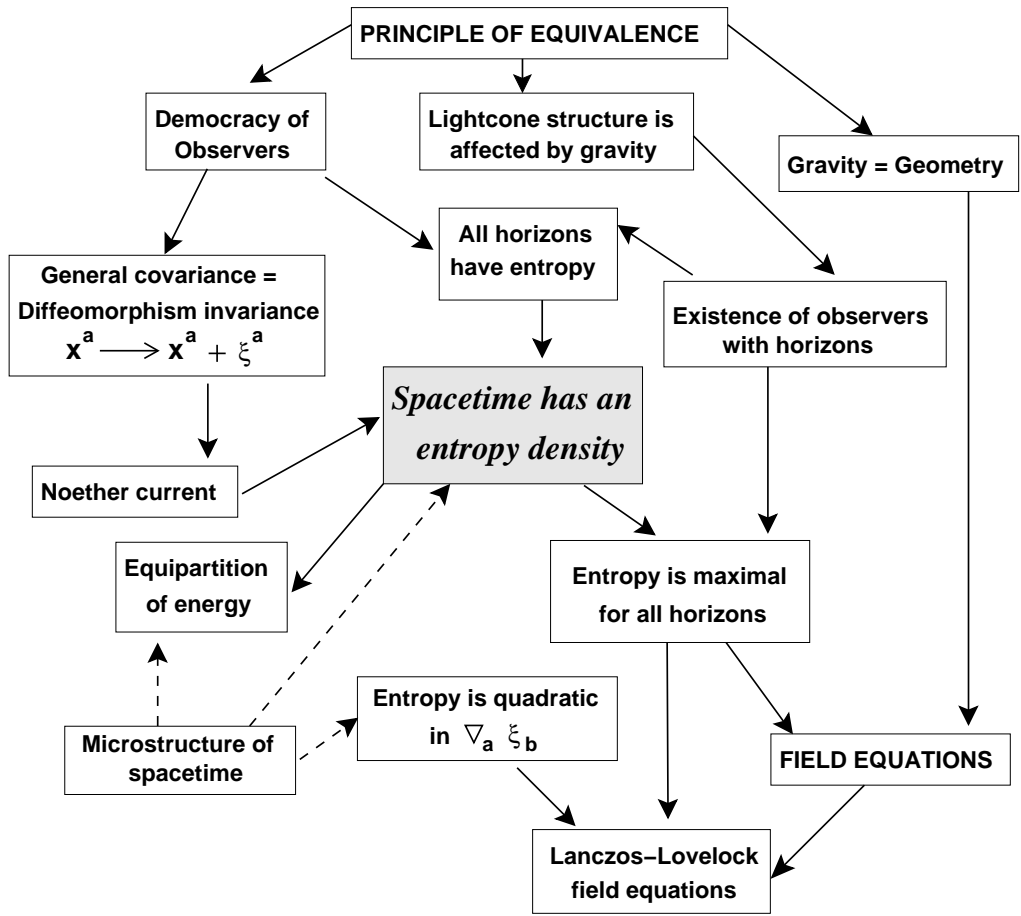}
	\end{center}	
\caption{Summary of the paradigm which interprets gravity as an emergent phenomenon; see text for discussion.}
\label{fig:synthesis}
\end{figure}  

\subsection{Thermodynamic route to gravity: Summary of the paradigm}

The paradigm described in this review is summarized in \fig{fig:synthesis}. The key idea is that the behaviour of bulk spacetime is similar to the behaviour of a macroscopic body of, say, gas and can be usefully described through thermodynamic concepts --- even though these concepts may not have any meaning in terms of the true microscopic degrees of freedom. This is exactly similar to the fact that, while one cannot  attribute entropy, pressure or temperature to a single molecule of gas, they are useful quantities to describe the bulk behaviour of large number of gas molecules. In such a paradigm, the field equations can be obtained by extremising the entropy expressed in terms of suitable variables. The motivation for such a thermodynamic route to gravity is amply demonstrated by the existence of local Rindler observers who perceive local horizons and thermal behaviour around \textit{any} event in spacetime and is summarised in the boxes in \fig{fig:synthesis} leading to the central theme: ``Spacetime has an entropy density" from the top.

The form of the entropy function encodes the information about dynamics and its extremisation leads to the field equations. We have described the specific forms of this function in different contexts along with their physical meaning and inter-relationship in the previous sections (see e.g.,  \eq{Sgrav}, \eq{altS} and \eq{gradentropy}). In the context of quadratic functionals, we are led to \LL\ model of gravity in general and  to Einstein gravity (uniquely) in $D=4$. This is indicated in the boxes at the bottom of \fig{fig:synthesis}.

In the complete description (statistical mechanics of the `atoms of spacetime') one should be able to obtain the form of the entropy in terms of the microscopic degrees of freedom (as indicated by dashed arrows in \fig{fig:synthesis}).  In the absence of such a theory, we are relying on a thermodynamic description in which the form of the entropy function that leads \cite{equipartition} to equipartition of energy among the degrees of freedom and to acceptable field equations. The leading term for entropy function in the correct theory is likely to be quadratic, thereby giving rise to \LL\ models, but in a full description we will also be able to calculate further corrections.

\section{Conclusions and outlook}\label{sec:conclusions}

This review concentrated, true to the title, several new insights which  have been gained regarding the thermodynamical aspects of gravity. It presented a case arguing that:
(a) The conventional approach in which one considers thermodynamical aspects of gravity as an interesting but subsidiary results of, say, doing quantum field theory in a spacetime with horizon, is fundamentally flawed. 
(b) There are several features of the theory which should be considered as strong hints favouring
a fundamental revision of our approach towards gravity and spacetime. These hints appear in the form of peculiar relationships, especially in the structure of the action functionals describing gravity, and in the generality of the thermal phenomena they represent.
(c) The last part of the review presented an alternate perspective which holds the promise for providing a more natural backdrop for understanding the relationship between gravitational dynamics and horizon thermodynamics.

It is useful to distinguish clearly (i) the mathematical results which can be rigorously proved from (ii) interpretational ideas which might evolve when our understanding of these issues deepen.

(i) From a purely algebraic point of view, without bringing in any physical interpretation or motivation, we can prove the following mathematical results: 
\begin{itemize}
\item
Consider a functional of null vector fields $n^a(x)$ in an arbitrary spacetime given by
\eq{ent-func-2} [or, more generally, by \eq{generalS}]. Demanding that this functional is an extremum for all null vectors $n^a$ leads to the field equations for the background geometry given by $(2\mathcal{G}_{ab}-T_{ab})n^an^b=0$ where $\mathcal{G}_{ab}$ is given by \eq{genEab1} [or, more generally, by \eq{genEab}]. Thus field equations in a wide class of theories of gravity can be obtained from an extremum principle without varying the metric as a dynamical variable.
\item
These field equations are invariant under the transformation 
$T_{ab} \to T_{ab} + \rho_0 g_{ab}$, which 
relates to the freedom of introducing a  \cc\ as an integration constant in the theory. Further, this symmetry forbids the inclusion of a cosmological constant  term in the variational principle by hand as a low energy parameter. \textit{That is, we have found a symmetry which makes the bulk \cc\ decouple from the gravity.} When linearized around flat spacetime, the graviton inherits this symmetry and does not couple to the \cc .
\item
On-shell, the functional in \eq{ent-func-2} [or, more generally, by \eq{generalS}] contributes only on the boundary of the region. When the boundary is a horizon, this terms gives precisely the Wald entropy of the theory.
\end{itemize}
 It is remarkable that one can derive not only Einstein's theory uniquely in $D=4$ but even \LL\ theory in $D>4$ from an extremum principle involving the null normals \textit{without varying $g_{ab}$ in an action functional!}.

(ii) To provide a physical picture behind these mathematical results, it is necessary to invoke certain effective degrees of freedom which can participate in the (observer dependent) thermodynamic interactions near any local patch of a null surface 
that acts as a horizon for certain class of observers. At present we have no deep understanding of how this comes about but, at a qualitative level, the physical picture is made of the following  ingredients:
\begin{itemize}
\item
Assume that the spacetime is endowed with certain microscopic degrees of freedom capable of 
exhibiting thermal phenomena. This is just the Boltzmann paradigm: \textit{If one can heat it, it
must have microstructure!}; and one can heat up a spacetime. 
\item
Whenever a class of observers perceive a horizon, they are ``heating up the spacetime''
and the  degrees of freedom close to a horizon  participate  
in a very \textit{observer dependent} thermodynamics. 
Matter which flows close to the horizon
(say, within a few Planck lengths of the horizon) transfers energy to these microscopic, near-horizon, degrees of freedom \textit{as far as the observer who sees the horizon is concerned}. Just as  entropy of a normal system at temperature $T$ 
will change by $\delta E/T$ when we transfer to it an energy $\delta E$, here also an entropy change will occur. (A freely falling observer in the same neighbourhood, of course, will deny all these!)
\item
We proved that when the field equations of gravity hold, one can interpret this entropy change in  a purely geometrical manner involving the Noether current.
From this point of view, the  normals $n^a$ to local patches of null surfaces are related to the (unknown) degrees of freedom that can participate in the thermal phenomena involving the horizon. These degrees of freedom seem to obey standard rules of thermodynamics, including equipartition.
\item
Just as demanding the validity of special relativistic laws with respect to all freely falling observers leads to the kinematics of gravity, demanding the local entropy balance in terms of the thermodynamic variables, as perceived by local Rindler observers, leads to the field equations of gravity in the form $(2\mathcal{G}_{ab}-T_{ab})n^an^b=0$. 
 
\end{itemize}

As stressed in earlier sections, this involves a new layer of observer dependent thermodynamics. 
In particular, since observers in different states of motion will have different regions of spacetime accessible to them --- for example, an observer falling into a black hole will not perceive a horizon in the same manner as an observer who is orbiting around it --- we are forced to accept that the notion of entropy is an observer dependent concept. 
 Fundamentally,  this is no different from the fact different freely falling observers will measure physical quantities differently comapared to non-geodesic observers; but in this case standard rules of special relativity allow us to translate the results between the observers. We do not yet have  a similar set of rules for quantum field theory in noninertial frames.
 It seems necessary to  integrate the entire thermodynamic machinery (involving what we usually consider to be the `real' temperature)
 with this notion of LRFs having their own (observer dependent) temperature. 
 
 This requires an intriguing relationship between quantum fluctuations and thermal fluctuations. As an illustration, consider the 
 relation $\delta E=T\delta S$ obtained in Sec. \ref {sec:obsdependence} for the one-particle excited state which has a curious consequence when we take the nonrelativistic limit \cite{equipartition}.   
 The mode function $e^{imc^2 t/\hbar} \langle 0| \phi(x)|1_{\bf k}\rangle$ corresponding to
a one-particle state in either inertial frame or Rindler frame goes over to a wave function $\psi(x)$ in the  nonrelativistic ($c\to \infty$), quantum mechanical, limit such that $\psi(x)$
 satisfies a Schrodinger equation with an accelerating potential $V=mgx$ when viewed from the second frame. In describing the motion of a wave packet corresponding to such a particle,
  the quantum mechanical averages will satisfy the relation $\langle \delta E \rangle = mg \langle \delta x\rangle = F \langle \delta x\rangle$. On the other hand, given the thermal description in the local Rindler frame,
  we would expect a relation like $\langle \delta E \rangle = T\Delta S$ to hold suggesting that the entropy gradient $\Delta S$ (due to the gradient  $\Delta n$ in the microscopic degrees of freedom) present over a region $\langle \delta x\rangle$  to give rise to a force $F = T \Delta S/\langle \delta x\rangle$. If one assumes that (i) $\Delta S/k_B$ has to be quantized (based on the results of ref.\cite{entropyideas}) in units of $2\pi$ and (ii) $\langle \delta x\rangle \approx \hbar/mc$ for a particle of mass $m$, then we reproduce $F=mg$ on using the Rindler temperature $k_BT = \hbar g/2\pi c$. Alternatively, if one assumes that the force $F = T \Delta S/\langle \delta x\rangle$
should be equal to $mg$, then the universality of the Rindler temperature for bodies with different $m$ arises if we use $\langle \delta x\rangle = \hbar / mc$. In this case --- which involves  
 the quantum mechanical limit of a one-particle state in a non-inertial frame ---
 we need to handle simultaneously both quantum and thermal fluctuations. The expression $F = T \Delta S/\langle \delta x\rangle$  
 demands
  an intriguing interplay between thermal fluctuations (in the numerator, $T\Delta S$, arising from the non-zero temperature and entropy in local Rindler frame) and the quantum fluctuations (in the denominator, $\langle \delta x\rangle$, related to the intrinsic position uncertainty $\hbar/mc$) for the theory to be consistent, including the choice of numerical factors.

At a conceptual level, this may be welcome when we note that 
every key progress in physics involved realizing that something we thought as absolute is  not absolute. With special relativity it was the flow of time and with general relativity it was the concept of global inertial frames and when we brought in quantum fields in curved spacetime it was the notion of particles and temperature.

 Many of these technical issues possibly  can be tackled in more or less straightforward manner, 
though the mathematics can be fairly involved. But  they may not be crucial to the alternative perspective or its further progress. The latter will depend on more serious conceptual issues, some of which are the following: 

(i) How come the microstructure of spacetime  exhibits itself indirectly through the 
horizon temperature even at scales much larger than Planck length and obeys an equipartition law (see e.g. \eq{idn} and ref. \cite{TPgravcqg04,equipartition})? This is possibly 
because the event horizon works as some kind of magnifying glass allowing us to probe
trans-Planckian physics \cite{Padmanabhan:1998jp,Padmanabhan:1998vr} but this notion needs to be made more precise. 

(ii) How does one obtain the expression for spacetime entropy density from some microscopic model? In particular, such an analysis  --- even with a toy model --- should throw more light on why normals to local patches of null surfaces play such a crucial role as effective degrees of freedom in the long wavelength limit. Of course, such a model
should also determine the expression for $P^{abcd}$ and get the metric tensor and spacetime as derived concepts - a fairly tall order!. (This is somewhat like obtaining theory of elasticity starting from a microscopic model for a solid, which, incidentally, is not a simple task either.)

\section*{Acknowledgments}

I thank
 Sunu Engineer, Dawood Kothawala, Aseem Paranjape,
 Apoorva Patel, Sudipta Sarkar    and    Kandaswamy Subramanian
 for several rounds of discussions over the past many years. 
Part of this review was written while I was visiting Department of Physics, University of Geneva in Oct 2009. I thank my host Ruth Durrer for hospitality.

\end{document}